\documentclass[12pt]{article}

\usepackage{amsmath}
\usepackage{amsfonts}
\usepackage{graphicx}
\usepackage{natbib}
\usepackage{url} 

\newcommand{\blind}{0}

 \usepackage{booktabs}
 \usepackage{subfig}
\usepackage{float}
\RequirePackage[colorlinks,citecolor=blue,urlcolor=blue]{hyperref} 
\usepackage{longtable,booktabs}
\usepackage[toc,page,title]{appendix}
\usepackage{multirow}
\usepackage{multicol}
\usepackage{lscape}
\usepackage{apalike}
\usepackage{epsfig, url}
\usepackage[super]{cite}

\newtheorem{theorem}{Theorem}

{\bf}{\it}

{\bf}{\it}
\newtheorem{corollary}{Corollary}

\newcommand{\E}{\mathrm{E}}
\newcommand{\Var}{\mathrm{Var}}
\newcommand{\Cov}{\mathrm{Cov}}
\newcommand{\Cor}{\mathrm{Cor}}
\renewcommand{\P}{\mathbb{P}}

\newcommand{\beginSupplFigTable}{
        \setcounter{table}{0}
        \renewcommand{\thetable}{S\arabic{table}}%
        \setcounter{figure}{0}
        \renewcommand{\thefigure}{S\arabic{figure}}%
     }

\begin{document}

\def\spacingset#1{\renewcommand{\baselinestretch}%
{#1}\small\normalsize} \spacingset{1}


\if0\blind
{
  \title{\bf    Accurate $p$-Value Calculation for Generalized Fisher's Combination Tests Under Dependence}
  \author{Hong Zhang \\
    Biostatistics and Research Decision Sciences, Merck Research Laboratories\\
    and \\
    Zheyang\ Wu \thanks{
    The authors gratefully acknowledge partial support by the NSF grants DMS-1309960 and DMS-1812082.}\\
    Department of Mathematical Sciences, Worcester Polytechnic Institute}
  \maketitle
} \fi

\if1\blind
{
  \bigskip
  \bigskip
  \bigskip
  \begin{center}
    {\LARGE\bf Title}
\end{center}
  \medskip
} \fi

\begin{abstract}
Combining dependent tests of significance has broad applications but the $p$-value calculation is challenging. Current moment-matching methods (e.g., Brown's approximation) for Fisher's combination test tend to significantly inflate the type I error rate at the level less than 0.05. It could lead to significant false discoveries in big data analyses. This paper provides several more accurate and computationally efficient $p$-value calculation methods for a general family of Fisher type statistics, referred as the GFisher. The GFisher covers Fisher's combination, Good's statistic, Lancaster's statistic, weighted Z-score combination, etc. It allows a flexible weighting scheme, as well as an omnibus procedure that automatically adapts proper weights and degrees of freedom to a given data. The new $p$-value calculation methods are based on novel ideas of moment-ratio matching and joint-distribution surrogating. Systematic simulations show that they are accurate under multivariate Gaussian, and robust under the generalized linear model and the multivariate $t$-distribution, down to at least $10^{-6}$ level. We illustrate the usefulness of the GFisher and the new $p$-value calculation methods in analyzing both simulated and real data of gene-based SNP-set association studies in genetics. Relevant computation has been implemented into R package {\it GFisher}. 
\end{abstract}

\noindent%
\vfill

\newpage
\spacingset{1.5} 

\section{Introduction}\label{sec:intro}

Combining tests of significance is a fundamental statistical procedure in global hypothesis testing. Such a procedure inputs a group of $p$-values, $P_1, \cdots, P_n$, which measure the significance of multiple tests. It forms a summary statistic and outputs a summary {\it test $p$-value} to determine the overall evidence against a global null hypothesis. For example, in meta-analysis each $P_i$ measures the statistical significance for the $i$th study. A $p$-value combination test summarizes the evidence from all $n$ studies against a global null hypothesis that none of the studies result a real positive outcome \citep{tseng2012comprehensive}. In the scenario of signal detection, each $P_i$ gives the significance from the $i$th potential ``signal" source; a $p$-value combination test can be used to test the global null that no real signals emerge. In the field of communication engineering, signal sources could be electronic sensors \citep{rago1996censoring}. In genetic studies, signal sources could be genetic variants that may or may not be associated with a trait \citep{hZhang2019TFisher}. 

Fisher's combination test is one of the oldest and most broadly applied $p$-value combination test \citep{Fisher1925}. Let $F_{d}$ be the cumulative distribution function (CDF) of $\chi^2_d$ distribution. Since $F^{-1}_{2}(1-x) =-2\log(x)$, Fisher's combination statistic can be viewed as the summation of $p$-values transformed by the chis-square inverse CDF:
\begin{equation}
\label{equ.TF}
T_F=\sum_{i=1}^n-2\log(P_i)=\sum_{i=1}^nF^{-1}_{2}(1-P_i). 
\end{equation} 
Under continuity and independence of $P_i$'s, $T_F$'s null distribution is simply $\chi^2_{2n}$. Besides the simplicity, $T_F$ enjoys many good properties. For example, it is asymptotically optimal (in the sense of Bahadur efficiency) when signals are homogenous over $i=1, \cdots, n$ under the alternative hypothesis \citep{littell1973asymptotic}. 

A challenge in real data analysis is that the $P_i$'s are often not independent. A classic model for describing such dependence is the Gaussian mean model (GMM) \citep{brown1975400, Hall2010}, which assumes that under the null hypothesis $P_i$'s come from a group of multivariate normal {\it input statistics} $Z_1, \cdots, Z_n$: 
\begin{equation}
\label{equ.GMM}
H_0: \mathbf{Z} = (Z_1, \cdots, Z_n)' \sim N(\mathbf{0}, \mathbf{\Sigma}). 
\end{equation}
This assumption is reasonable because in real data analysis the input statistics are often normal or asymptotically normal. Without loss of generality, we assume $Z_i$'s are standardized so that $\mathbf{\Sigma}_{n\times n} = (\sigma_{ij})_{1\leq i, j \leq n}$ is the correlation matrix with  $\sigma_{ij}=\Cor(Z_i,Z_j)$  and $\sigma_{ii}=1$. $\mathbf{\Sigma}$ is assumed known or estimable, but otherwise arbitrary in terms of its structure and value. Depending on the relevant scientific studies, the input $p$-values could be either one-sided or two-sided:
\begin{equation}
\label{equ.pvalues}
\text{ one-sided: }P_i=1-\Phi(Z_i); \text{ two-sided: } P_i=2\Phi(-|Z_i|) = 1-F_{1}(Z_i^2),
\end{equation}
where $\Phi(x)$ denotes the CDF of standard normal distribution. 

To address this correlated data, one could first de-correlate the input statistics to be $\mathbf{\Sigma}^{-\frac{1}{2}}\mathbf{Z}$, which goes back to independency. However, de-correlation often reduces ``signal strength" and thus lose power under a variety of alternative hypotheses 
\citep{zhang2018generalized, Hall2010}. Therefore, it is important to approximate the complicated null distribution of the test statistic under arbitrary correlation. The aim is to accurately calculate the test $p$-value for properly controlling the type I error rate $\alpha$. This paper targets on the statistical computation problem for calculating the distribution, not the re-sampling (e.g., simulation or permutation) based methods to generate empirical distribution. The latter is computationally expensive as well as innately limited in accuracy for controlling small $\alpha$. In particular, the smoothness of the empirical distribution curve is restricted by data variation \citep{routledge1997p}. For example, when the data have very small variation (e.g., in the scenario of analyzing rare genetic variants), the empirical distribution curve may not be smooth enough for accurate control of small $\alpha$. 

\begin{figure}[ht]
\centering
\includegraphics[width=0.49\textwidth]{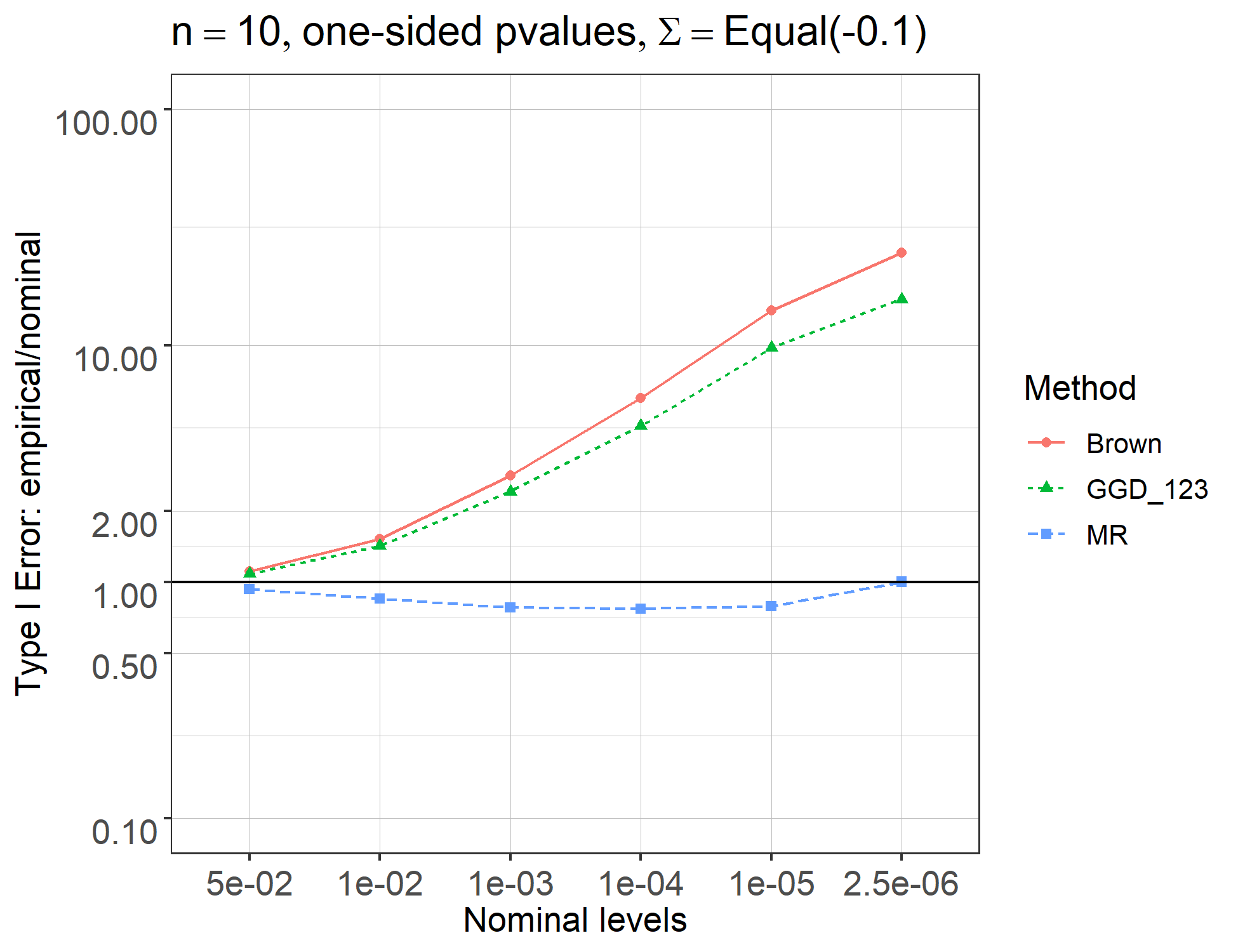}
\includegraphics[width=0.49\textwidth]{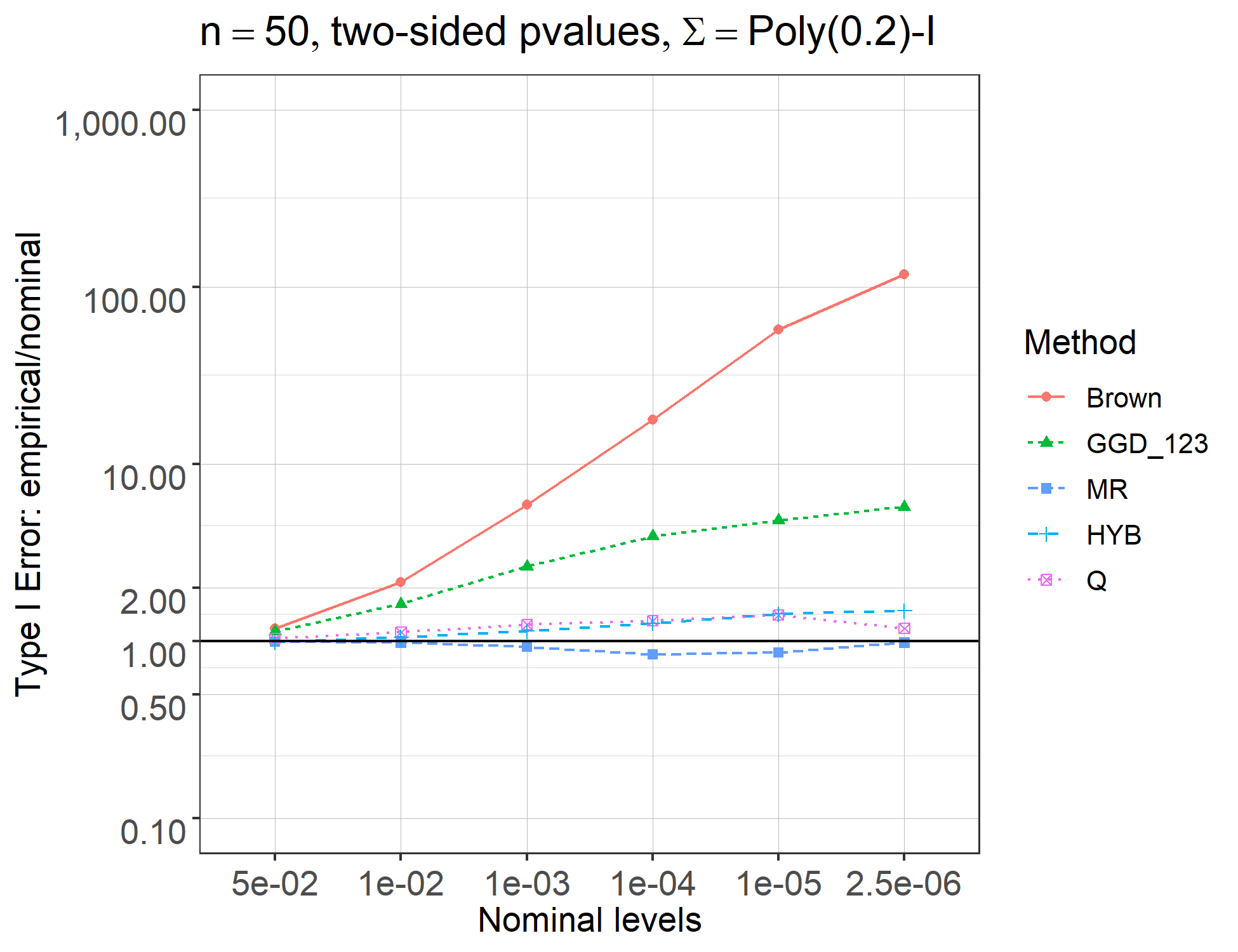}
\caption{Ratios between the empirical and the nominal type I error rates of $T_F$ by different methods.  Brown: Brown's approximation method; GGD\_123: Matching first three moments of the generalized gamma distribution; MR: The proposed moment-ratio matching method; Q: The proposed Q-approximation; HYB: The proposed hybrid method. The empirical type I error rates were obtained by $2\times 10^7$ simulations of GMM in (\ref{equ.GMM}). Left penal: $\mathbf{\Sigma}$ contains equal correlations of -0.1; Right: $\mathbf{\Sigma}$ is polynomial decaying in (\ref{equ.B}) with rate $\kappa=0.2$ in the upper left block.}
\label{fig.tie_brown}
\end{figure}

To calculate the distribution of $T_F$ under GMM with finite $n$, current methods are based on Brown's approximation \citep{brown1975400}, which is essentially a moment matching method. Specifically, assuming $T_F$ follows gamma distribution (GD) under $H_0$, the shape and scale parameters of GD are obtained by matching the mean $\E(T_F) = 2n$ and the variance $\Var(T_F)$. Various strategies have been proposed to approximate the value of $\Var(T_F)$ \citep{brown1975400, kost2002combining, poole2016combining, yang2016efficient}. However, even if $\Var(T_F)$ is perfectly obtained (actually its exact value can be efficiently calculated, as to be shown later), Brown's approximation fails to control small $\alpha$. Figure \ref{fig.tie_brown} illustrates the ratios between the empirical type I error rate (representing the true control level of false discoveries) and the nominal $\alpha$ in two examples. Brown's approximation consistently leads to inflated type I error because it yields significantly liberal test $p$-values (smaller than they should be). As to be shown in a systematic simulation study in Section \ref{sec:accu.GMM}, this problem is consistent over various settings, especially when the input $p$-values are two-sided. The inflation is also inherited when Brown's method is adopted to other Fisher type statistics such as Good's statistic \citep{good1955weighted, hou2005simple} and Lancaster's statistic \citep{lancaster1961combination, dai2014modified}.

A natural extension from the GD model is to assume that $T_F$ follows a more general distribution model, e.g., the generalized gamma distribution (GGD), in order to gain more flexibility in fitting the distribution curve \citep{li2014fisher}. 
However, as illustrated in Figure \ref{fig.tie_brown}, GGD approximation by matching its first three moments is still inadequate. Moreover, GGD requires solving more equations for more parameters, which not only demands a more complicated computation but also often ends up with no solutions.  

The limitations of the current methods raise two critical concerns in real applications. First, in the new era of big data it is very common that $\alpha$ needs to be small. For example, in the gene-based genome-wide association studies (GWAS), we need to simultaneously test the genetic association of about 20,000 human genes, the genome-wide significance level would be $0.05/20000 = 2.5\times 10^{-6}$. 
Second, two-sided input $p$-values are very commonly desired in real data analysis because the potential true signals often have unknown directionality.  

Before addressing these problems, this paper starts from an even broader perspective by extending the Fisher's combination statistic to a more general family of statistics in (\ref{equ.T}), referred as the GFisher. GFisher family covers many classic statistics, including Good's statistic \citep{good1955weighted}, Lancaster's statistic \citep{lancaster1961combination}, Z-score combinations \citep{wu2011rarevariant, chen2014optimally}, the Gamma Methods \citep{zaykin2007combining, chen2014new}, etc. GFisher statistics share the similar style of combining $p$-values by weighted summation over the inverse-chi-square transformations with arbitrary degrees of freedom. Therefore, they share the similar distribution computing problem, which can be resolved in a unified framework. From statistical application perspective, GFisher family provides a flexible and powerful tool for data analysis. The weights and degrees of freedom allow a general framework for incorporating useful prior information into the hypothesis testing procedure. Such data integrative analysis is particularly attractive to studying scientific problems from a system perspective, e.g., to promote biological discoveries by combining prior information from different sources \citep{tyekucheva2011integrating, chen2013hybrid, dai2016optimal}. 
At the same time, if no prior information is available, two omnibus tests are proposed to automatically adapt the weights and degrees of freedom to a given data. Such a data adaptive procedure likely helps to gain high and robust statistical power in various scientific researches. 

In this paper we aim at more accurately calculating the test $p$-values and thus properly controlling the type I error of the GFisher statistics.
Two main novel strategies are introduced. First, instead of treating GD or GGD as the ``true" distribution model (as was assumed in literature \citep{li2014fisher}), we consider them as a ``surrogate" distribution. We emphasize the tail property that is more important than the whole distribution for accurately calculating the test $p$-values. In the traditional methods, matching the first two (or three) moments is sufficient to get the whole distribution, but it is insufficient to control the tail behavior since the truth is likely not GD (or GGD). 
It is well-known that higher moments could better regulate the tail property \citep{jarque1980efficient, thadewald2007jarque}.
However, directly matching higher moments requires more complicated surrogate and involves difficulties in solving for more parameters. Instead, we propose to utilize moment-ratios \citep{vargo2010moment} to increase the versatility in distribution estimation. A moment-ratio matching approach is designed to engage both skewness and kurtosis. It significantly improves the accuracy while keeping computation simple (see Figure \ref{fig.tie_brown}). Interestingly, with this approach GD is shown sufficient to model the needed tail behavior; GGD does not provide extra improvement. Therefore, the computational complexity of dealing with GGD can be avoid with no loss of accuracy. 

As for the second novel strategy, instead of directly addressing the distribution of a GFisher statistic, we surrogate the joint distribution of the transformed input $p$-values (i.e., the summands of the GFisher statistic). A good approximation at the joint distribution level surely guarantees the distribution of the sum. Because it involves information matching in a higher dimension, in theory it provides a finer approximation. Specifically, we approximate the joint distribution of the summands by component-wisely matching their individual means and covariances.  In the scenario of two-sided input $p$-values, we deduce a quadratic form of Gaussian vectors to obtain a surrogate joint distribution. This Q-approximation method is fully analytical and thus computationally efficient. Moreover, higher moments of Q-approximation can be calculated by closed forms. Therefore, a hybrid method that combines the moment-ratio matching approach with the Q-approximated moments gives a computationally efficient solution. 

In order to cater for applications we illustrate the procedure of applying GFisher in the generalized linear model (GLM) based data analysis. Furthermore, we evaluate the robustness of the relevant $p$-value calculation methods under two scenarios. First, under GLM by either least-squares estimation or maximum-likelihood estimation, the input statistics may not be exactly normal as in (\ref{equ.GMM}), even though they are asymptotically so under some weak assumptions. The correlation matrix $\mathbf{\Sigma}$ also often needs to be estimated (they often converges in probability). The second robustness study is under the multivariate $t$-distributions with finite degrees of freedom \citep{kost2002combining}. This is a non-asymptotic scenario, and is more fundamentally different from GMM. 

The remainder of the paper is organized as follows. The GFisher family and its connection to classic statistics are introduced in Section \ref{sec:GFisher}. We discuss the existing and new approximation methods in Section \ref{sec:distn}. Section \ref{sec:accu.GMM} provides systematic simulations to evaluate the accuracy of the approximation methods. Section \ref{sec:appli} describes the linear-model based practical application procedure, and show the robustness of relevant methods. A genome-wide gene-based association study is illustrated in Section \ref{sec:GWAS} as a real data analysis example. Extra simulation results can be found in the Appendix.

\section{The GFisher}\label{sec:GFisher}

We extend Fisher combination statistic in (\ref{equ.TF}) to a more general family of statistics, referred as the GFisher. A GFisher statistic is a weighted sum of transformed input $p$-values by chi-square inverse CDF. Let $F_{d_i}^{-1}$ be the inverse CDF of $\chi_{d_i}^2$ with any given degrees of freedom $d_i > 0$. The statistic is defined by
\begin{equation}
\label{equ.T}
T=\sum_{i=1}^nw_iF_{d_i}^{-1}(1-P_i) = \sum_{i=1}^nw_iT_i,
\end{equation} 
where $T_i \equiv F_{d_i}^{-1}(1-P_i)$ denotes each transformed $P_i$. We assume that the weights $w_i\geq 0$, $i=1,...,n$, to avoid potential signal cancellation, and that the average weight $\bar{w} = \sum_i w_i/n=1$ without loss of generality. 
The smaller the $P_i$'s, the larger the statistic, and the more significant evidence against $H_0$. 

The GFisher is a rather broad family including many well known statistics. Clearly, Fisher's combination test statistic is a special case with fixed $d_i=2$ and $w_i =1$, $i = 1, \cdots, n$. Good's statistic is a weighted Fisher combination \citep{good1955weighted} with fixed $d_i=2$ and arbitrary $w_i$'s:
\begin{equation*}
T_G=\sum_{i=1}^n-2w_i\log(P_i) =\sum_{i=1}^n w_i F_{2}^{-1}(1-P_i).
\end{equation*}  
Lancaster's statistic \citep{lancaster1961combination} has arbitrary $d_i$'s and fixed $w_i=1$:
\begin{equation*}
T_L=\sum_{i=1}^nF_{d_i}^{-1}(1-P_i).
\end{equation*}  
When $d_i=1$, for the two-sided $P_i$'s in (\ref{equ.pvalues}), we have $F^{-1}_1(1-P_i) = F^{-1}_1(F_1(Z_i^2)) = Z_i^2$. Therefore, GFisher reduces to the statistics that combine the squared Z-scores of two-sided input $p$-values. These statistics include the unweighted version with $w_i=1$ \citep{chen2014optimally}, or the weighted version of arbitrary $w_i$'s: the Sequence Kernel Association Test (SKAT) under the default linear kernel \citep{wu2011rarevariant}. When $d_i\to\infty$, $(F_{d}^{-1}(x)-2d)/\sqrt{2d}\to\Phi^{-1}(x)$. For the one-sided $P_i$'s in (\ref{equ.pvalues}) we have $(F_{d}^{-1}(1-P_i)-2d)/\sqrt{2d} \approx \Phi^{-1}(\Phi(Z_i)) = Z_i$ for large $d$. Therefore, asymptotically GFisher can approach to the combinations of Z-scores of one-sided input $p$-values \citep{lancaster1961combination}. Again, these statistics include the unweighted version with $w_i=1$ (i.e., Stouffer's statistic \citep{Stouffer1949}), or the weighted version of arbitrary $w_i$ (i.e., Lipt\'{a}k's statistic \citep{liptak1958combination}).

Furthermore, GFisher can be considered as a combination of inverse gamma transformations with arbitrary parameters. This is because $w_iF_{d_i}^{-1}(x)=F_{G(d_i/2, 2w_i)}^{-1}(x)$, where $G(d_i/2, 2w_i)$ denotes a gamma distribution with shape parameter $d_i/2$ and scale parameter $2w_i$. Therefore, GFisher covers the Gamma Methods, which have $2w_i=1$ fixed and either fixed $d_i/2 = a$ \citep{zaykin2007combining} or varying $d_i/2=a_i$ \citep{chen2014new}. 
In fact, the Gamma Methods are essentially the Lancaster's test. This is because fixing scale parameters $2w_i=1$ means fixing weights $w_i = 1/2$, which is equivalent to setting constant $w_i=1$ in terms of the testing procedure.  

The practical benefit of GFisher is that it allows each input $p$-values being weighted by both $w_i$ and $d_i$. In many studies, e.g., for the problems of statistical experimental designs, 
it is desired to have proper weights instead of treating all input $p$-values equally \citep{berk1979asymptotically}. Furthermore, in the era of big data, it is an important strategy to combining information from heterogeneous data sources. The given $w_i$'s and $d_i$'s in the GFisher can be used to incorporate important prior information on the input $p$-values, which may correspond to different studies or signal sources. Applicational studies have shown that such strategy can significantly improve statistical power of the test. For example, in Lancaster's statistic the $d_i$'s have been used to incorporate genotype information to increase the power detecting novel disease genes \citep{dai2016optimal}. 

If the prior information about the choice of $d_i$ or $w_i$ is not available, the data-adaptive omnibus test is a good strategy to choose the ``best" weights over a variety of candidates for a given data. Omnibus test generally provides high and robust statistical power. For example, \citep{li2011adaptively} proposed to adapt $w_i$ in Good's statistic to increase statistical power of detecting differential gene expressions. They used re-sampling based method to control type I error rate $\alpha$. In the following we present two omnibus tests within the GFisher family. Both of their null distributions can be efficiently calculated based on the methods given in the next section. 
 
Let $\{(d_{1j},...,d_{nj}; w_{1j},...,w_{nj}), j=1,...,m\}$ be the set of $m$ candidate weighting schemes. An omnibus GFisher test statistic, referred as oGFisher, would select among the corresponding statistics: 
\begin{equation}
\label{equ.Tj}
T(j) = \sum_{i=1}^nw_{ij}F_{d_{ij}}^{-1}(1-P_i), \quad j=1, \cdots, m.
\end{equation}
A good choice for a given data should give a strong statistical significance measured by its test $p$-value. Denote $P(j)$ the test $p$-value of $T(j)$, $j=1, \cdots, m$. 
The traditional omnibus test takes the minimum test $p$-value as statistics (which we denote oGFisher\_minp):
\begin{equation}
\label{equ.minP}
minP = \min_{j=1,...,m}P(j).  
\end{equation}
The second omnibus test statistic follows the Cauchy combination of $P(j)$'s (which we denote oGFisher\_cc)  \citep{liu2018cauchy}: 
\begin{equation}
\label{equ.ccP}
ccP = \sum_{j=1}^m\tan\left(\left(\frac{1}{2}-P(j)\right)\pi\right)/m.
\end{equation}
Each summand of $ccP$ are the transformed $P(j)$'s by the inverse Cauchy CDF. Due to the heavy tail of Cauchy distribution, $ccP$ performs similar as $minP$. Meanwhile, $ccP$ possesses a significant computational advantage because its distribution depends little on the correlations among $T(j)$'s as long as they are pairwise normal. Since $T(j)$'s are in the format of summation, this condition is justifiable by the Central Limit Theorem (CLT) for large $n$.

\section{Distribution approximation methods}\label{sec:distn}

Under independence of input $p$-values, the null distribution of GFisher corresponds to the summation of independent gamma random variables with potentially different parameters. The calculation of such distribution has been developed \citep{mathai1982storage, moscuoroums1985distrlbution}. In this section we discuss approximation methods to compute the null distribution of GFisher under the dependence defined by (\ref{equ.GMM}) and (\ref{equ.pvalues}).

\subsection{Calculating covariances}

Since the GFisher statistic $T$ in (\ref{equ.T}) is a linear combination of $T_i$, $i=1,...,n$, it is important to obtain the covariances $\Cov(T_i, T_j)$, $1 \leq i,j \leq n$, which capture the dependence information among $P_i$'s. It is worth noting that under independence, the null distribution of $T$ is exactly the same for either one- or two-sided $P_i$'s in (\ref{equ.pvalues}). This is because in either way $P_i \sim \text{Uniform}(0, 1)$ for all $i$ under the null and $T_i$'s are independent. However, under correlated data the distribution of $T$ significantly differs for one- or two-sided $P_i$'s. This is because $T_i$'s would have different correlations and thus different joint distributions, even though their marginal distributions remain the same. 

Literature papers have devoted to estimate the covariances by resample \citep{poole2016combining} or by scatterplot fitting \citep{brown1975400, kost2002combining, yang2016efficient}. For example, for $T_F$ with one-sided $P_i$'s,  \citep{kost2002combining} fitted a cubic regression on $\Cov(-2\log P_i, -2\log P_j)$ (which is obtained by numerical double integration) 
at a grid values of $-0.98 \leq \sigma_{ij} \leq 0.98$ with the grid step 0.02. They recommended the formula 
\begin{equation}
\label{equ.cov.kost}
\Cov(-2\log P_i, -2\log P_j) \approx 3.263\sigma_{ij} + 0.710\sigma_{ij}^2 + 0.027\sigma_{ij}^3,
\end{equation}
which improved \citep{brown1975400}'s original stage-wise quadratic formula. A further refinement was provided by the same scatterplot fitting approach on the basis of $-0.99 \leq \sigma_{ij} \leq 0.99$ with grid steps of 0.01 (cf. \citep{yang2010distribution} equation (3)). Similarly, for $T_F$ with two-sided input $p$-values, \citep{yang2016efficient} 
recommended the formula:
\begin{equation}
\label{equ.cov.yang}
\Cov(-2\log P_i, -2\log P_j) \approx 3.9081\sigma_{ij}^2 + 0.0313\sigma_{ij}^4 + 0.1022\sigma_{ij}^6 - 0.1378\sigma_{ij}^8 + 0.0941\sigma_{ij}^{10}.
\end{equation} 
These analytical formulas directly calculate the covariances among $T_i = -2\log P_i$, $i = 1, \cdots, n$, for each given $\sigma_{ij}$ and thus is more computationally efficient than resampling based method. However, it requires carefully redo the fitting process for each different $d_i$ in $T_i=F_{d_i}^{-1}(1-P_i)$. 

In Theorem \ref{thm.cov} we provide a unified exact formula to efficiently calculate the covariances among any $T_i$'s. The proof follows Mehler's theorem \citep{patel1996handbook}. 
\begin{theorem}
\label{thm.cov}
Under (\ref{equ.GMM}) and (\ref{equ.pvalues}), let $T_i=F^{-1}_{d_i}(1-P_i)$ with $P_i=1-F(Z_i)$, $i=1,...,n$, where $F(x)\equiv \Phi(x)$ for one-sided $P_i$'s or $F(x)\equiv F_1(x^2)$ for two-sided $P_i$'s.  Then
\begin{align}
\label{equ.cov}
\Cov(T_i,T_j) = \sum_{k=1}^{\infty}\frac{\sigma^k_{ij}}{k!}I_i(k)I_j(k), \quad  i,j=1, \cdots, n, 
\end{align}
where $I_i(k)=\int_{-\infty}^{\infty} F^{-1}_{d_i}(F(z))H_k(z)\phi(z)d z$, $H_k$ denotes the $k$th order Hermite polynomial.
\end{theorem}

Note that $H_k(x)$ is an odd or even function when $k$ is an odd or even number, respectively. For two-sided $P_i$'s, $F(x)=F_1(x^2)$ is an even function, and therefore the covariances $\Cov(T_i,T_j)\geq 0$ always hold. However, for one-sided $P_i$'s, $F(x)=\Phi(x)$ is a monotone function, $\Cov(T_i,T_j)$ could be positive or negative depending on $\sigma_{ij}$. 

\begin{figure}[ht]
\includegraphics[width=0.5\textwidth]{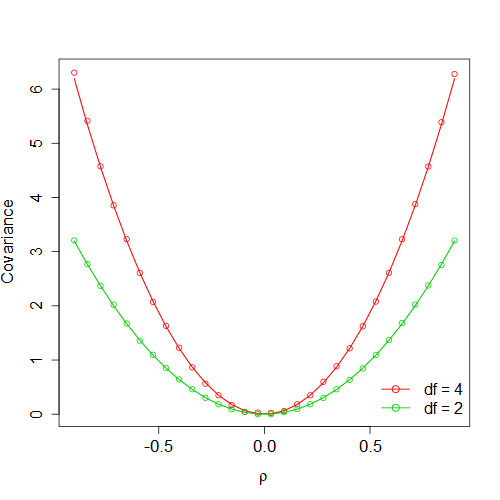}
\includegraphics[width=0.5\textwidth]{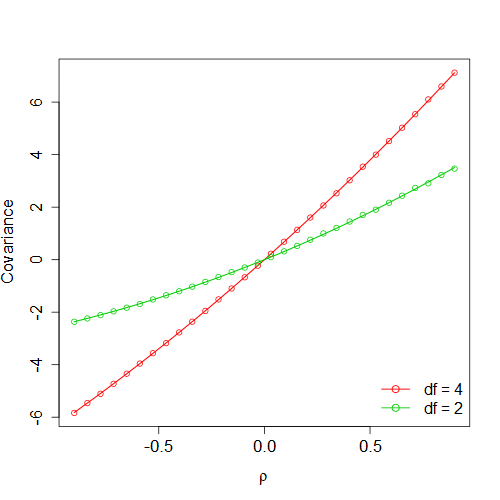}
\caption{$\Cov(T_i,T_j)$ as the function of $\sigma_{ij}=\rho$. Left: two-sided $P_i$'s. Right: one-sided $P_i$'s. Dots are obtained by $10^6$ simulations; curves are calculated by truncating equation (\ref{equ.cov}) with $k^*=2$.}
\label{fig.cov}
\end{figure}

Also note that as $k$ increases the summands in (\ref{equ.cov}) become quickly negligible. In practice we can safely ignore the summands with $k > k^*$ for some cutoff $k^*$. Figure \ref{fig.cov} shows that $k^*=2$ already gives satisfactory accuracy for $d_i=2$ and $4$ and both one- and two-sided $P_i$'s. 

When setting $d_i=2$ for $T_F$, Equation (\ref{equ.cov}) provides a theoretical justification to above formulas obtained by scatterplot fitting. In particular, the equation says that the true covariances should be polynomial instead of the original stage-wise formula given by \citep{brown1975400}. For one-sided $P_i$'s, setting the number of polynomial terms $k^*=3$, the coefficients obtained by (\ref{equ.cov}) agree with (\ref{equ.cov.kost}) up to three decimal digits. It indicates \citep{kost2002combining} has already done a great job and the space for further improvement, e.g., by \citep{yang2010distribution}, is very limited. Setting $k^*=5$ for two-sided $p$-values, the exact formula (\ref{equ.cov}) leads to  
\begin{align*}
\Cov(-2\log P_i, -2\log P_j) \approx 3.9068\sigma_{ij}^2 + 0.0506\sigma_{ij}^4 + 0.0173\sigma_{ij}^6 + 0.0082\sigma_{ij}^8 + 0.0046\sigma_{ij}^{10},
\end{align*}
which is somewhat different from to the fitted formula in (\ref{equ.cov.yang}) for the higher order terms. 
Meanwhile, as discussed above, it is often sufficient to set a smaller $k^*$. Therefore, the differences in coefficients of the higher order terms did not cause (\ref{equ.cov.yang}) being significantly inaccurate, even though its negative coefficient may not be theoretically justified. 

Equation (\ref{equ.cov}) is computational efficient. For getting all covariances $\Cov(T_i,T_j)$, $1 \leq i, j \leq n$, we only need to calculate $n^*k^*$ of $I_i(k)$ terms (i.e., univariate integrals), where $n^*$ is the number of distinct values among $d_1,...,d_n$. For any GFisher statistic with fixed $d_i = d$, e.g., Fisher's combination $T_F$ or Good's statistic $T_G$, we have $n^*=1$ and thus only need to calculate $k^*$ of univariate integrals.  

\subsection{Generalized Brown's method}\label{sec:BrownMeth}

Following the Brown's essential idea we term the generalize Brown's method (GB) as a unified approach to approximate the null distribution of GFisher statistics under (\ref{equ.GMM}) and (\ref{equ.pvalues}).  The method fits GD by matching the first two moments (mean and variance) of any given GFisher statistic. The GB covers literature works for $T_F$\citep{brown1975400, kost2002combining, poole2016combining, yang2016efficient},  $T_G$ \citep{hou2005simple}, $T_L$ \citep{dai2014modified}, etc. 


Specifically, for any GFisher statistic $T$ in (\ref{equ.T}), we have the mean $\mu_T = \E(T)=\sum_{i=1}^n w_i d_i$ and the variance $\sigma^2_T = \Var(T)$ calculated by Corollary \ref{coro.var}. 
\begin{corollary}
\label{coro.var}
Let $T$ be the GFisher statistic defined in (\ref{equ.T}). 
Following the same notations in Theorem \ref{thm.cov},
\begin{align}
\label{equ.var}
\sigma^2_T &= \sum_{ij}\Cov(w_iT_i,w_jT_j) = \sum_{ij}\sum_{k=1}^{\infty}\frac{\sigma^k_{ij}}{k!}w_iI_i(k)w_jI_j(k).
\end{align}
\end{corollary}

The null distribution of $T$ is then approximated by gamma distribution $G(a, \theta)$, 
which are calculated by matching the first two moments $\mu_T = a\theta$ and $\sigma^2_T = a\theta^2$:
\begin{equation}
\label{equ.GD-1-2}
a=\frac{\mu_T^2}{\sigma^2_T},\text{ }\theta=\frac{\sigma^2_T}{\mu_T}.
\end{equation}
The test $p$-value of $T$ is then approximated by 
\begin{equation}
\label{equ.pval.Brown}
\text{$p$-value }\approx1 - F_{G(a, \theta)}\left(T\right),
\end{equation}
where, $F_{G(a, \theta)}$ denotes the CDF of $G(a, \theta)$. Note that since $\theta$ is a scale parameter, the equation on $\theta$ in (\ref{equ.GD-1-2}) is actually redundant (i.e., no need to calculate $\theta$). That is, with matching the first two moments, an equivalent calculation to (\ref{equ.pval.Brown}) is 
\begin{equation}
\label{equ.pval.scale}
\text{$p$-value }\approx1 - F_{G(a, 1)}\left(\frac{T-\mu_T}{\sigma_T}\sigma_F + \mu_F\right), 
\end{equation}
where $\mu_F=a$ and $\sigma_F=\sqrt{a}$ are the mean and the standard deviation of $G(a, 1)$, respectively. 

As a natural extension one could assume $T$ follows a more general family of distributions, such as GGD with probability density function $f(x;a,\theta,p)=\frac{p/\theta^a}{\Gamma(a/p)}x^{a-1}e^{-(x/\theta)^p}$ with three parameters $a$, $\theta$, and $p$  \citep{stacy1962generalization}. 
Equipped with more parameters GGD should provide more flexibility to better fit the null distribution of $T$.
\citep{li2014fisher} proposed applying GGD through the maximum likelihood estimation (MLE) of related parameters based on resampling many $T_F$ values under $H_0$. This method is computationally very intensive, and in our experience the MLE process often fails to converge, especially when the input $p$-values are two-sided, $n$ is big, and/or the number of simulations is not considerably large. 
In order to exam the adequacy of GGD, in this paper we obtain the parameter values of $a$, $\theta$, and $p$ by matching the first three moments:
\begin{equation}
\label{equ.BM.GGD}
\E(T^k) = \theta^k\frac{\Gamma((a+k)/p)}{\Gamma(a/p)}, \text{ } k=1,2, \cdots .
\end{equation}
With $k = 1, 2, 3$, the moment matching method is equivalent to the MLE method if GGD is the true distribution of $T$ because both methods would reveal the right parameters. Regarding computation and implementation, moment matching is relatively easier than MLE. However, the GGD approximation is still computationally more challenging than the GD-based method. The roots of (\ref{equ.BM.GGD}) have no closed form; they require computationally expensive numerical solutions. Moreover, because the true null distribution of $T$ is likely not GGD in general, sometimes the roots are not obtainable (see more on the simulation results in Section \ref{sec:accu.GMM}). 

\subsection{Moment-ratio matching method}\label{sec:skratio}

To obtain more accurate small test $p$-values of $T$'s, we emphasize the tail behavior of its null distribution. 
The true distribution of $T$ are not GD nor GGD but they can be utilized to construct a surrogate distribution that has similar tail probability as $T$ does.   
For this purpose, we need to engage more and higher moments for better control of the tails. At the same time, a simpler distribution model is preferred because fewer parameters ease computation. 
To address this ``conflict", we propose a moment-ratio matching method that satisfied both desires -- using more moments while allowing the number of distribution parameters smaller than the number of moments. 

The moment-ratio matching method has two steps. First, we obtain the parameters of the surrogate distribution model by matching the ratio(s) of higher moments. 
Specifically, let $\gamma_T=E(T-\mu_T)^3/\sigma_T^3$ and $\kappa_T=E(T-\mu_T)^4/\sigma_T^4$ be the skewness and kurtosis of the targeting distribution of $T$'s, and $\gamma_F$ and $\kappa_F$ be the skewness and kurtosis of the surrogate distribution $G(a, \theta)$, respectively. By matching the ratio between skewness and excess kurtosis:
\begin{equation}
\label{equ.MR34}
\frac{\gamma_T}{\kappa_T-3} = \frac{\gamma_F}{\kappa_F-3}, 
\end{equation}
we get a closed forms for $a$:
\begin{equation}
\label{equ.MR34.GD}
a = \frac{9\gamma_T^2}{(\kappa_T-3)^2}. 
\end{equation}
The second step 
still follows the matching of the first two moments, and thus the $p$-value of $T$ is approximated by (\ref{equ.pval.scale}) with the $a$ value obtained in (\ref{equ.MR34.GD}). Note that essentially the moment-ratio matching method allow us to use a linear transformation of gamma distribution to estimate the distribution of $T$. That is, instead of using $G(a, \theta)$ to approximate $T$, we can use $bG(a, 1) + c$, where the parameters $a, b, c$ are estimated by involving all four moments. 

The moment-ratio method is a general idea that can be applied to any surrogate distribution, with potential adjustments.  For example, we can use GGD as the surrogate, except its three parameters need three equations to solve for.  We have implemented a GGD-based moment-ratio matching method, denoted by GGD\_MR, which solves for the parameters by equations $\mu_T = \mu_F$, $\sigma_T=\sigma_F$, and (\ref{equ.MR34}), where $\mu_F$ and $\sigma_F$ denote the mean and standard deviation of GGD, respectively. However, because no closed forms of the GGD parameters are available, more computation is needed to get numeric roots. Also, in general it does not provide noticeable improvement (and sometimes the performance could be even worse, see Section \ref{sec:accu.GMM} for more details). Therefore, for addressing GFisher in this paper we prefer GD rather than GGD due to simpler computation and satisfiable performance. It is also worth mentioning that different moment ratios could be applied. After considering possible combinations we got the best results by the ratios in (\ref{equ.MR34}) for our problem in hand. 

One computational challenge of this method is to obtain $\gamma_T$ and $\kappa_T$. Exact calculation is difficult because we don't yet have closed forms for the higher moments of $T_i$'s (like what we got in Theorem \ref{thm.cov}). Even if the higher moments of $T_i$'s are available (e.g., by computationally expensive high order integrations), following the similar type of summation as indicated by (\ref{equ.var}), calculating $\gamma_T$ and $\kappa_T$ will involve summations of a huge amount of high-order cross-product moments of $T_i$'s. The number of these summands is in the order of $O(n^m)$ for the $m$-th order of moments. For example, if $n=100$ input $p$-values are combined, the exact calculation of the $\kappa_T$ involves summation of tens of millions of the 4th-order cross-product moments of $T_i$'s. 

We propose two strategies to address this computational challenge. The first strategy is the traditional resample-based method -- the empirical estimate of $\gamma_T$ and $\kappa_T$ are to plug in (\ref{equ.MR34.GD}). Empirical estimates are obtained from randomly generated $T$ values by either simulation (based on the known GMM) or by permutation (e.g., in linear-model based real data analysis). This strategy is to get empirical estimate of parameters; we still rely on the GD model to approximate the test $p$-values. Therefore, it tolerates a smaller number of resampling processes than the strategy of getting the empirical $p$-value itself. The difference can be significant when the true $p$-value is very small. 
The second strategy is to estimate $\gamma_T$ and $\kappa_T$ by another easy-to-handle surrogate distribution of $T$. Follow this idea we will present a hybrid method after introducing the Q-approximation as the second surrogate under proper conditions.  

\subsection{Quadratic approximation}

We propose a new strategy for approximating the distribution of $T$ through the joint distribution of its summands $T_i$, $i=1,...,n$. 
That is, instead of directly surrogating the distribution of $T$ as a whole, we component-wisely surrogate the distributions of $T_i$'s, while matching the covariances among them. Since this approach involves information matching in a higher dimension, it provides a finer solution than Brown's approximation. 

Following this idea, here we propose a quadratic form approximation, referred as the Q-approximation, for GFisher statistics with integer $d_i$'s and two-sided $P_i$'s. 
Specifically, considering $T_i$'s jointly follow a sort of `multi-variate' chi-squared distribution, we construct a random vector $(Q_1,...,Q_n)'$ such that marginally $Q_i\overset{d}{=}T_i$ and pair-wisely $\Cov(Q_i,Q_j)=\Cov(T_i,T_j)$, $i,j=1,...,n$. Then the distribution approximation is 
$$
T = \sum_{i=1}^nw_iT_i \overset{d}{\approx}\sum_{i=1}^nw_iQ_i \equiv Q.
$$

In order to construct $Q_i$'s, we are motivated by a few special cases. 
First, when $d_i=1$, GFisher gives the weighted Z-squared test when $P_i$'s are  two-tailed:
$$
T_i = F^{-1}_{\chi^2_1}(1-P_i) = F^{-1}_{\chi^2_1}(F_{\chi^2_1}(Z_i^2)) = Z_i^2 \sim \chi^2_1. 
$$
If we define $Q_i = Z_i^2 = T_i$, then $Q=T$ exactly. 
Meanwhile, we can also construct $Q_i = Z_{i,(1)}^2 \sim \chi^2_1$, where $\mathbf{Z}_{(1)} = (Z_{1,(1)}, \cdots, Z_{n,(1)})' \sim N(0, \mathbf{M})$ denotes a random normal vector with correlation matrix $\mathbf{M}$ (the diagonal are 1's).  By matching the pair-wise covariance 
$$
\Cov(Q_i, Q_j) = 2M_{ij}^2 = \Cov(T_i, T_j) = 2\Sigma_{ij}^2,
$$
and keeping the same signs among the coordinate-wise elements in $\mathbf{M}$ and $\mathbf{\Sigma}$, we have $\mathbf{M}=\mathbf{\Sigma}$. Thus we have exactly the same distribution $Q\overset{d}{=}T$. 
 
Furthermore, considering Fisher's combination statistic $T_F$ with $d=2$, for the two-sided $P_i$'s we have
\begin{align*}
T_i = F^{-1}_{\chi^2_2}(1-P_i) = F^{-1}_{\chi^2_2}(F_{\chi^2_1}(Z_i^2))  \sim \chi^2_2.
\end{align*}
Based on the marginal chi-square distribution, we construct $Q_i = Z_{i,(1)}^2 + Z_{i,(2)}^2$, where $\mathbf{Z}_{(1)}$ and $\mathbf{Z}_{(2)}$ are iid multivariate Gaussian $N(\mathbf{0}, \mathbf{M})$. 
Similarly, for any GFisher statistic $T$ in (\ref{equ.T}) in general, to marginally surrogate $T_i \sim \chi^2_{d_i}$, $d_i \in \mathbb{N}^+$, $i=1, \cdots, n$, we construct 
\begin{equation*}
Q_i=\sum_{k=1}^{d_i}Z_{i,(k)}^2 \sim \chi^2_{d_i},
\end{equation*} 
where $\mathbf{Z}_{(1)}, \mathbf{Z}_{(2)}, \cdots, \mathbf{Z}_{(d_i)}$ are iid $N(\mathbf{0}, \mathbf{M})$. 

Straightforward calculation gives that 
\begin{equation*}
\Cov(Q_i, Q_j) = \sum_{k=1}^{\min\{d_i,d_j\}}\Cov(Z_{i,(k)}^2, Z_{j,(k)}^2) = 2\min\{d_i,d_j\}M_{ij}^2. 
\end{equation*}
By matching $\Cov(Q_i, Q_j)$ and $\Cov(T_i, T_j)$, we estimate the correlation matrix $\mathbf{M}$ by
\begin{equation*}
M_{ij}=sgn(\sigma_{ij}) \min \{ \sqrt{\frac{\Cov(T_i,T_j)}{2\min\{d_i,d_j\}}}, 0.99 \}, \quad i,j=1,...,n.
\end{equation*}
Note that $sgn(\sigma_{ij})$ guarantees $M_{ij}$ and $\sigma_{ij}$ have the same sign, so that $\mathbf{Z}_{(k)}$'s are as close to the original $\mathbf{Z}$ as possible. 
Furthermore, the fact that $\Cov(Q_i, Q_j) \geq 0$ is consistent to the fact that  $\Cov(T_i, T_j) \geq 0$  for two-sided $P_i$'s according to Theorem \ref{thm.cov}. However, numerical results show that in some rather extreme cases it could happen $\Cov(T_i,T_j)\geq2\min\{d_i,d_j\}$ when $\sigma_{ij}$ is large. In this case we let $M_{ij}=0.99$. Also, in case if the resulting matrix $\mathbf{M}$ is not positive-definite, we will find the nearest correlation matrix in terms of Frobenius norm (e.g., by simply using \textit{nearPD} function in the R package \textit{Matrix}) \citep{higham2002computing}. 

The exact distribution of $Q=\sum_{i=1}^nw_iQ_i$ is obtained based on the quadratic form of iid standard normal variables. Specifically, we can rewrite $\mathbf{Z}_{(k)}=\mathbf{M}^{1/2}\mathbf{U}_{(k)}$, $k=1, \cdots, d^*$, where $\mathbf{U}_{(k)}$'s are iid standard normal random vectors, $d^*=\max_i d_i$ is the maximum degrees of freedom, and $\mathbf{M}^{1/2}$ is the lower-triangular matrix from Cholesky decomposition such that $\mathbf{M}^{1/2}(\mathbf{M}^{1/2})^\prime =\mathbf{M}$. Let $\mathbf{W}=\text{diag}\{w_1,...,w_n\}$ be the diagonal matrix of the weights. Then 
\begin{equation*}
Q=\sum_{i=1}^nw_i \sum_{k=1}^{d_i}Z_{i,(k)}^2 = \sum_{k=1}^{d^*}\mathbf{U}_{(k)}^\prime \mathbf{A}_k \mathbf{U}_{(k)},
\end{equation*}
where for each $k = 1, \cdots, d^*$, 
\begin{align*}
\mathbf{A}_k = \mathbf{W}^{1/2}(\mathbf{M}^{1/2})^\prime \text{diag}\{b_k(1), \cdots, b_k(n)\}\mathbf{M}^{1/2}\mathbf{W}^{1/2},
\end{align*}
with $b_k(l)=I_{\{d_l\geq k\}}$, $l=1,...,n$, $I_{\{\cdot\}}$ is the indicator function of an event $\{\cdot\}$. 

By eigendecomposition, the distribution of $Q$ is
\begin{equation}
\label{equ.QinLambda}
Q \overset{D}{=} \sum_{i=1}^n\sum_{k=1}^{d_i}\lambda_{ik} U^2_{i, (k)},
\end{equation}
where $U^2_{i, (k)} \overset{iid}{\sim} \chi_1^2$ for all $i=1, \cdots, n, k=1, \cdots, d^*$. 
For each $k$, $\lambda_{1k}, \cdots, \lambda_{nk} \geq 0$ are the eigenvalues of $\mathbf{A}_k$, and the number of non-zero eigenvalues is equal to the number of non-zero elements in $\{b_k(1), \cdots, b_k(n)\}$. Numerically, the CDF of (\ref{equ.QinLambda}) can be found by inverting its characteristic function \citep{davies1980algorithm,imhof1961computing}.  

It is worth mentioning that the Q-approx is not proper for the scenario of one-sided input $p$-values. For example, when $d=1$ we have
$T_i = F^{-1}_{\chi^2_1}(1-P_i) = F^{-1}_{\chi^2_1}(\Phi(Z_i)) \neq Z_i^2$. Even if their marginal distribution still follows the same $\chi^2_1$ (since $\Phi(Z_i) \sim \text{Uniform}(0, 1)$), their joint distributions are quite different. Moreover, when $P_i$'s are one-sided $\Cov(T_i, T_j)$ could be negative, which is no longer consistent with the fact that $\Cov(Q_i, Q_j) \geq 0$.
Overall, the correlation structure among $T_i$'s are quite different when the input $p$-values are one-sided from that when they are two-sided. 

\subsection{Hybrid approximation} 

When the input $p$-values are two-sided, we can apply a hybrid method that combines the moment-ratio matching method and the Q-approximation. The higher moments of $Q$ can be analytically calculated. Therefore, in the moment-ratio matching procedure we can surrogate $\gamma_T$ and $\kappa_T$ by $\gamma_Q$ and $\kappa_Q$, respectively. This hybrid approximation is fully analytical and thus is efficient. 
 
Specifically, the $t$th cumulant of $Q$ can be written as 
\begin{equation*}
c_t = 2^{t-1}(t-1)!\sum_{i=1}^n\sum_{k=1}^{d_i}\lambda_{ik}^t.
\end{equation*}
Accordingly, the higher moments of $Q$ can be calculated in closed form, i.e.,
\begin{align*}
\gamma_Q &=c_3/c_2^{3/2}; \quad \kappa_Q=c_4/c_2^2 + 3.
\end{align*}
Based on the GD model $G(a, \theta)$,  following the moment-ratio matching in (\ref{equ.MR34.GD}) we get  
\begin{equation*}
a=\frac{(\sum_{ik}\lambda_{ik}^2)(\sum_{ik}\lambda_{ik}^3)^2}{2(\sum_{ik}\lambda_{ik}^4)^2}. 
\end{equation*}

The idea can be straightforwardly extended to the GGD model based on in (\ref{equ.BM.GGD}). However, GD is still preferred because of its satisfiable accuracy and easier computation.  Again, the hybrid method is not proper for one-sided input $p$-values due to the limitation of Q-approximation. 

\subsection{Distribution of oGFisher}

The oGFisher tests utilize $T(j)$ in (\ref{equ.Tj}), $j = 1, \cdots, m$,  to get the summary test $p$-value. 
The covariances among $T(j)$'s are given in Corollary \ref{coro.var.T(j)} following Theorem \ref{thm.cov}. 
\begin{corollary}
\label{coro.var.T(j)}
Let $T(l), T(r)$ defined in (\ref{equ.Tj}), $1\leq l\leq r\leq m$. Following the same notations in Theorem \ref{thm.cov}, we have
\begin{align}
\label{equ.Tcov}
\Cov(T(l), T(r)) &= \sum_{ij}\Cov(w_iT_i(l),w_jT_j(r)) = \sum_{ij}\sum_{k=1}^{\infty}\frac{\sigma^k_{ij}}{k!}w_{il}I_{il}(k)w_{jr}I_{jr}(k), 
\end{align}
where $I_{il}(k)=\int_{-\infty}^{\infty} F^{-1}_{d_{il}}(F(z))H_k(z)\phi(z)d z$, and $I_{jr}(k)$ is similarly defined. 
\end{corollary}

For the omnibus statistic $minP$ in (\ref{equ.minP}), we can apply asymptotic distribution to approximate its test $p$-value. This analytical approximation avoids computationally intensive simulation or permutation that was often implemented in literature \citep{li2011adaptively}. 
Specifically, since $T(j)$'s are all summations of $T_i$'s, $i=1, \cdots, n$, for large $n$ by the CLT they are asymptotically multivariate normal distribution:
\begin{equation}
\label{equ.T(j)normal}
(T(1), \dots, T(m))' \overset{D}{\approx} N(\mathbf{\mu}, \mathbf{\Omega}), 
\end{equation}
where $\mu_j = \sum_{i=1}^n w_{ij}d_{ij}$ and $\mathbf{\Omega}_{lr} = \Cov(T(l), T(r))$ is given in (\ref{equ.Tcov}). 
At any fixed value $p_o \in (0, 1)$, as $n \to \infty$, the null CDF of $minP$ statistic is 
$$
\P(minP \geq p_o | H_0) = (1+o(1)) \P(\frac{T(j) - \mu_j}{\sqrt{\mathbf{\Omega}_{jj}}} \leq \bar{\Phi}^{-1}(p_o), \text{ for all } j=1,\dots,m),
$$
where $\bar{\Phi}(x) = 1 - \Phi(x)$ is the survival function of $N(0, 1)$. 
Therefore, the test $p$-value of oGFisher\_minp at an observed statistic $minp_o$ is 
\begin{equation*}
\P(minP < minp_o | H_0) \to 1 - \Phi_{\mathbf{R}}(\bar{\Phi}^{-1}(minp_o), \cdots, \bar{\Phi}^{-1}(minp_o)),
\end{equation*}
where $\Phi_{\mathbf{R}}$ denotes the CDF of a multivariate normal distribution with mean zero and correlation matrix $\mathbf{R}=\mathbf{\Lambda\Omega\Lambda}$, where the diagonal matrix $\mathbf{\Lambda}=\text{diag}(1/\sqrt{\mathbf{\Omega}_{ii}})_{1\leq i\leq n}$. The multivariate normal probabilities can be efficiently computed, e.g., by \citep{genz1992numerical}. 
Note that in theory normal distribution can also be used to calculate $p$-values $P(j)$'s and the observed $minp_0$. However, under finite $n$ the $minp_o$ obtained by normal approximation will be almost certainly smaller than the value obtained by the exact CDF of GFisher (because the normal tail reduces faster). Therefore, the normal approximation approach could be more inflated than our proposed methods to get $P(j)$'s. 

As for the Cauchy combination omnibus test (oGFisher\_cc) statistic in (\ref{equ.ccP}), because of the asymptotic normality of $T(j)$'s in (\ref{equ.T(j)normal}), we can directly apply the result by \citep{liu2018cauchy} and approximate its test $p$-value by 
\begin{equation*}
\P(ccP > ccp_o | H_0) \approx \frac{1}{2} - \tan^{-1}(ccp_o)/\pi. 
\end{equation*}

\section{Accuracy under GMM}\label{sec:accu.GMM}

In this section we systematically compare relevant $p$-value calculation methods through simulations of GMM in (\ref{equ.GMM}). A summary of all settings and methods are given first, followed by results and observations.  

Regarding the correlation matrix $\mathbf{\Sigma}$, two correlation patterns are considered: equal and polynomial-decaying correlations, for representing dense and sparse correlation patterns, respectively. Specifically, define $m\times m$ equal correlation matrix and polynomial-decaying correlation matrix, respectively:
\begin{align}
\label{equ.equ-decay-matrix}
\mathbf{A}_m(\rho) &: \mathbf{A}_m(i,j)=\rho, \quad{} 1\leq i\neq j\leq m \text{ and }  0\leq\rho<1, \\
\label{equ.B}
\mathbf{B}_m(\kappa) &: \mathbf{B}_m(i,j)=1/|i-j|^\kappa, \quad{} 1\leq i\neq j\leq m \text{ and } \kappa>0.
\end{align} 
The parameters $\rho$ and $\kappa$ control correlation strength.  
We also consider that $\mathbf{\Sigma}$ may follow certain block-wise structures, which are often interested in practice (e.g., the haplotype blocks in genetics). Let $n$ be the number of input statistics, $\mathbf{\Sigma}$ be a $2\times 2$ block matrix 
\[
\mathbf{\Sigma} = 
\begin{bmatrix}
\mathbf{\Sigma}_{11} & \mathbf{\Sigma}_{12} \\
\mathbf{\Sigma}'_{12} & \mathbf{\Sigma}_{22} 
\end{bmatrix},
\]
where each block is a $(n/2)\times (n/2)$ matrix. Besides the identity matrix (for independence case), totally 12 structures of $\mathbf{\Sigma}$ in Table~\ref{tbl.sigma} are considered. Parameters $\rho=0.1,0.5,0.9$ and $\kappa=0.2, 1, 3$ are used in the simulations. The numbers of input $p$-values (i.e., the correlation dimensions) are $n=10$, $20$ and $50$. Both one- and two-sided input $p$-values in (\ref{equ.pvalues}) are simulated. 

\begin{table}[]
\centering
\caption{Correlation structures involved in GMM simulations based on (\ref{equ.equ-decay-matrix}) and (\ref{equ.B}). }
\label{tbl.sigma}
\begin{tabular}{@{}llll@{}}
\toprule
Type                 & I (Upper Left) & II (Diagonal Blocks) & III (All Blocks) \\ \midrule
Equal($\rho$)      & $\mathbf{\Sigma}_{11}= \mathbf{A}_{n/2}(\rho)$           & $\mathbf{\Sigma}_{11} = \mathbf{\Sigma}_{22}= \mathbf{A}_{n/2}(\rho)$                    & $\mathbf{\Sigma}= \mathbf{A}_{n}(\rho)$                \\
Poly($\kappa$)     & $\mathbf{\Sigma}_{11}= \mathbf{B}_{n/2}(\kappa)$              & $\mathbf{\Sigma}_{11} = \mathbf{\Sigma}_{22}= \mathbf{B}_{n/2}(\kappa)$                   & $\mathbf{\Sigma}= \mathbf{B}_{n}(\kappa)$               \\
Inv-Equal($\rho$)*  & $\mathbf{\Sigma}_{11}= \mathbf{A}^{-1}_{n/2}(\rho)$             & $\mathbf{\Sigma}_{11} = \mathbf{\Sigma}_{22}= \mathbf{A}^{-1}_{n/2}(\rho)$                    & $\mathbf{\Sigma}= \mathbf{A}^{-1}_{n}(\rho)$                \\
Inv-Poly($\kappa$)* & $\mathbf{\Sigma}_{11}= \mathbf{B}^{-1}_{n/2}(\kappa)$              & $\mathbf{\Sigma}_{11} = \mathbf{\Sigma}_{22}= \mathbf{B}^{-1}_{n/2}(\kappa)$                    & $\mathbf{\Sigma}= \mathbf{B}^{-1}_{n}(\kappa)$ \\ \bottomrule
\multicolumn{4}{l}{\footnotesize{*In these cases $\mathbf{\Sigma}$ is standardized to become a correlation matrix.}}
\end{tabular}
\end{table}

The following GFisher statistics are implemented: 1) $d_i\equiv 1$, i.e. SKAT with linear kernel; 2) $d_i \equiv 2$, i.e. Fisher's combination and 3) $d_i\equiv 3$. The oGFisher\_minp and oGFisher\_cc adapt to these three $d_i$ values. A case of varying weights is also considered: $d_i=i$ and $w_i=2i/(n+1)$, $i=1, \cdots, n$. 

The following $p$-value calculation methods are compared: generalized Brown's method (GB for short); the moment-ratio matching method (MR) with empirical skewness and kurtosis obtained by $10^5$ simulated statistics; the Q-approximation (Q); the hybrid method (HYB); and the GGD based moment-matching methods. We denote GGD\_123 the method that matches the first three moments following equations in (\ref{equ.BM.GGD}) with $k = 1, 2, 3$. We further considered matching even higher moments: GGD\_234 denotes such a method that solves for the GGD parameters by matching the variance, skewness and kurtosis, following (\ref{equ.BM.GGD}), and adjusted the mean similarly in (\ref{equ.pval.scale}). GGD\_234 is equivalent to matching the first four moments of GGD$(a,\theta,p)+c$. GGD\_MR denotes the moment-ratio matching method based on a GGD surrogate that uses matching equations $\mu_T = \mu_F$, $\sigma_T=\sigma_F$, and (\ref{equ.MR34}).

We illustrate the accuracy from two different perspectives. First, the survival curve, i.e., the curve of right tail probability, demonstrates the closeness of the overall distributions over the majority the domain. We considered the range from 0 to 0.9999 quantiles. Second, the empirical type I error rate is defined as the proportion of the calculated $p$-values of $2\times 10^7$ simulated $T$ statistics that are smaller than the nominal levels $\alpha$ = 5e-2, 1e-2, 1e-3, 1e-4, 1e-5, and 2.5e-6. 

\begin{figure}
\subfloat[Fisher; $\mathbf{\Sigma}=$ Equal(0.7)-I]{\includegraphics[width=0.5\textwidth]{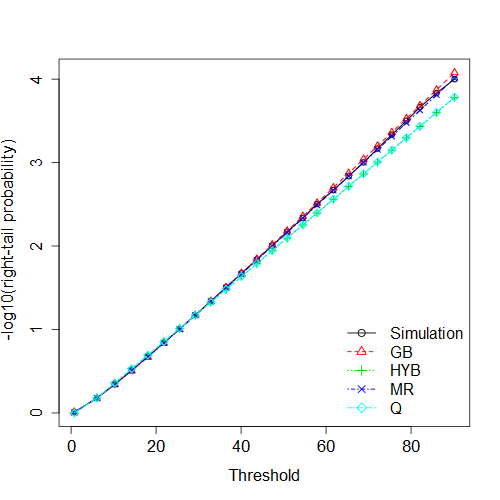}}
\subfloat[Fisher; $\mathbf{\Sigma}$= Equal(0.7)-III]{\includegraphics[width=0.5\textwidth]{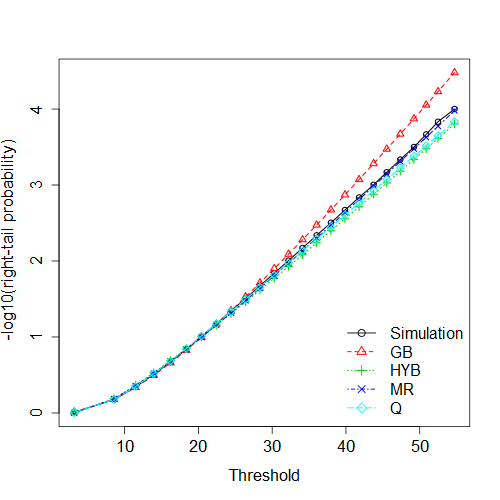}}\\
\subfloat[GFisher $d_i=i$, $w_i=2i/7$; $\mathbf{\Sigma}=$ Equal(0.7)-I]{\includegraphics[width=0.5\textwidth]{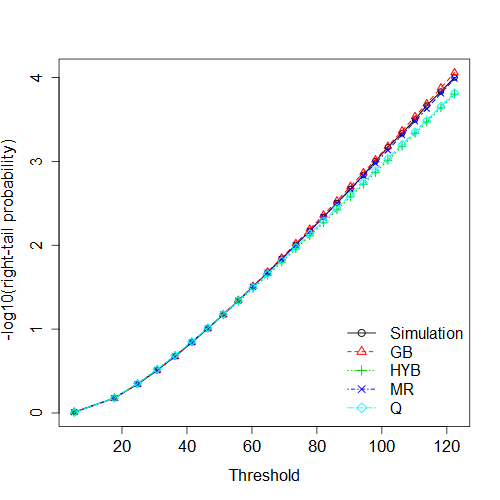}}
\subfloat[GFisher $d_i=i$, $w_i=2i/7$; $\mathbf{\Sigma}$= Equal(0.7)-III]{\includegraphics[width=0.5\textwidth]{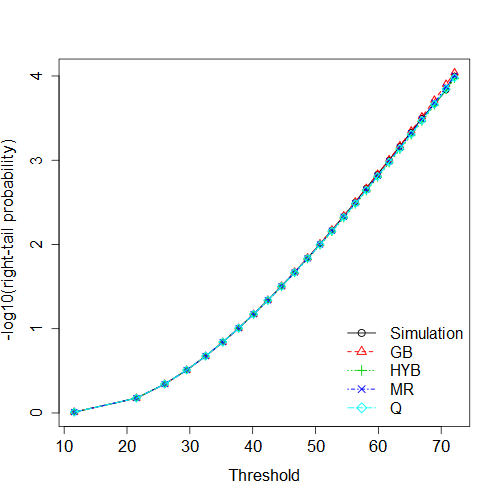}}
\caption{Right-tail probability ($-\log_{10}$) of $T$ when input $p$-values are one-sided. $n=6$. Two $\mathbf{\Sigma}$ patterns with $\rho=0.7$ are defined in Table \ref{tbl.sigma}.
}
\label{fig.survival_oneside}
\end{figure}

\begin{figure}
\subfloat[Fisher; $\mathbf{\Sigma}=$ Equal(0.7)-I]{\includegraphics[width=0.5\textwidth]{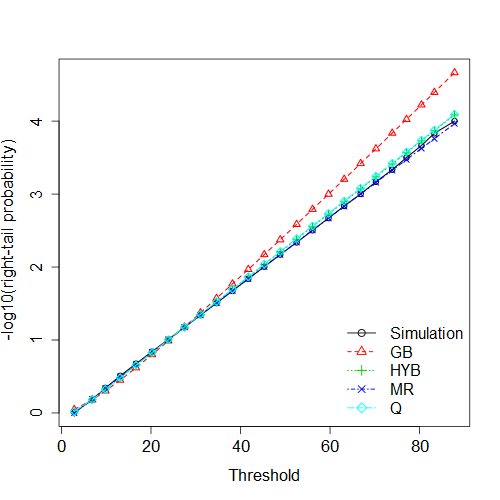}}
\subfloat[Fisher; $\mathbf{\Sigma}$= Equal(0.7)-III]{\includegraphics[width=0.5\textwidth]{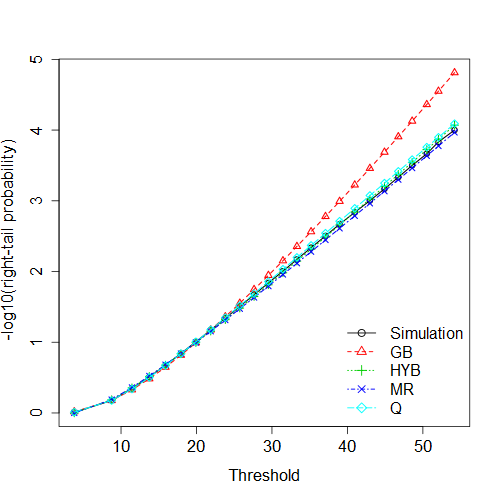}}\\
\subfloat[GFisher $d_i=i$, $w_i=2i/7$; $\mathbf{\Sigma}=$ Equal(0.7)-I]{\includegraphics[width=0.5\textwidth]{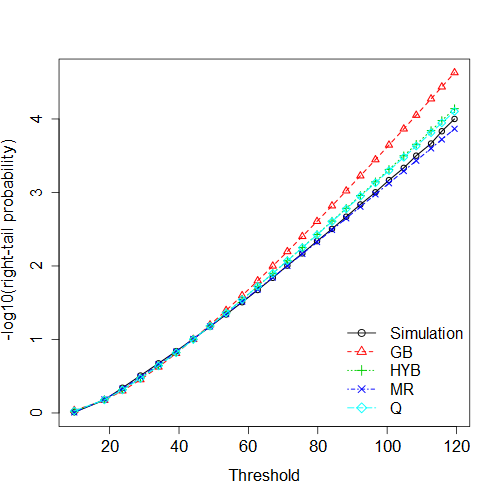}}
\subfloat[GFisher $d_i=i$, $w_i=2i/7$; $\mathbf{\Sigma}$= Equal(0.7)-III]{\includegraphics[width=0.5\textwidth]{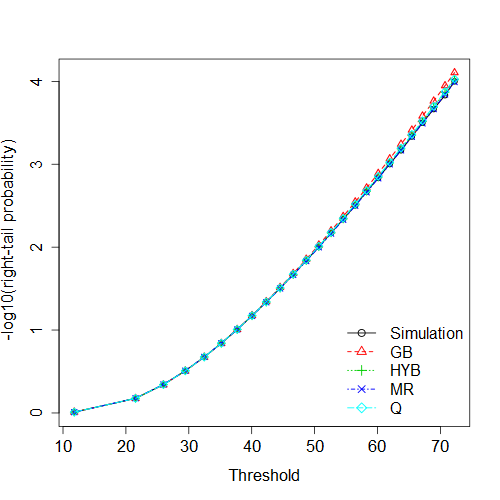}}
\caption{Right-tail probability ($-\log_{10}$) of $T$ when input $p$-values are two-sided. $n=6$. Two $\mathbf{\Sigma}$ patterns with $\rho=0.7$ are defined in Table \ref{tbl.sigma}.
}
\label{fig.survival_twoside}
\end{figure}

The results of survival curve are demonstrated in Figures \ref{fig.survival_oneside} and \ref{fig.survival_twoside} for one- and two-sided input $p$-values, respectively. The calculated survival curves by various approximation methods are compared to the empirical survival curves of the statistics, the latter are obtained by simulations of $10^6$ replicates and are treated as the ``gold standard". The statistics that we considered are Fisher's statistic, i.e. $d_i=2$, $w_i=1$, and a GFisher statistic, $d_i=i$, $w_i=2i/7$, $i=1,...,6$. Two types of correlation matrix, $\mathbf{\Sigma}=$ Equal(0.7)-I or Equal(0.7)-III, are defined in Table \ref{tbl.sigma}. More results are also given in Supplementary Figures \ref{fig.survival_oneside_supp} and \ref{fig.survival_twoside_supp} for $\mathbf{\Sigma}=$ Equal(0.3)-I and Equal(0.3)-III, respectively. These figures show that all methods are very consistent to the gold standard in the range of lower quantiles, and they start to differ in the range of higher quantiles. Overall, the moment-ratio matching method is the most accurate approximation to the survival curves of GFisher statistics in all simulation scenarios. For two-sided input $p$-values, the Q-approximation and the hybrid methods provide significantly higher accuracy than the generalized Brown's method, which could be inflated as early as $0.99$ quantile. For one-sided $p$-values, these two methods are generally conservative because they are not designed for this scenario. The inflation of the generalized Brown's method could be smaller for one-sided input $p$-values than the two-sided depending on the correlation matrix. 


\begin{figure}
\includegraphics[width=0.5\textwidth]{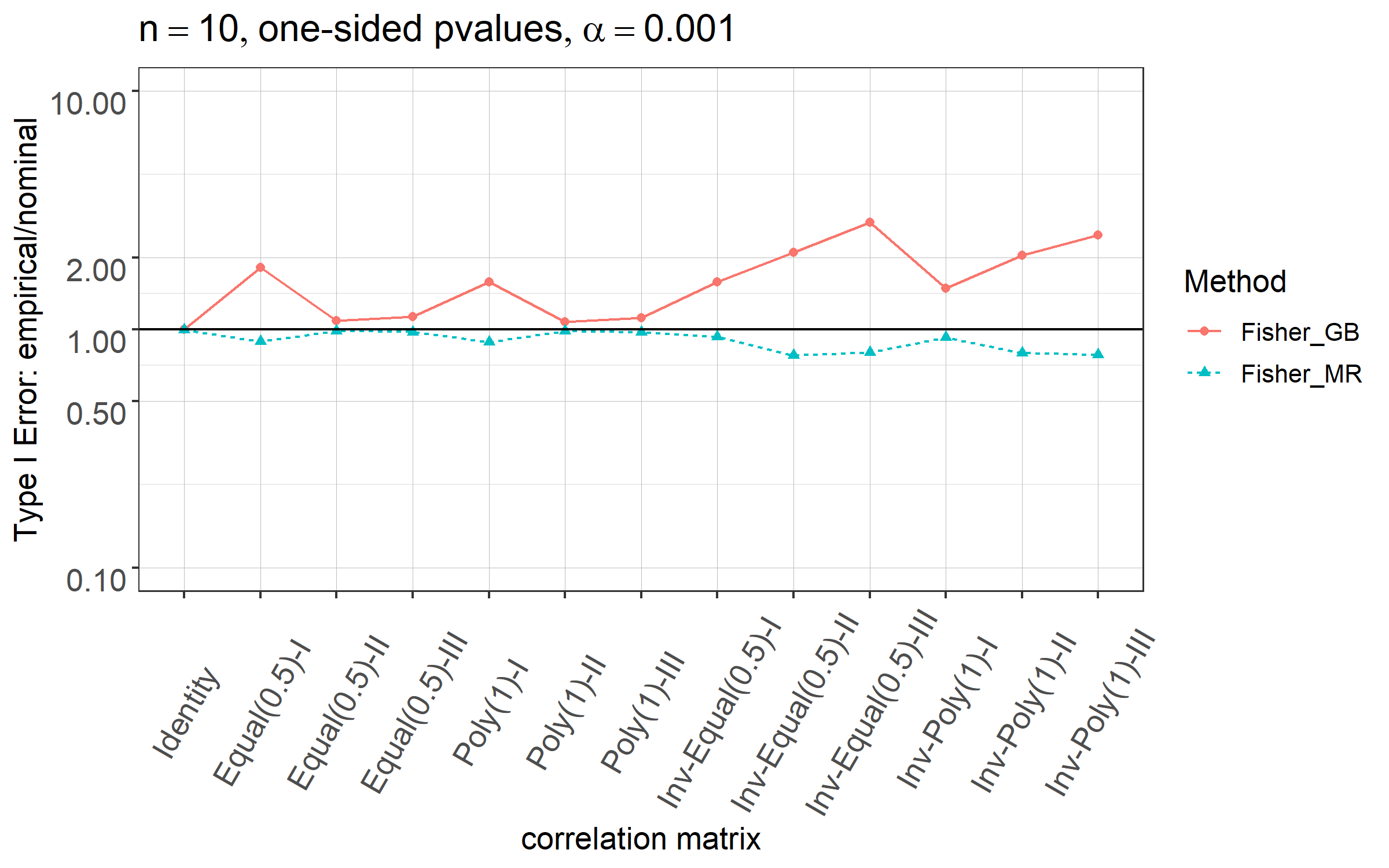}
\includegraphics[width=0.5\textwidth]{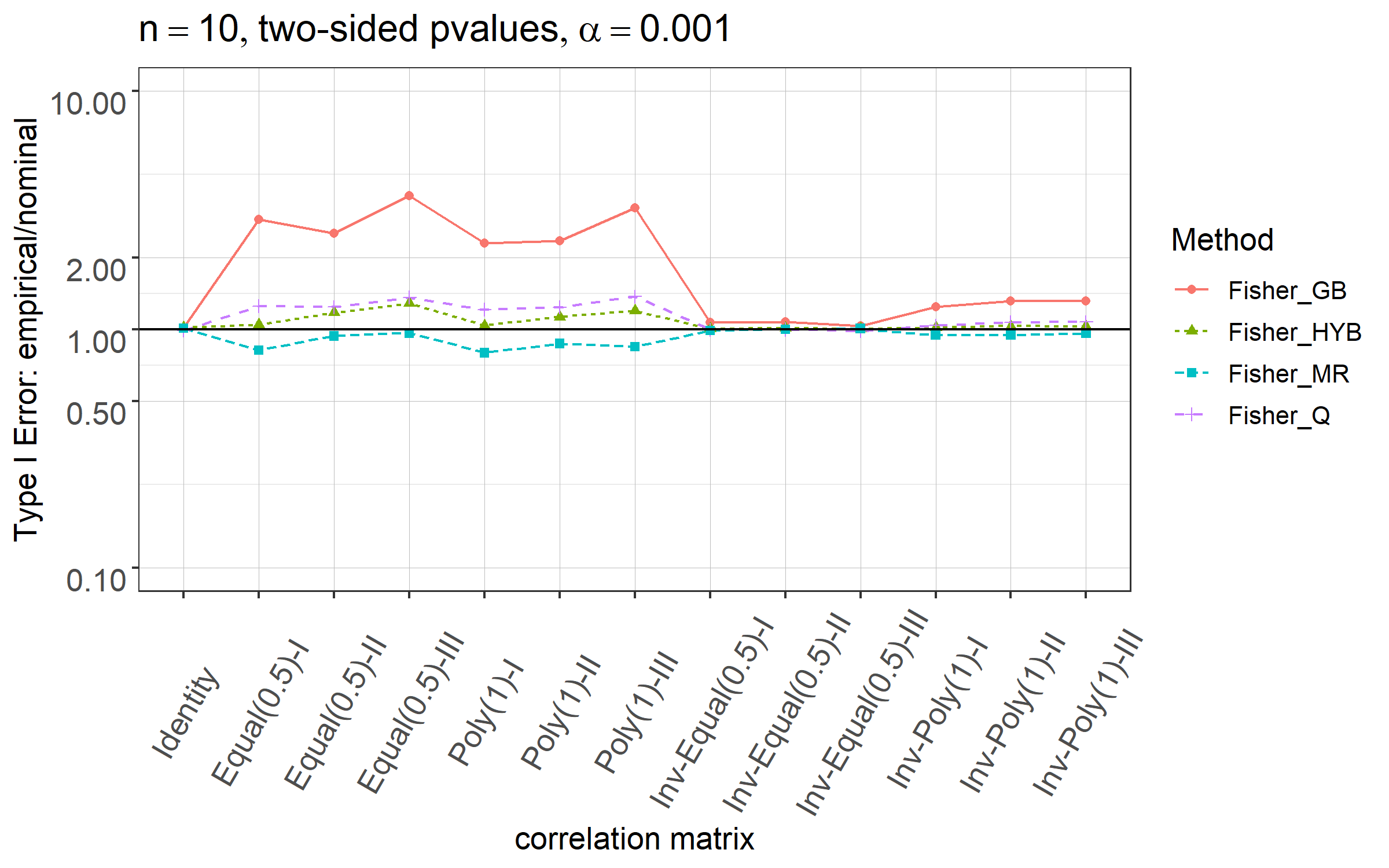}\\
\includegraphics[width=0.5\textwidth]{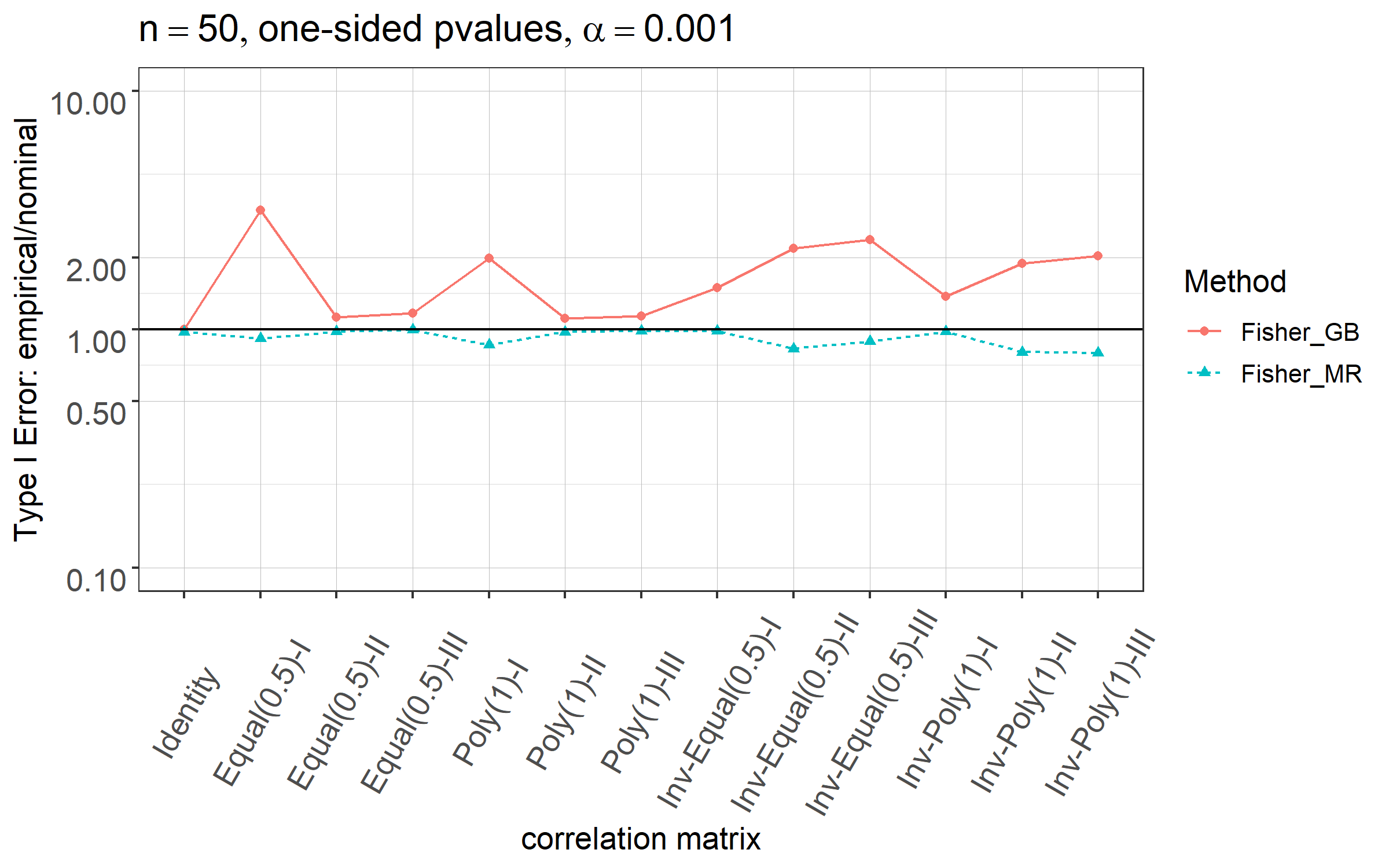}
\includegraphics[width=0.5\textwidth]{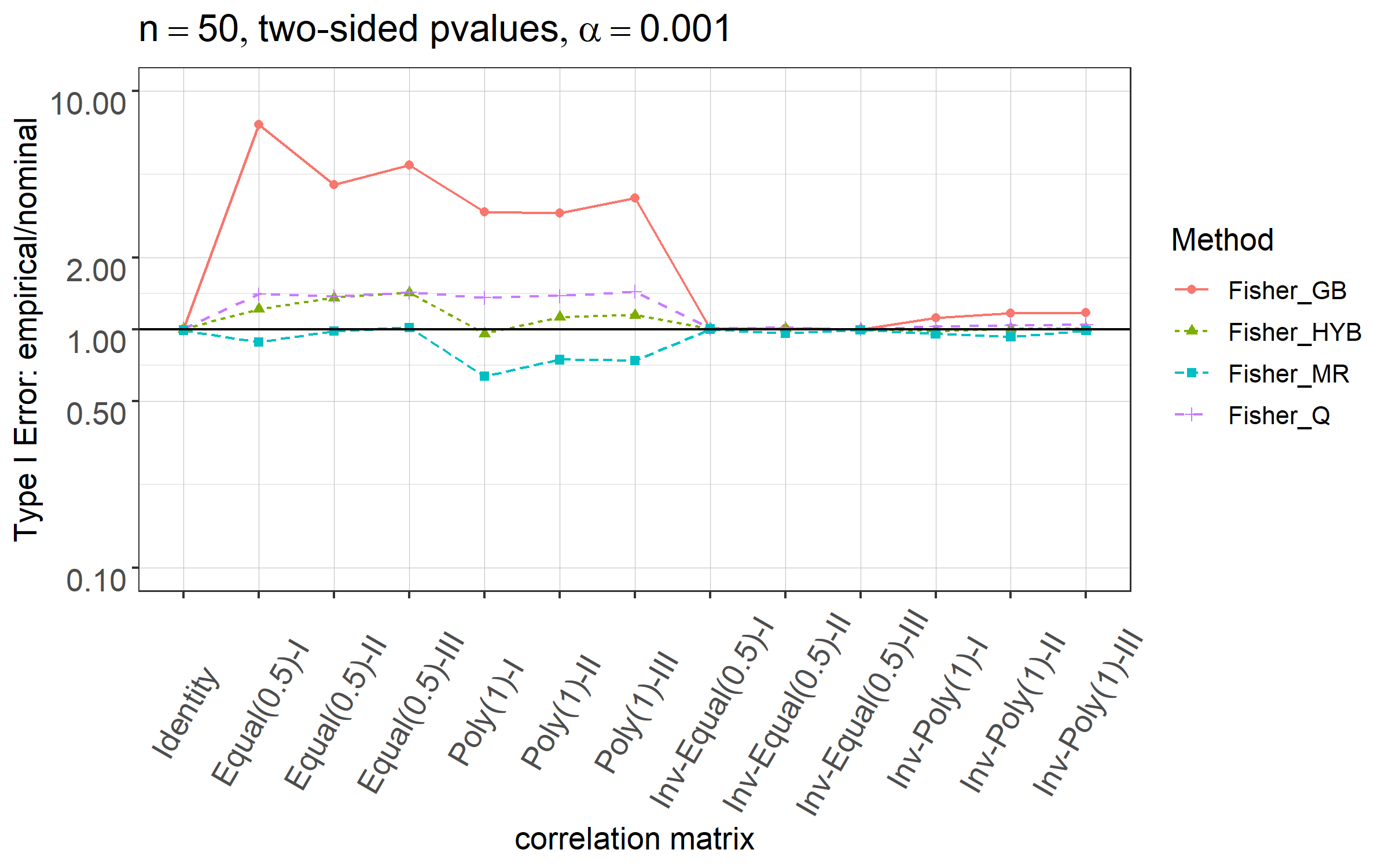}\\
\includegraphics[width=0.5\textwidth]{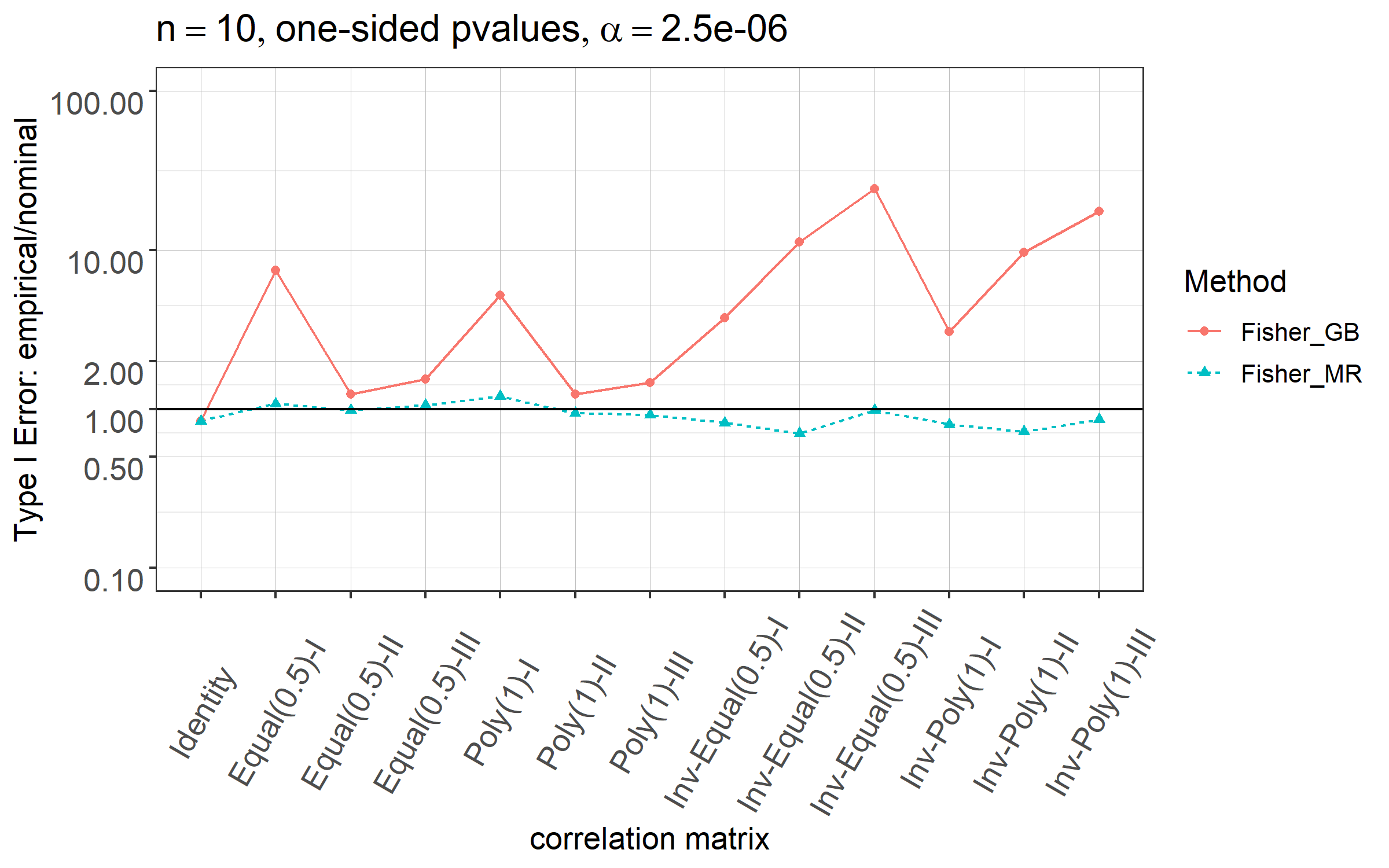}
\includegraphics[width=0.5\textwidth]{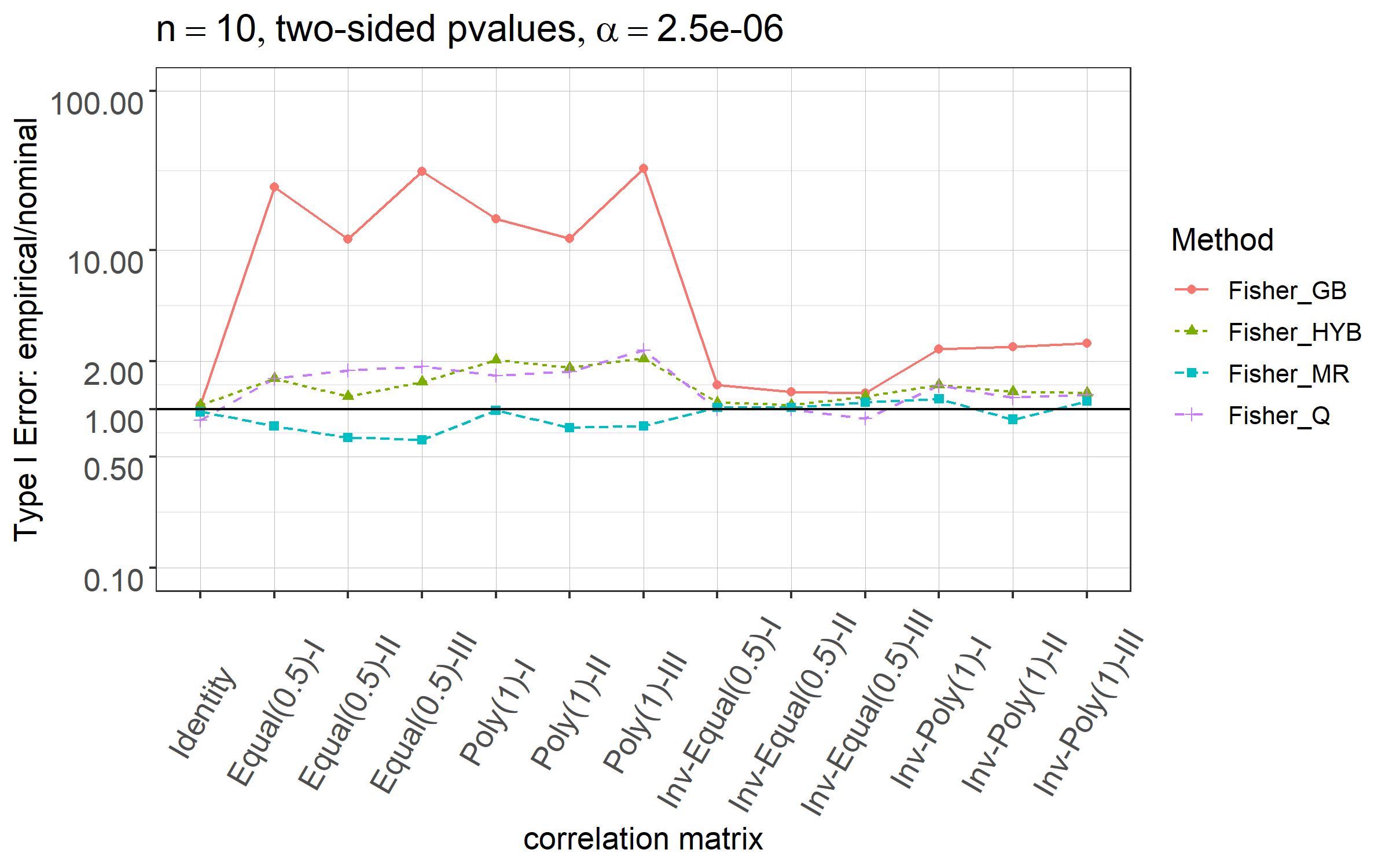}\\
\includegraphics[width=0.5\textwidth]{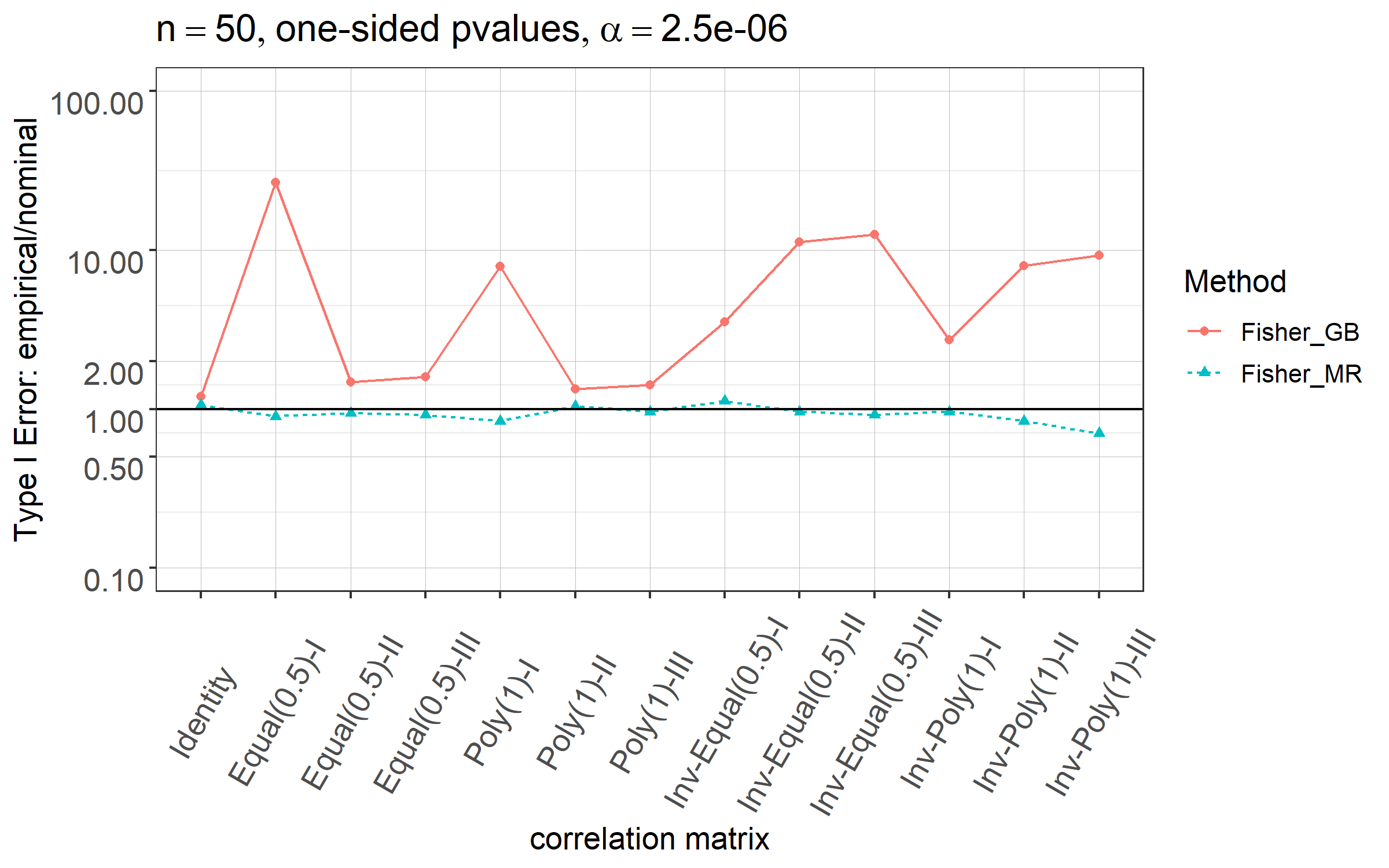}
\includegraphics[width=0.5\textwidth]{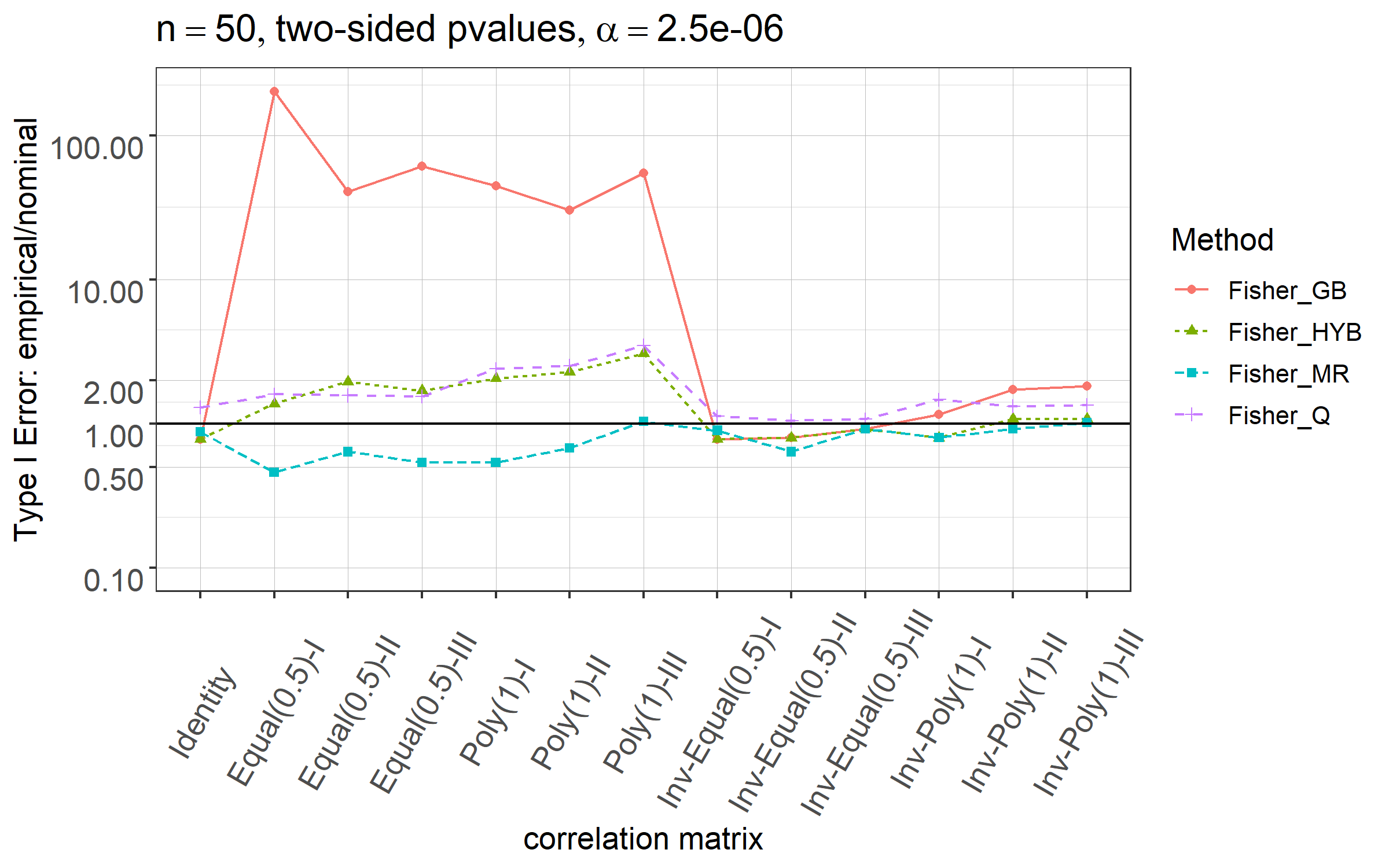}
\caption{Ratios between empirical type I error rates and nominal $\alpha=10^{-3}$ or $2.5\times 10^{-6}$. Fisher's combination test under independence and the 12 correlation structures in Table~\ref{tbl.sigma}. 
GB: generalized Brown's method. HYB: the hybrid method. MR: moment-ratio matching method. Q: Q-approximation. }
\label{fig.tie_GMM_cases}
\end{figure}

Now we demonstrate the ratio of the empirical type I error rate and the nominal level $\alpha$. A ratio of 1 indicates a perfect control, while a ratio larger (or smaller) than 1 indicates an inflated (or conservative) result because more (or fewer) rejections would be made than they should be. 
Figures~\ref{fig.tie_GMM_cases} summarizes the ratios over various correlation cases including the independent case and the 12 structures in Table \ref{tbl.sigma}. 
At a small but not most stringent level $\alpha=0.001$, for either one-sided or two-sided $p$-values, the generalized Brown's method already yields significantly inflated type I error rates. The inflation is even more striking at the $\alpha=2.5\times 10^{-6}$, where it could be as high as $200$ times of the nominal level.
The moment-ratio matching method controls the type I errors very well across all correlation patterns. It may occasionally slightly conservative, which discourages false discoveries.
The Q-approximation and hybrid method performs similarly very well for two-sided $p$-values at $\alpha=0.001$. At $\alpha=2.5\times 10^{-6}$, they could be mildly inflated under certain correlation matrices, although the hybrid method is a little better. Moreover, the value $n=10$ or $50$ may slightly change the ratio under the same setting, but overall the results are consistent over $n$ values. For comparing the same methods Supplementary Figures \ref{fig.tie_GMM_cases_supp1} and \ref{fig.tie_GMM_cases_supp2} give the ratios between type I error rates and nominal $\alpha = 0.05, 0.01, 10^{-4}$ and $10^{-5}$. As expected, different calculation methods are similar at large $\alpha$, e.g., at 0.05. 

The results of 1,404 settings of $n$, $\mathbf{\Sigma}$, $\rho$, $\kappa$, and $\alpha$ can be found in Supplementary Table 1. Figure \ref{fig.tie_GMM_box} summarizes the ratios by box-plots over levels of $\alpha$, which illustrates the overall performances of the approximating methods under GMM. 
In general, for both one-sided and two-sided input $p$-values, the type I error rates of the generalized Brown's method seem to be accurate at $\alpha=0.05$ but increasingly inflated as $\alpha$ decreases. In particular, at $\alpha=2.5\times 10^{-6}$ the generalized Brown's method could generate at least $10$ times inflated type I errors in more than $25\%$ of all the settings considered (including the less challenging cases of independence and weak correlations). The moment-ratio matching method, on the other hand, has accurate type I error rates at all $\alpha$ levels. The hybrid method and Q-approximation perform fairly well overall if the $p$-values are two-sided. 
\begin{figure}
\includegraphics[width=0.5\textwidth]{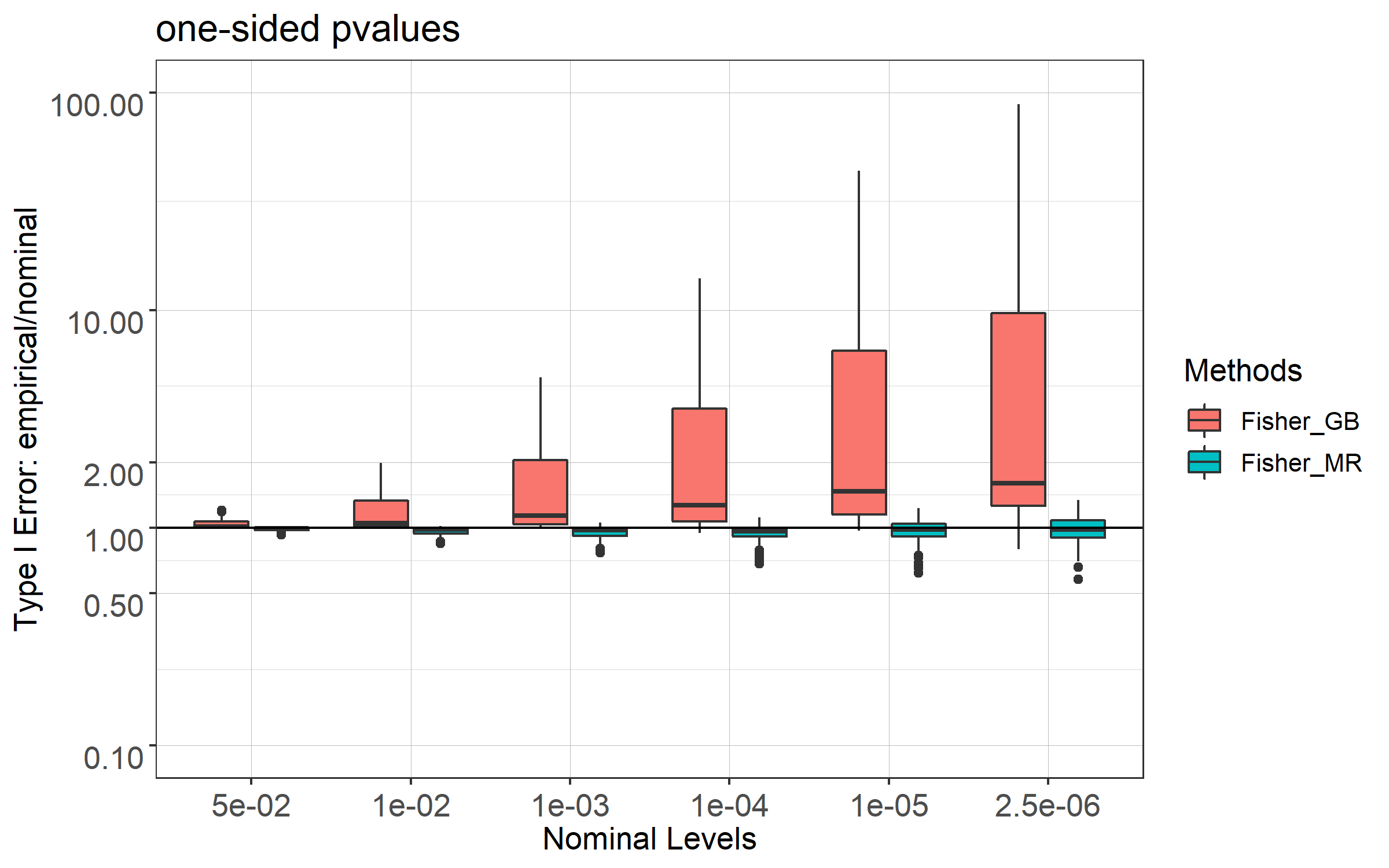}
\includegraphics[width=0.5\textwidth]{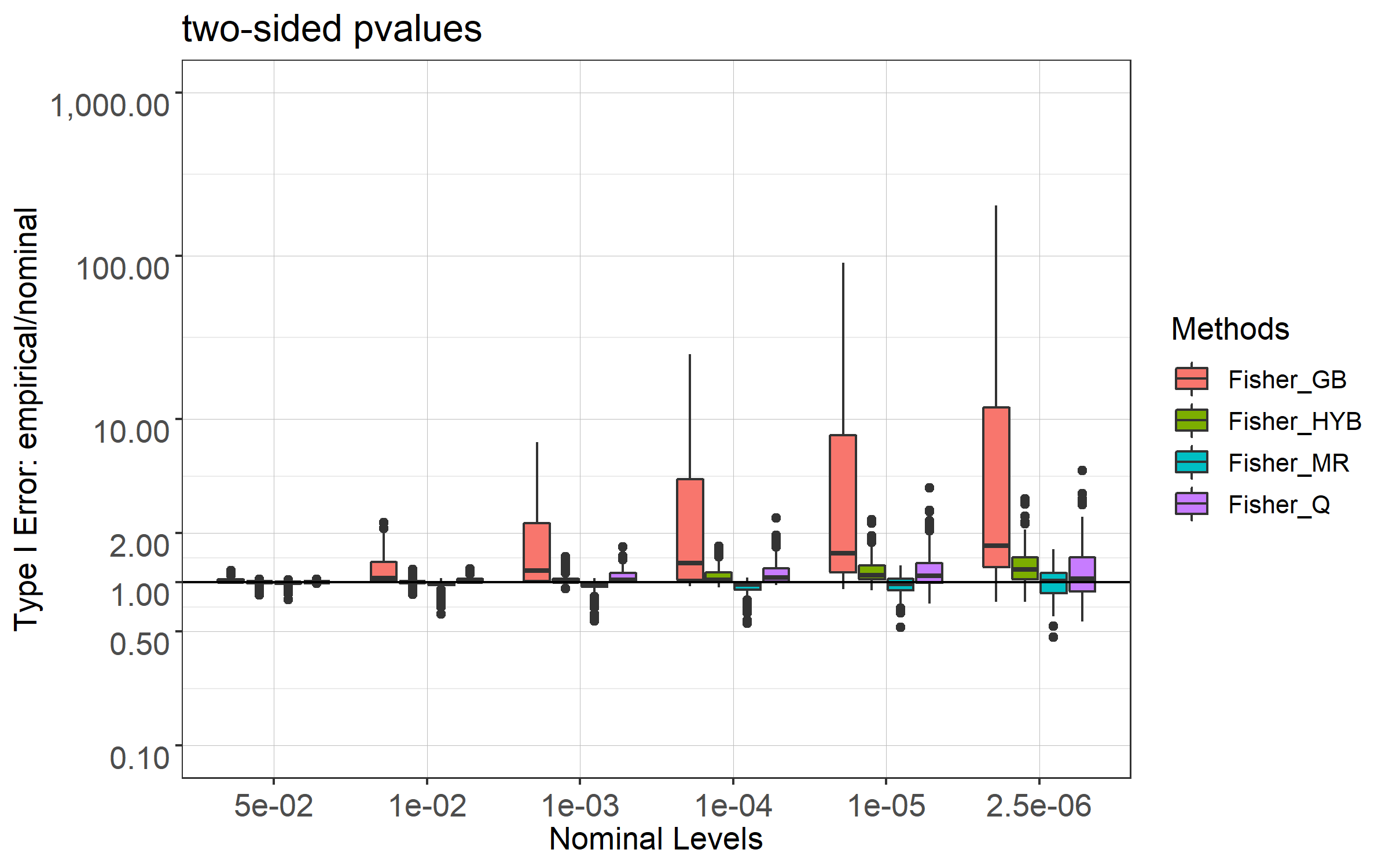}
\caption{Box-plots summarizing the ratios between empirical type I error rates and the nominal levels over all settings under GMM. Fisher's combination test is studied. GB: generalized Brown's method. HYB: the hybrid method. MR: moment-ratio matching method. Q: Q-approximation.} 
\label{fig.tie_GMM_box}
\end{figure}

Additionally, we took a careful look at the type I error controls of GGD-based methods (complete results of 936 settings are given in Supplementary Table 2). Figure~\ref{fig.tie_GGD} illustrates the results for Fisher's combination test under $n=10, 50$, and $\alpha=2.5\times 10^{-6}$.  It shows that GGD\_123, as a natural extension of the Brown's approximation, is overall inadequate. Involving skewness and kurtosis, GGD\_234 and GGD\_MR improve the accuracy but are still not as good as the GD-based MR method overall. Moreover, as we discussed in the previous section, GGD-based methods are much more computationally challenging. They not only take more computational time, but may not have solutions to the moment or moment-ratio matching equations under some correlation structures. This problem is demonstrated in Figure~\ref{fig.tie_GGD} by their discontinuous curves. Under the similar settings Supplementary Figure \ref{fig.tie_GGD_supp} gives the results at $\alpha=10^{-4}$ and $10^{-5}$; similar performances are observed. 
\begin{figure}
\includegraphics[width=0.5\textwidth]{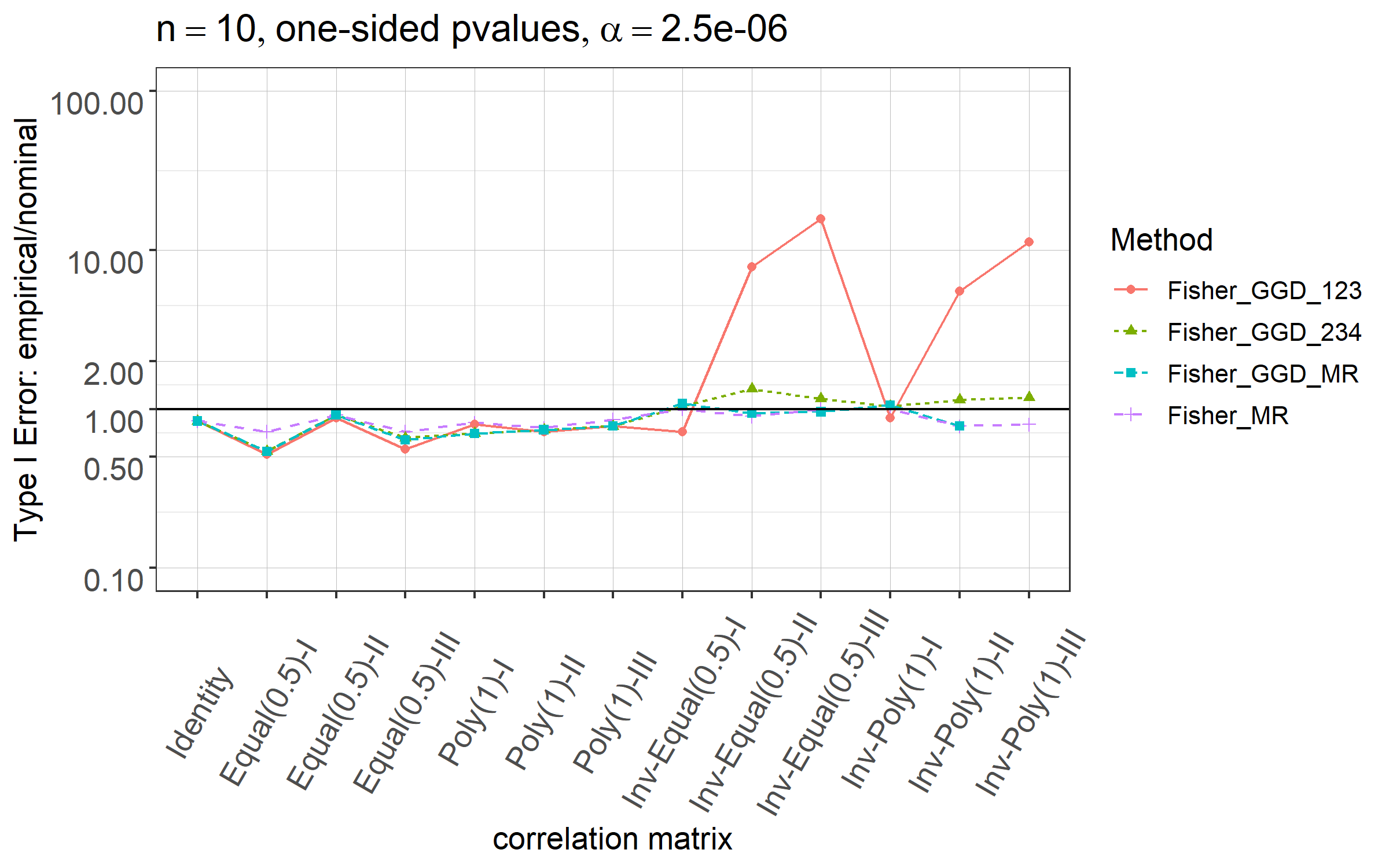}
\includegraphics[width=0.5\textwidth]{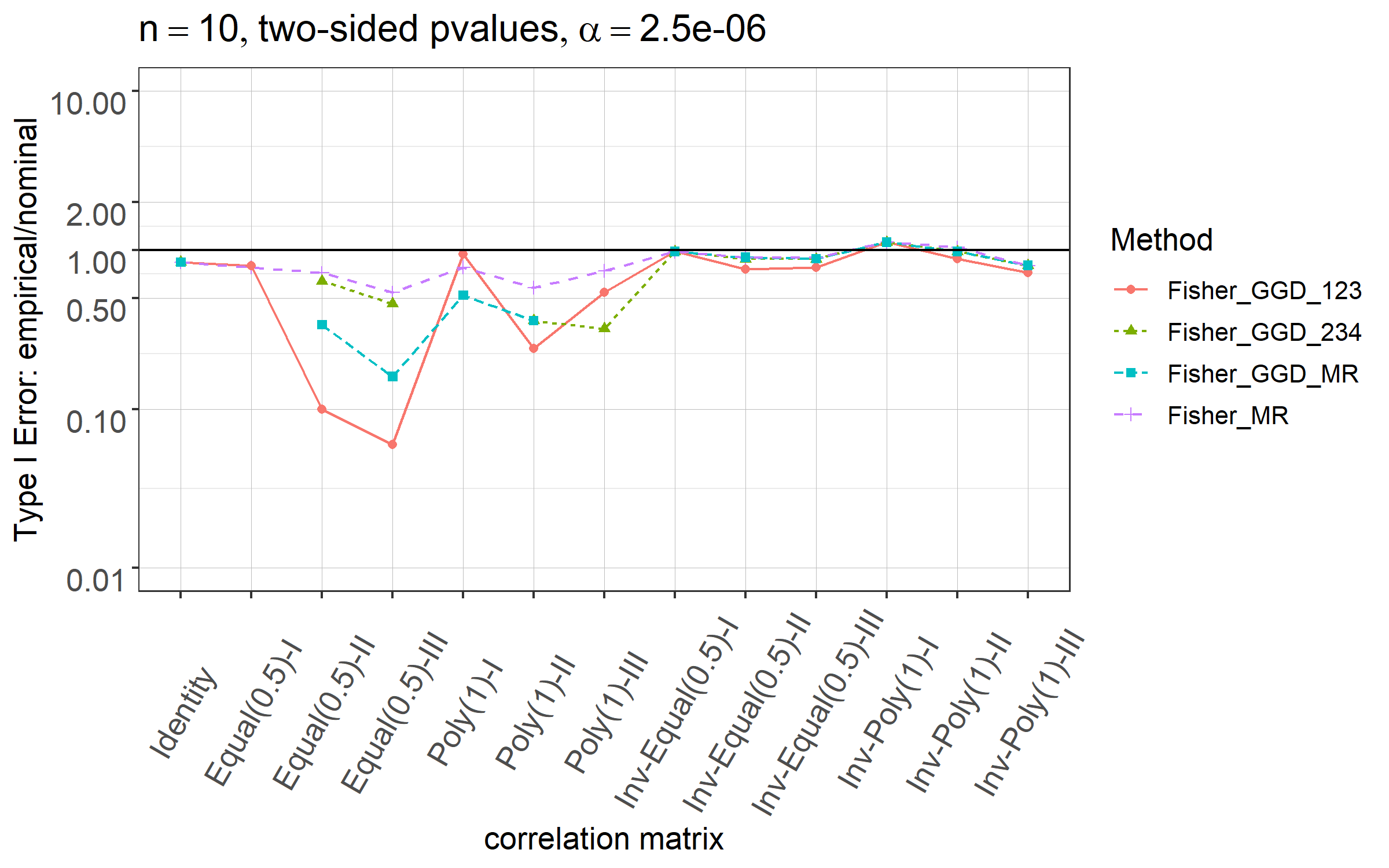}\\
\includegraphics[width=0.5\textwidth]{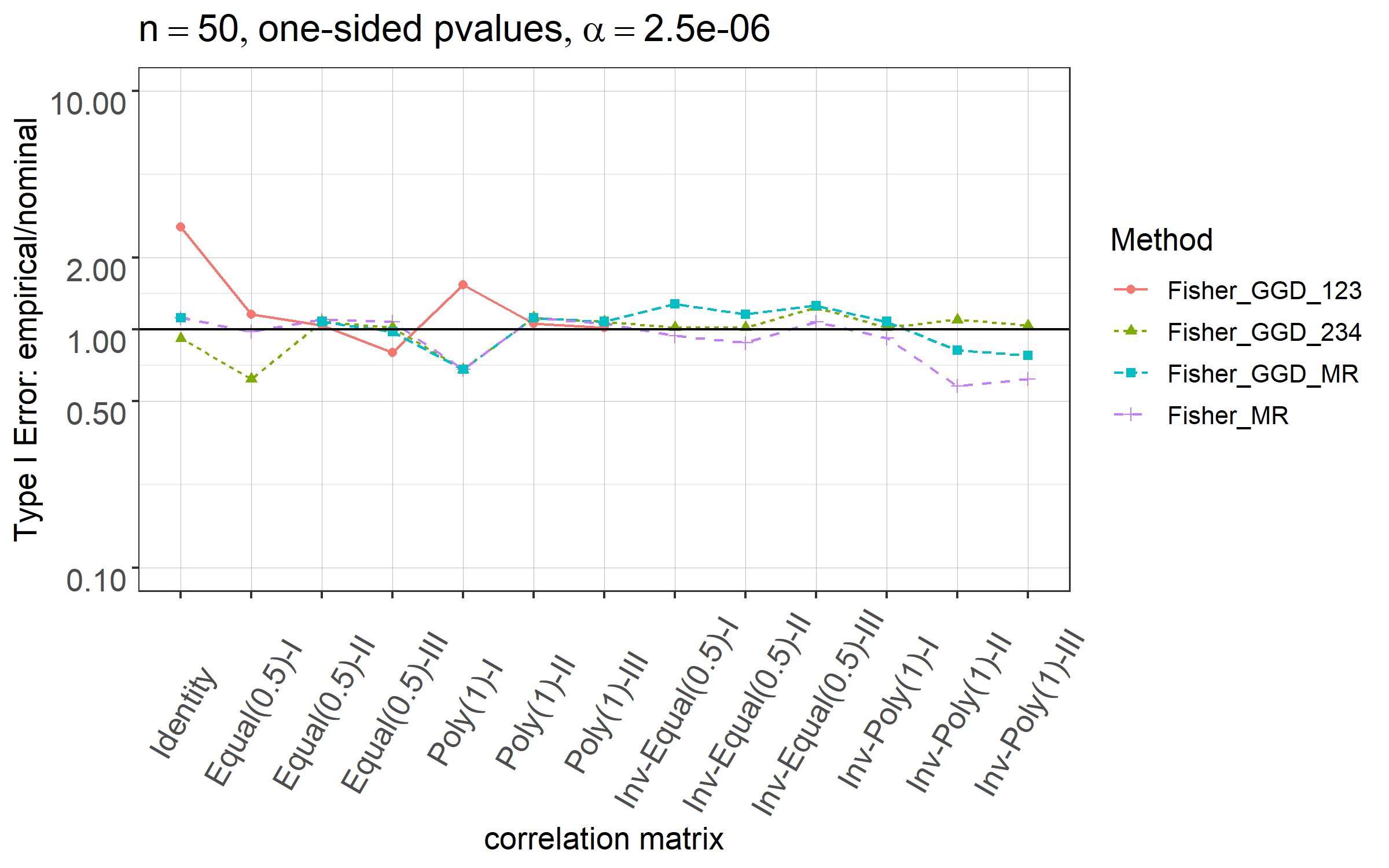}
\includegraphics[width=0.5\textwidth]{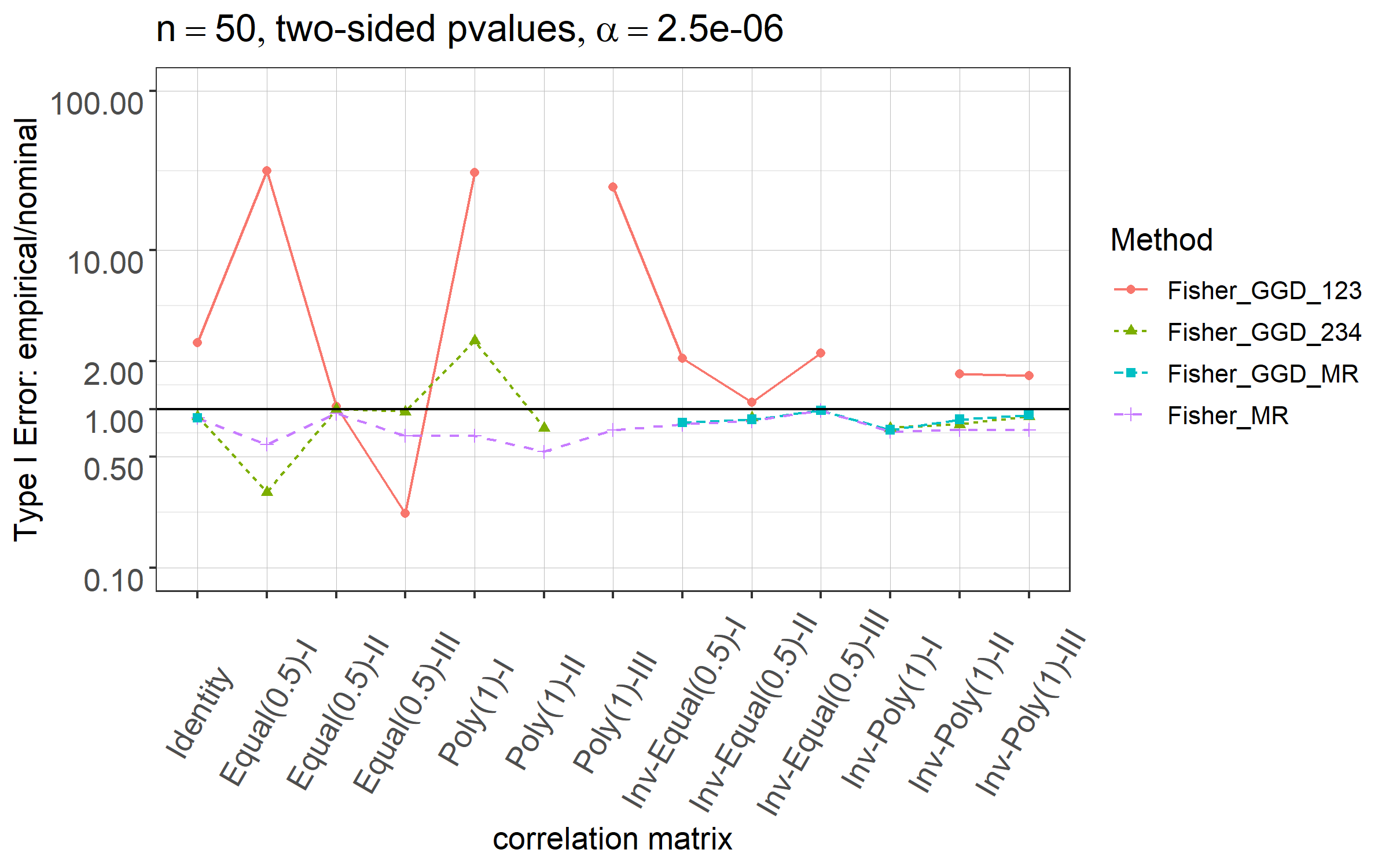}
\caption{Ratios between empirical type I error rates and the nominal $\alpha=2.5\times10^{-6}$ using GGD-based methods. Fisher's combination test under independence and the 12 correlation structures in Table~\ref{tbl.sigma}. GGD\_123: matching the first three moments of GGD. GGD\_234: matching the variance, skewness and kurtosis of GGD. GGD\_MR: GGD-based moment-ratio matching method. MR:  GD-based moment-ratio matching method. Missing values indicate moment-matching equations don't have a solution. 
}
\label{fig.tie_GGD}
\end{figure}

\begin{figure}
\includegraphics[width=0.5\textwidth]{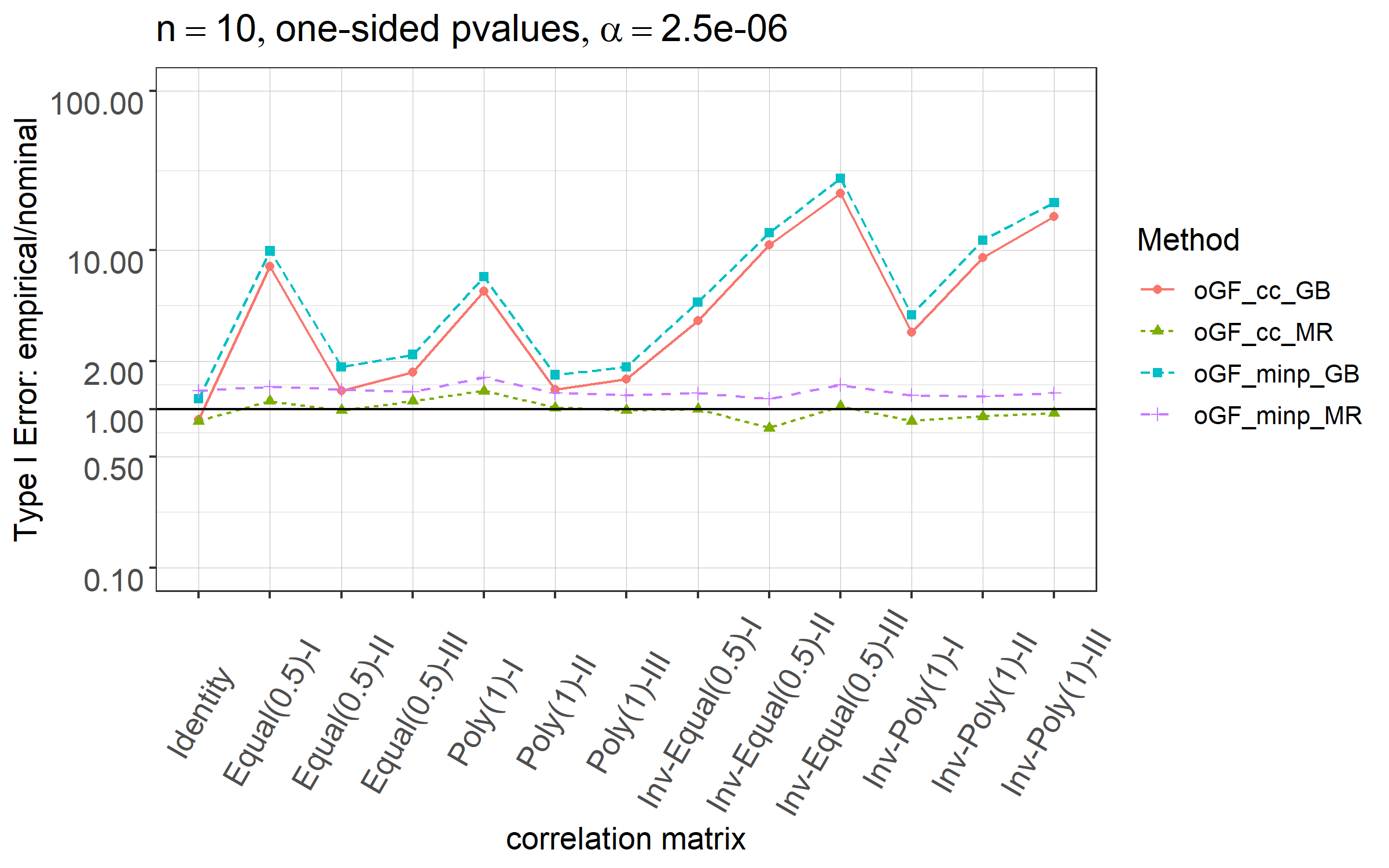}
\includegraphics[width=0.5\textwidth]{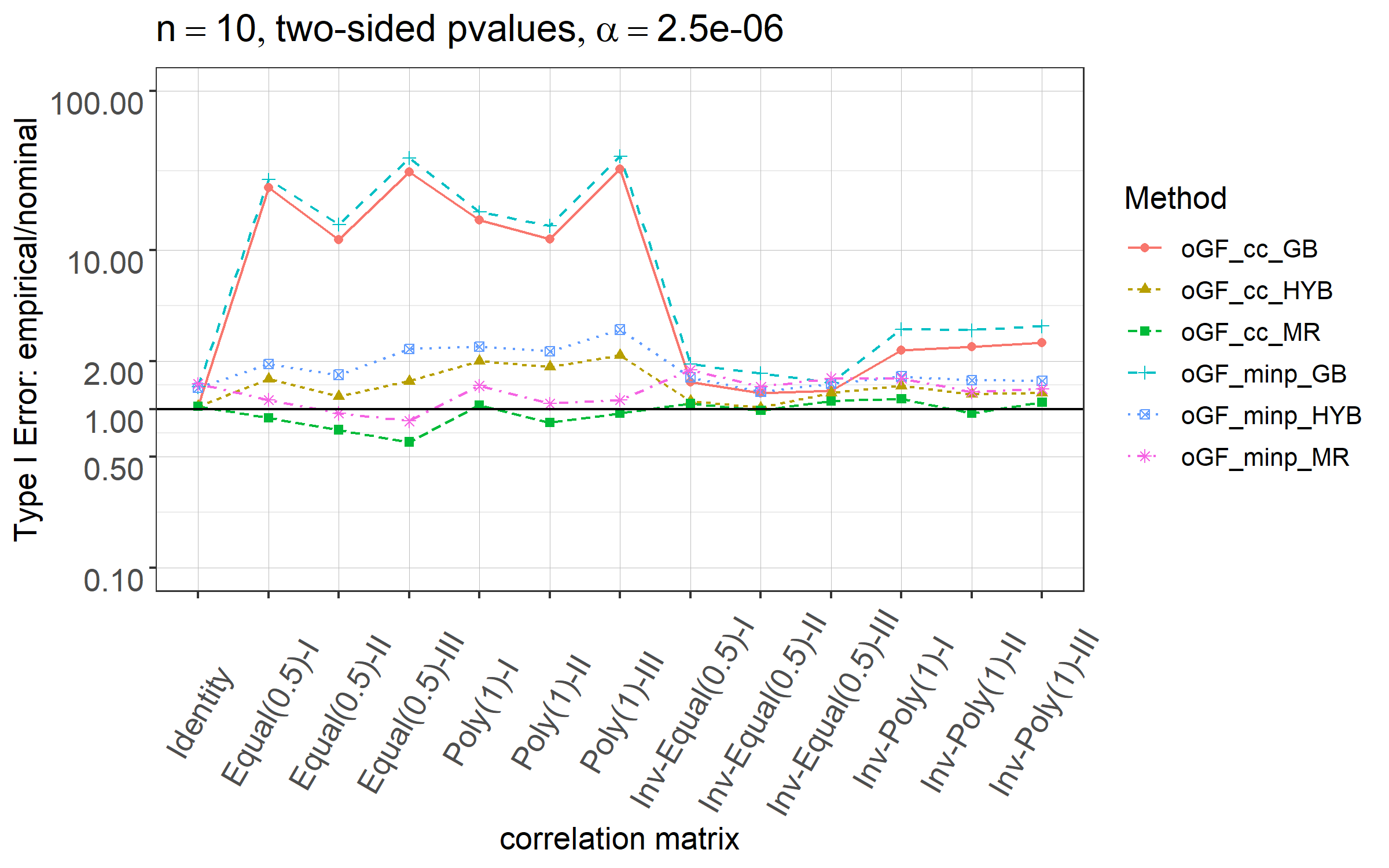}\\
\includegraphics[width=0.5\textwidth]{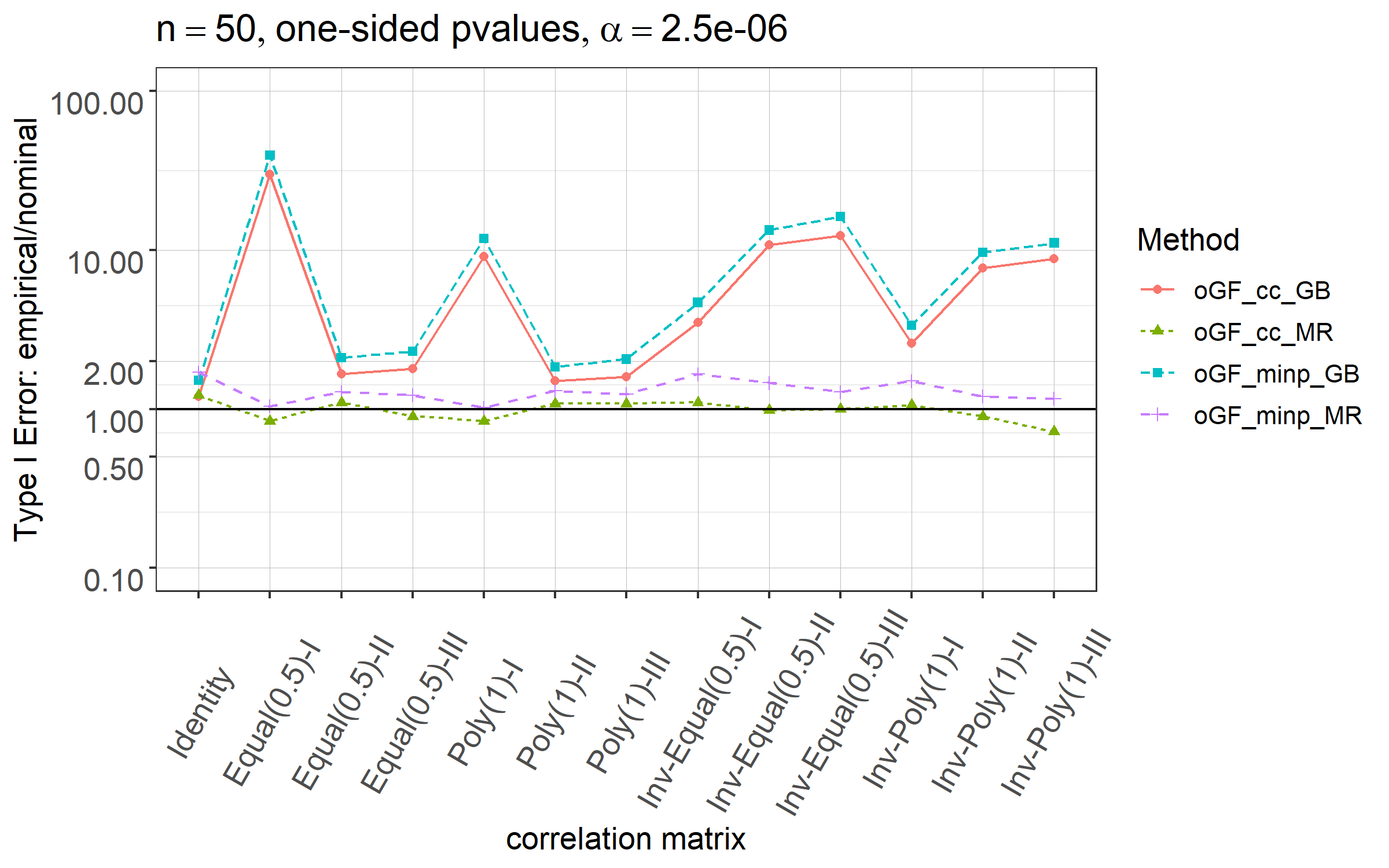}
\includegraphics[width=0.5\textwidth]{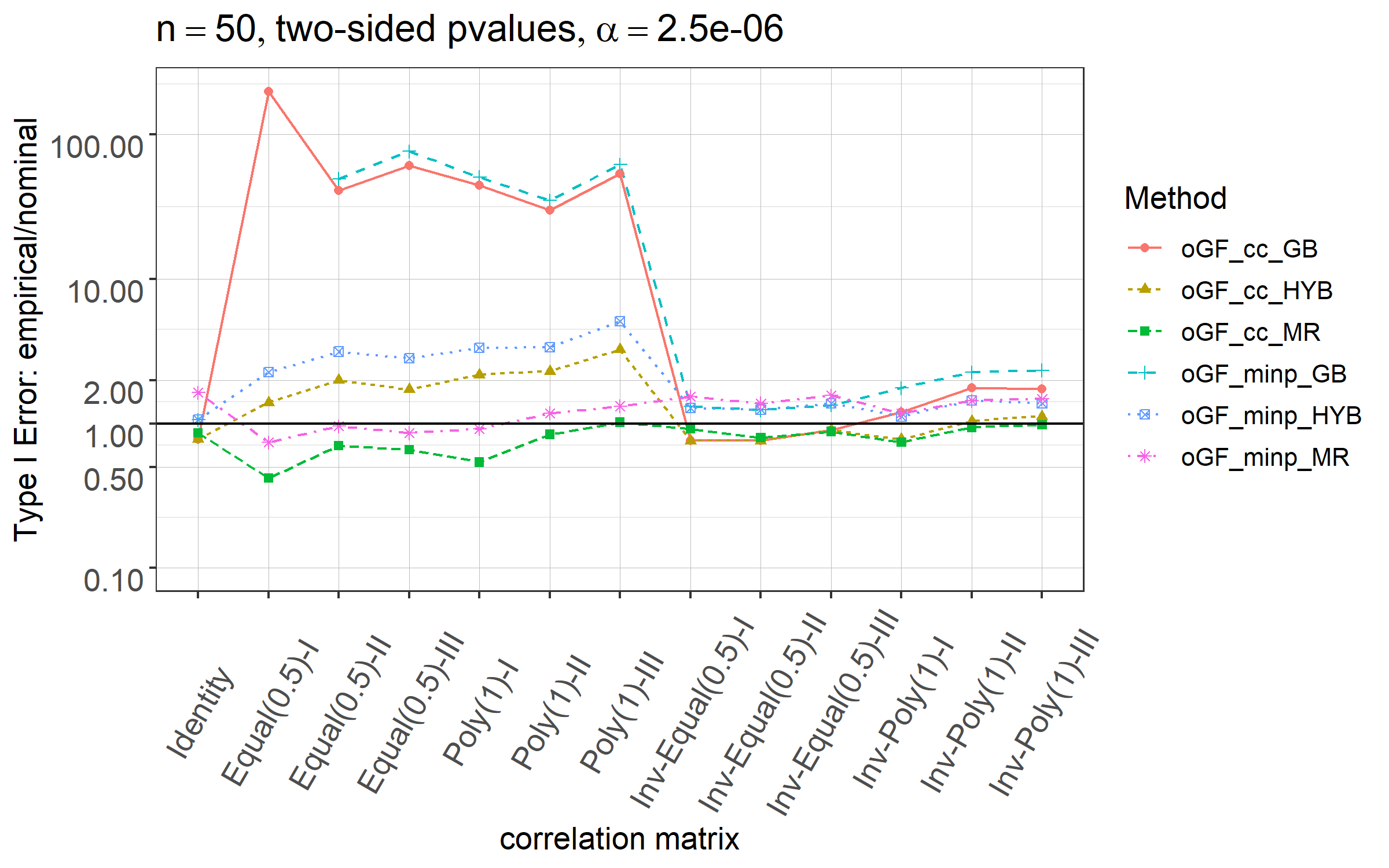}
\caption{Ratios between empirical type I error rates and the nominal $\alpha=2.5\times 10^{-6}$ for oGFisher tests under independence and the 12 correlation structures in Table~\ref{tbl.sigma}. Omnibus methods: Cauchy combination (cc); minimal $p$-value (minp). The $p$-value calculation methods: generalized Brown's method (GB); hybrid method (HYB); moment-ratio matching method (MR). 
}
\label{fig.tie_GMM_omni_2.5e-6}
\end{figure}

Lastly, we show the accuracy for the oGFisher tests. As evidenced in Figures~\ref{fig.tie_GMM_omni_2.5e-6}, at the $\alpha=2.5\times 10^{-6}$, the type I error rates of the oGFisher tests are highly inflated if the generalized Brown's approximation is applied to calculate individual GFisher $p$-values. When the moment-ratio matching method is applied, both oGFisher\_cc and oGFisher\_minp are well controlled across various correlation structures. When the hybrid method is applied, the type I error rates of oGFisher\_minp could be moderately inflated.  

\section{Application and robustness}\label{sec:appli}

The GMM in (\ref{equ.GMM}) is a reasonable baseline assumption for studying the distributions of GFisher because in practice the input statistics $\mathbf{Z}$ are often close to GMM under the null. This is the case when correlated data is analyzed by linear models, where $\mathbf{Z}$ asymptotically satisfy GMM as sample size $N\to\infty$ under some weak regulatory conditions. To cater for applications here we illustrate a few widely used statistics of such type, for which the estimated correlation matrices are given so that GFisher procedure can be applied accordingly. Simulations were carried out to show the robustness of relevant calculation methods in controlling $\alpha$ under 
typical $N$ in genetic association studies. Simulations also illustrate the robustness when $\mathbf{Z}$ follows the multivariate $t$-distribution with small degrees of freedom, a scenario further departure from GMM. 

Consider a generalized linear model (GLM) that contains inquiry covariates to be tested conditional on controlling covariates:
\begin{equation}
g(E(Y_k | \mathbf{X}_{k\cdot}, \mathbf{C}_{k\cdot})) = \mathbf{X}'_{k\cdot}\mathbf{\beta} + \mathbf{C}'_{k\cdot} \mathbf{\gamma},
\label{equ.GLM}
\end{equation}
where for the $k$th subject, $k = 1, \ldots, N$, $Y_k$ denotes the random response with a distribution in the exponential family, $\mathbf{X}'_{k\cdot} = (X_{k1}, \ldots, X_{kn})$ denotes the values of $n$ inquiry covariates, and $\mathbf{C}'_{k\cdot} = (C_{k1}, \ldots, C_{km})$ denotes the values of $m$ control covariates. The link function $g$ is assumed the canonical link function of the given distribution of $Y_k$. A special case of the GLM is the linear model (LM) assuming Gaussian $Y_k$: 
\begin{equation*}
	\mathbf{Y} = \mathbf{X}\mathbf{\beta} + \mathbf{C} \mathbf{\gamma} + \mathbf{\epsilon},  
\end{equation*}
where $\mathbf{X}_{N\times n}$ and $\mathbf{C}_{N\times n}$ are the inquiry and controlling design matrices with their $k$th row vectors being  $\mathbf{X}_{k\cdot}^\prime$ and $\mathbf{C}_{k\cdot}^\prime$, respectively. The error term $\mathbf{\epsilon} \sim N(\mathbf{0}, \sigma^2 \mathbf{I}_{N \times N})$, where the variance $\sigma^2$ is often known and needs to be estimated in practice.

Based on this model, we consider global hypothesis testing of the null that none of the inquiry covariates are associated with the outcome conditioning on the control covariates: 
\begin{equation}
\label{equ.GLM.H0Ha}
H_0: \mathbf{\beta}=\mathbf{0}. 
\end{equation}

For testing (\ref{equ.GLM.H0Ha}) we can apply the GFisher procedure, where the $n$ input $p$-values come from the input statistics of the $n$ inquiry covariates. Many widely applied input statistics asymptotically satisfies GMM under weak or mild conditions as sample size $N\to\infty$ at fixed $n$. 
In the following we give a few examples of such statistics. With their correlation matrices estimated based on data, in practice our approximating methods can be applied for calculating the test $p$-value of any given GFisher or oGFisher statistic. The technical conditions for the asymptotic normality can be found in literature \citep{zhang2018generalized}. Here we focus on application, acknowledging the gap between theory and practice and thus providing a study of robustness afterward. 

In LM, let $\mathbf{H}=\mathbf{C}(\mathbf{C}^\prime \mathbf{C})^{-1}\mathbf{C}^\prime$ be the projection matrix onto the column space of $\mathbf{C}$. Denote $\mathbf{G} = \mathbf{X}^\prime  (\mathbf{I}-\mathbf{H})\mathbf{X}$. By joint least-squares estimation of all $\beta$ elements, the vector of statistics are 
\begin{equation*}
	\mathbf{Z}_J  = \mathbf{\Lambda}_J\hat{\mathbf{\beta}}_J/\hat{\mathbf{\sigma}} \overset{D}{\to} N(\mathbf{\mu}_J, \mathbf{\Sigma}_J),
\end{equation*}
where $\hat{\mathbf{\beta}}_J=  \mathbf{G}^{-1}\mathbf{X}^\prime  (\mathbf{I}-\mathbf{H})\mathbf{Y}$, 
$\mathbf{\Lambda}_J =\text{diag}\left(1/ \sqrt{\left(\mathbf{G}^{-1}\right)_{ii}}\right)_{1 \leq i \leq n}$ is a diagonal matrix, 
$\hat{\sigma}^2$ is the mean squared error, 
$\mathbf{\mu}_J=\mathbf{\Lambda}_J\mathbf{\beta}/\sigma \overset{H_0}{=} \mathbf{0}$, and 
$\mathbf{\Sigma}_J=\mathbf{\Lambda}_J \mathbf{G}^{-1}\mathbf{\Lambda}_J$ has diagonal of 1's. 

By marginal least-squares estimation of individual $\beta$ elements, the vector of marginal statistics are 
\begin{equation}
\label{equ.Tmarg}
\mathbf{Z}_M = \mathbf{\Lambda}_M \mathbf{X}'(\mathbf{I}-\mathbf{H})\mathbf{Y}/\hat{\sigma} \overset{D}{\to} N(\mathbf{\mu}_M, \mathbf{\Sigma}_M),
\end{equation}
where $\mathbf{\Lambda}_M  = {\rm diag}\left( 1/\sqrt{\mathbf{G}_{ii}}\right)_{1 \leq i \leq n}$, 
$\mathbf{\mu}_M = \mathbf{\Sigma}_M \mathbf{\Lambda}_M^{-1}\mathbf{\beta}/\sigma \overset{H_0}{=} \mathbf{0}$, and 
$\mathbf{\Sigma}_M = \mathbf{\Lambda}_M \mathbf{G} \mathbf{\Lambda}_M$ has diagonal of 1's. 

Under the GLM, let $\mathbf{W}={\rm diag}(\Var(Y_k | \mathbf{X}_{k\cdot}, \mathbf{C}_{k\cdot}))_{1 \leq k \leq N}$. 
Denote $\tilde{\mathbf{X}} = \mathbf{W}^{1/2}\mathbf{X}$, $\tilde{\mathbf{C}} = \mathbf{W}^{1/2}\mathbf{C}$, $\tilde{\mathbf{H}}=\tilde{\mathbf{C}}(\tilde{\mathbf{C}}^\prime \tilde{\mathbf{C}})^{-1}\tilde{\mathbf{C}}^\prime$, and $\tilde{\mathbf{G}} = \tilde{\mathbf{X}}^\prime  (\mathbf{I}-\tilde{\mathbf{H}})\tilde{\mathbf{X}}$. 
By the joint maximum likelihood estimation (MLE) of all $\mathbf{\beta}$ elements, the vector of statistics are
\begin{equation*}
	\mathbf{Z}_{MLE} = \hat{\mathbf{\Lambda}}_{MLE} \hat{\mathbf{\beta}}_{MLE} \overset{D}{\to} N( \mathbf{\mu}_{MLE},  \mathbf{\Sigma}_{MLE}),
\end{equation*}
where $\hat{\mathbf{\beta}}_{MLE}$ is the MLE of $\mathbf{\beta}$, 
$\hat{\mathbf{\Lambda}}_{MLE}$ is the MLE (using the MLE of $\mathbf{W}$) for $\mathbf{\Lambda}_{MLE} = \text{diag}\left(1/\sqrt{(\tilde{\mathbf{G}}^{-1})_{ii}} \right)_{1 \leq i \leq n}$, 
$\mathbf{\mu}_{MLE} = \mathbf{\Lambda}_{MLE} \mathbf{\beta} \overset{H_0}{=} \mathbf{0}$, and 
$\mathbf{\Sigma}_{MLE} = \mathbf{\Lambda}_{MLE}\tilde{\mathbf{G}}^{-1} \mathbf{\Lambda}_{MLE}$.  
For the input to the GFisher tests we can use the MLE of $\mathbf{\Sigma}_{MLE}$, i.e., $\hat{\mathbf{\Sigma}}_{MLE} = \hat{\mathbf{\Lambda}}_{MLE} \hat{\tilde{\mathbf{G}}}^{-1} \hat{\mathbf{\Lambda}}_{MLE}$ as the correlation matrix.

Under GLM for marginal model fitting, to be consistent with literature we consider the marginal score test, which has been widely applied in data analysis \citep{barnett2016generalized, sun2017set}.  Specifically, denote $\mathbf{\mu}^{(0)} = (\mu^{(0)}_1, \cdots, \mu^{(0)}_N)'$, where $\mu^{(0)}_k= \E_{H_0} (Y_k | \mathbf{C}_{k\cdot}) = g^{-1}(\mathbf{C}_{k\cdot}^{\prime} \mathbf{\gamma})$, $k=1, \cdots, N$, be the null expectation of $Y_k$. Let $\hat{\mathbf{\mu}}^{(0)}$ be the MLE of $\mathbf{\mu}^{(0)}$, with $\hat{\mu}^{(0)}_k= g^{-1}(\mathbf{C}_{k\cdot}^{\prime} \hat{\mathbf{\gamma}}^{(0)})$, where $\hat{\mathbf{\gamma}}^{(0)}$ is the MLE estimator of $\gamma$ under $H_0$.  
For example, in the logit model, when $\mathbf{C}_k= 1$ for the intercept, we have $\hat{\mu}^{(0)}_k=\bar{y}$ and $\hat{\gamma}^{(0)}=\log(\frac{\bar{y}}{1-\bar{y}})$. 
Let $\mathbf{W}_0={\rm diag}(\Var(Y_k | \mathbf{C}_{k\cdot}))_{1 \leq k \leq N}$. 
Denote $\tilde{\mathbf{X}}_0 = \mathbf{W}_0^{1/2}\mathbf{X}_0$, $\tilde{\mathbf{C}}_0 = \mathbf{W}_0^{1/2}\mathbf{C}_0$, $\tilde{\mathbf{H}}_0=\tilde{\mathbf{C}}_0(\tilde{\mathbf{C}}_0^\prime \tilde{\mathbf{C}}_0)^{-1}\tilde{\mathbf{C}}_0^\prime$, and $\tilde{\mathbf{G}}_0 = \tilde{\mathbf{X}}_0^\prime  (\mathbf{I}-\tilde{\mathbf{H}}_0)\tilde{\mathbf{X}}_0$. 
The vector of marginal score test statistics are
\begin{equation}
\label{equ.Tmarg.GLM}
\mathbf{Z}_S = \hat{\mathbf{\Lambda}}_S\mathbf{X}^\prime (\mathbf{Y}-\hat{\mathbf{\mu}}^{(0)}) \overset{D}{\to} N(\mathbf{\mu}_S, \mathbf{\Sigma}_S),
\end{equation}
where  $\hat{\mathbf{\Lambda}}_S$ is the MLE (using the MLE of $\mathbf{W}_0$ under $H_0$) for $\mathbf{\Lambda}_S = \text{diag}\left( 1/ \sqrt{\tilde{\mathbf{G}}_{0ii}} \right)$,   
$\mathbf{\mu}_S = \mathbf{\Sigma}_S \mathbf{\Lambda}_S^{-1}\beta \overset{H_0}{=} \mathbf{0}$, and
$\mathbf{\Sigma}_S = \mathbf{\Lambda}_S \tilde{\mathbf{G}}_0 \mathbf{\Lambda}_S$. 
For the input to the GFisher tests we can use the MLE $\hat{\mathbf{\Sigma}}_S$ to replace $\mathbf{\Sigma}_S$.
Note that $\mathbf{Z}_S$ in (\ref{equ.Tmarg.GLM}) under the GLM reduces to $\mathbf{Z}_M$ in (\ref{equ.Tmarg}) under the LM. 


Now we exam the robustness of our methods in genetic association study based on GLM models. The aim of the study is to test the genetic associations of $n$ single nucleotide polymorphisms (SNPs). Following (\ref{equ.GLM}), $\mathbf{X}_{k\cdot}$ is the genotype vector of the $n$ SNPs of the $k$th individual;  $\mathbf{C}_{k\cdot}$ is the vector of $m$ control covariates (such as the intercept and other environmental and genetic variants). 
We simulated the genotype data by the genetic coalescent model \citep{shlyakhter2014cosi2}. Specifically, we first composed a database of 1,290 SNPs in a region of 250k base-pairs on Chromosome 1, for which a population of 10,000 haplotypes were generated based a linkage disequilibrium structure of European ancestry. Each simulation represents a separate associate study, in which the genotype data $\mathbf{X}$ of $n=20$ SNPs and $N=500$ individuals was randomly selected from the database. Phenotype trait values were calculated by linear models under the $H_0$ that no SNPs are associated but environmental factors are influential. Two types of phenotypes were considered: The quantitative traits were calculated under linear regression model:
\begin{align*}
  {Y}_k = 0.5 {C}_{1k} + 0.1 {C}_{2k} + \epsilon_k, \text{ }\epsilon_k \sim N(0,1), \text{ }k=1,...,N.
\end{align*}
The binary traits were calculated under the logit model:
\begin{align*}
  \text{logit}\left(\P({Y}_k=1)\right) = -1.25 + 0.5{C}_{1k} + 0.5{C}_{2k}, \text{ }k=1,...,N.
\end{align*}
The controlling covariates ${C}_{1k} \sim \text{Bernoulli}(0.494)$ and ${C}_{2k} \sim N(0,1)$ represent discrete and continuous environmental factors, respectively. Following the simulated data, marginal score statistics $\mathbf{Z}_S$ in (\ref{equ.Tmarg.GLM}) were calculated, and the two-sided $p$-values in (\ref{equ.pvalues}) were input into Fisher's combination test or the oGFisher test. $1\times 10^7$ simulations were used to evaluate empirical type I error rates up to $\alpha=2.5\times10^{-6}$.

\begin{figure}
\includegraphics[width=0.5\textwidth]{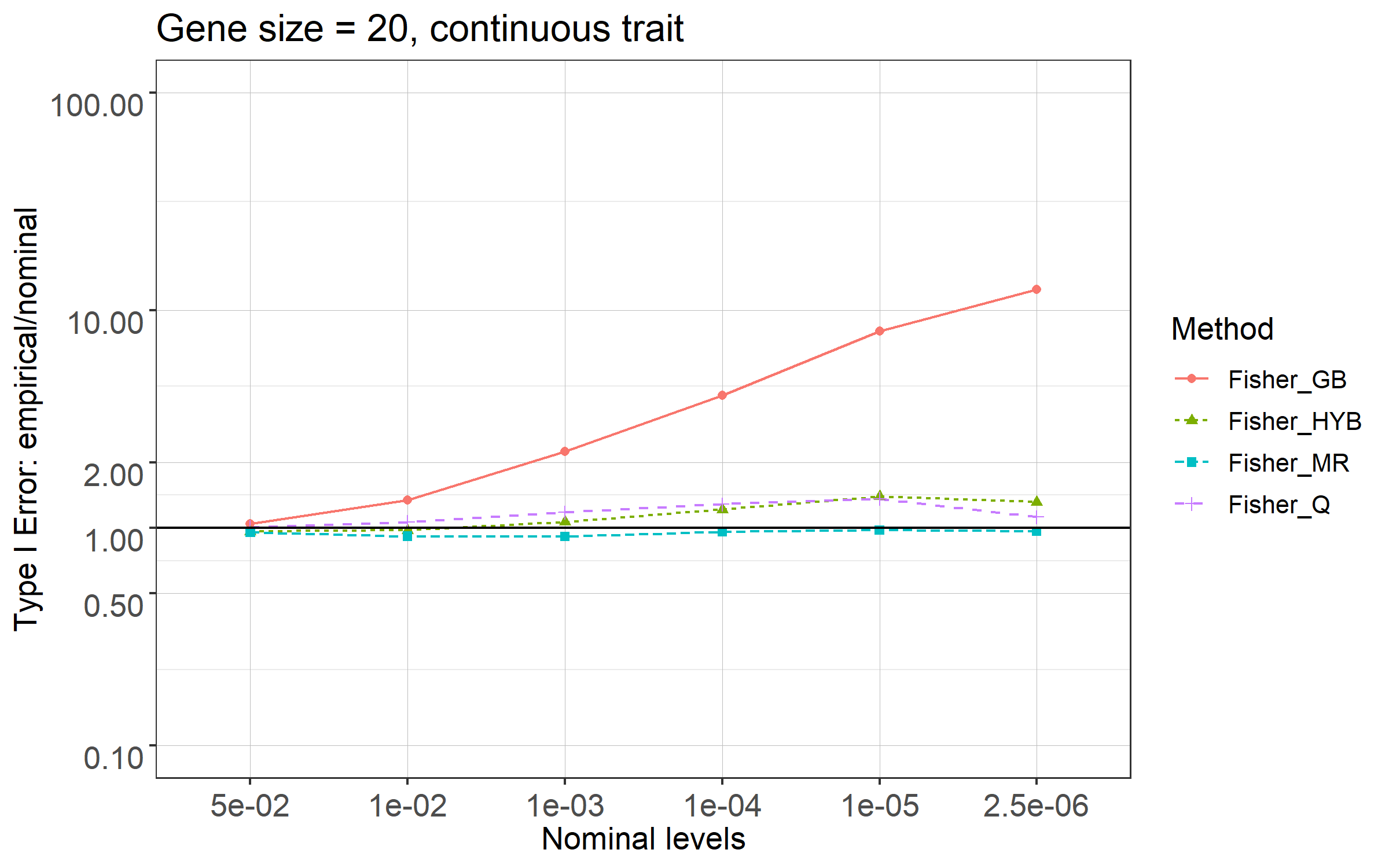}
\includegraphics[width=0.5\textwidth]{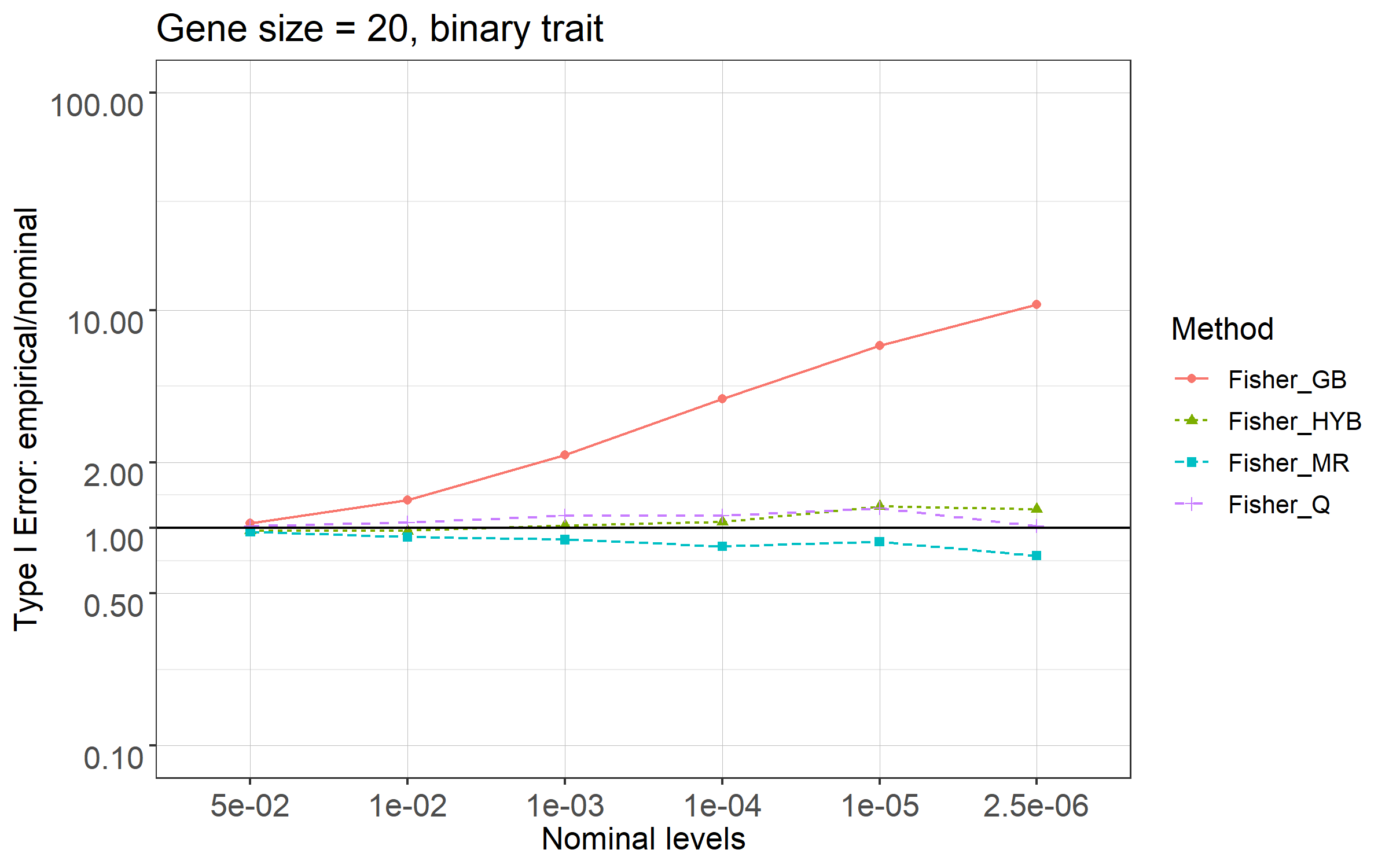}\\
\includegraphics[width=0.5\textwidth]{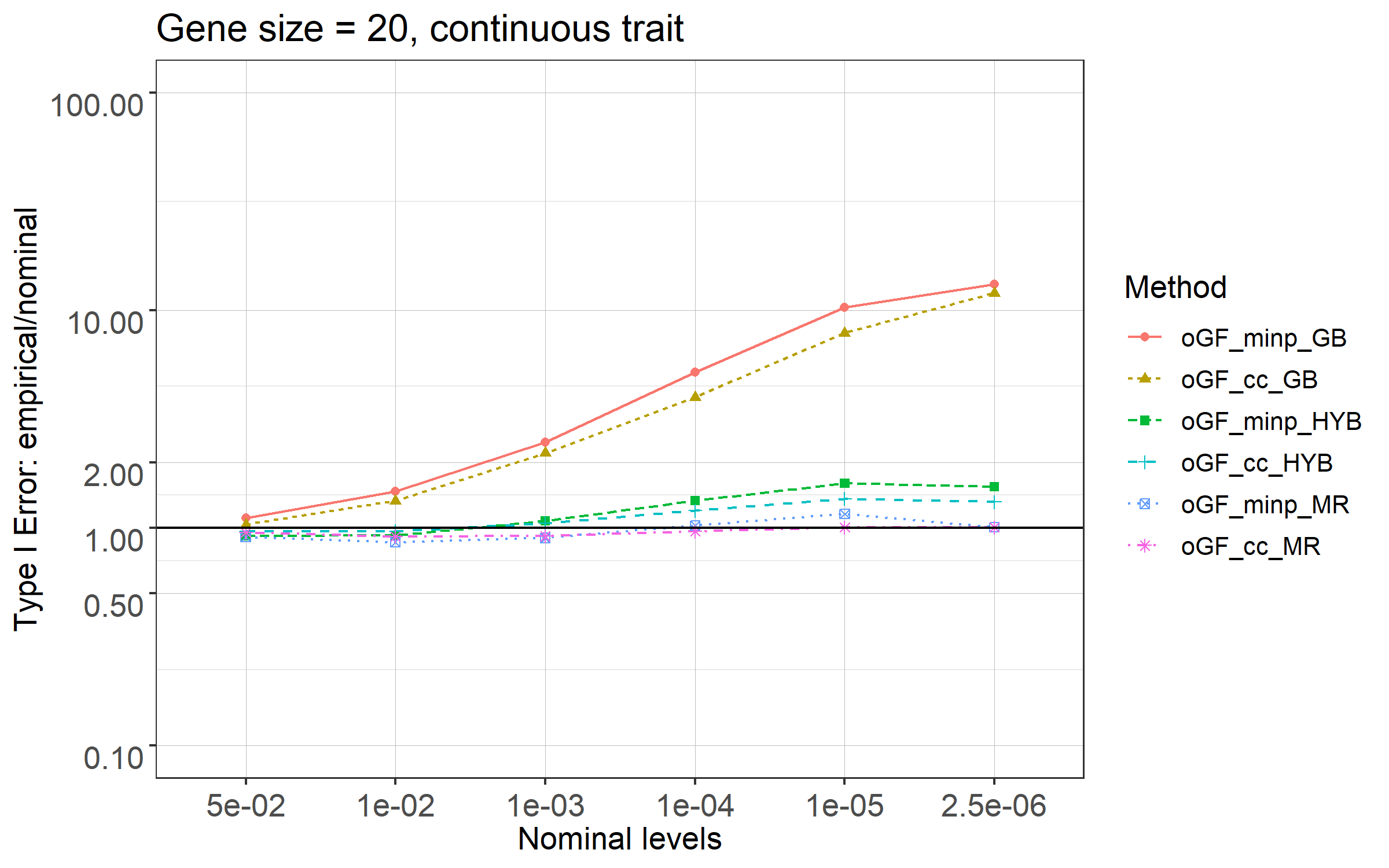}
\includegraphics[width=0.5\textwidth]{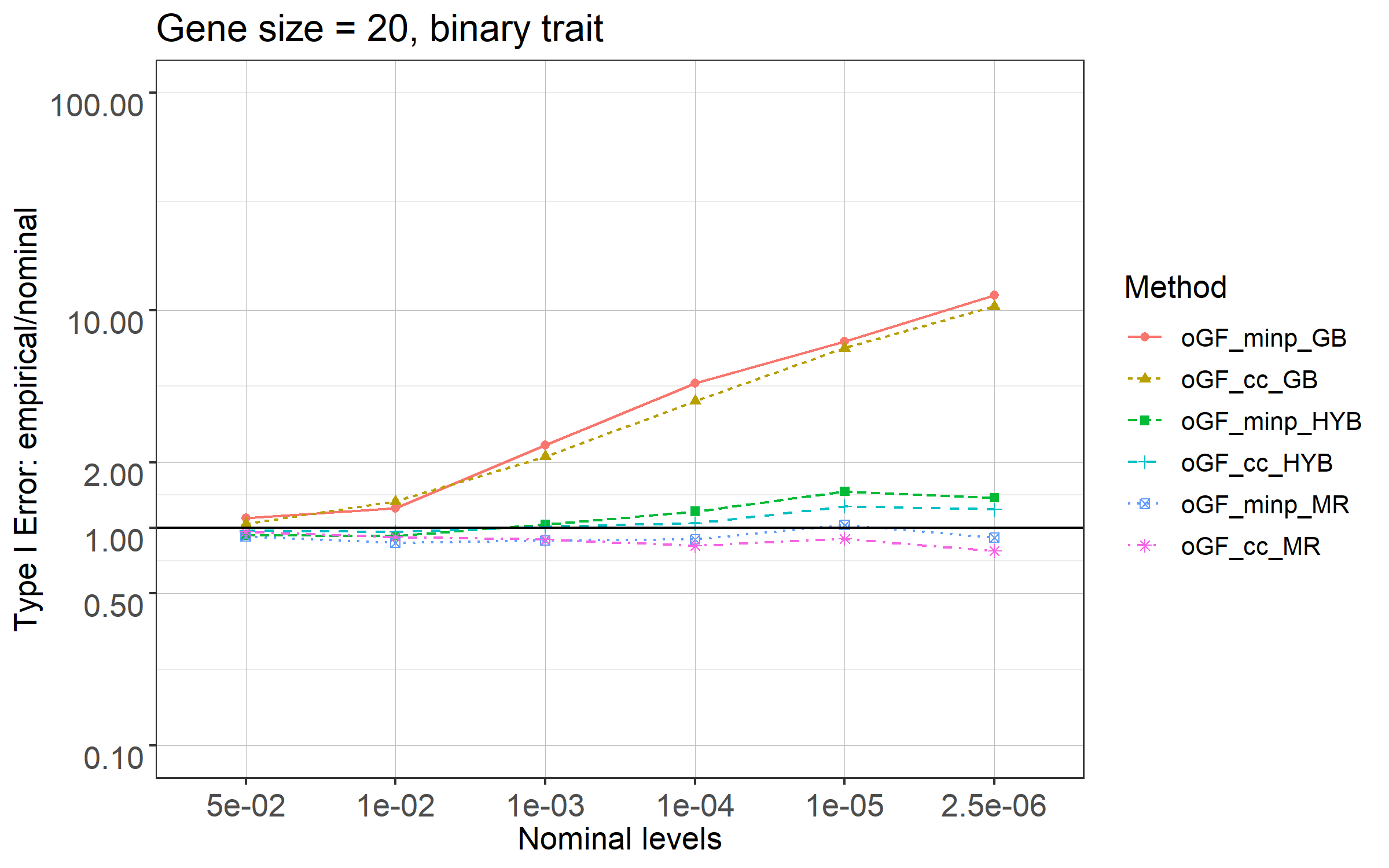}
\caption{Ratios between empirical type I error rates and the nominal $\alpha$ levels under GWAS settings. Row 1: Fisher's combination test; Row 2: oGFisher tests. GB: generalized Brown's method. HYB: the hybrid method. MR: moment-ratio matching method. Q: Q-approximation.}
\label{fig.tie_GWAS}
\end{figure}

Figure~\ref{fig.tie_GWAS} shows that the performances of the methods are consistent under linear regression (quantitative trait) and logit model (binary trait). The proposed methods are fairly robust under these applicational settings. Specifically, for Fisher's combination test, the first row of the figure shows that the generalized Brown's method starts to significantly inflate the type I error rates as early as $\alpha=0.01$. The proposed moment-matching, Q-approximation, and hybrid methods controls the type I error rates fairly well. 
In logit model the moment-ratio matching method is slightly conservative and the other two methods Q and hybrid methods are slightly liberal.  As for the oGFisher tests, the second row of the figure shows that 
the moment-ratio matching method keeps the type I error rates around the nominal levels very well. The hybrid method generates a slight inflation for oGFisher\_minp when $\alpha\leq 10^{-5}$ but it is better for oGFisher\_cc. Comparing the two omnibus tests, the minP approach tends to be slightly more liberal than the Cauchy combination approach.   

The robustness study for the multivariate $t$-distribution with small degrees of freedom is of interest because it is more fundamentally different from GMM. For example, $\mathbf{\Sigma}=\mathbf{I}$ does not indicate independence under the multivariate $t$-distribution \citep{nadarajah2005mathematical}. In the context of Fisher's combination test for dependent data, \citep{kost2002combining} approximated $\Var(T_F)$ through estimating $\Cov(-2\log P_i, -2\log P_j)$ by the scatterplot fitting strategy. Here we focus on a higher level concern -- when the moments of the statistic can be accurately obtained (e.g., by simulation), how robust the calculation methods designed for GMM would be if the input statistics were under the multivariate $t$-distribution. 

We simulated input statistics $\mathbf{Z}$ by the multivariate $t$-distribution with degree of freedom $\nu=10$, mean zero, and correlation matrix $\mathbf{\Sigma}$ defined in Table~\ref{tbl.sigma} with various $n$, $\rho$, and $\kappa$. The moments of $T_F$ were obtained by simulation to eliminate the influence of imperfect estimation. Figure~\ref{fig.tie_MVT_box} summarizes the results of 936 settings (details are given in Supplementary Table 3) in box-plots for generalized Brown's method, GGD\_123, and GD-based moment-ratio matching method. Similar to the GMM, the generalized Brown's method only works for $\alpha=0.05$, and drastically increases the type I error rates when $\alpha$ decreases.  GGD\_123 leads to some improvement but is still significantly inflated. The proposed moment-ratio matching method is not as accurate as under GMM (Figure \ref{fig.tie_GMM_box}), but overall it is still reasonably robust in controlling the type I error rates at the nominal levels till $\alpha=10^{-4}$, and only mildly inflated for smaller $\alpha$, e.g. $2.5\times 10^{-6}$. The GGD based higher moment methods, i.e. GGD\_234 and GGD\_MR, did not have solutions for most of the scenarios, and therefore they were excluded from the figure.  Supplementary Figure \ref{fig.tie_MVT_supp} gives the ratios at $\alpha=10^{-4}$ or $2.5\times10^{-6}$, showing that the Brown's method is significantly inflated with ratios up to hundreds. This is because the input $p$-values were obtained under the improper normal assumption in (\ref{equ.pvalues}). However, the moment-ratio method still largely corrected the ``mistake" and gave fairly robust results (it could mildly inflate the ratios to up to 2 at $\alpha=2.5\times 10^{-6}$). 
\begin{figure}
\includegraphics[width=0.5\textwidth]{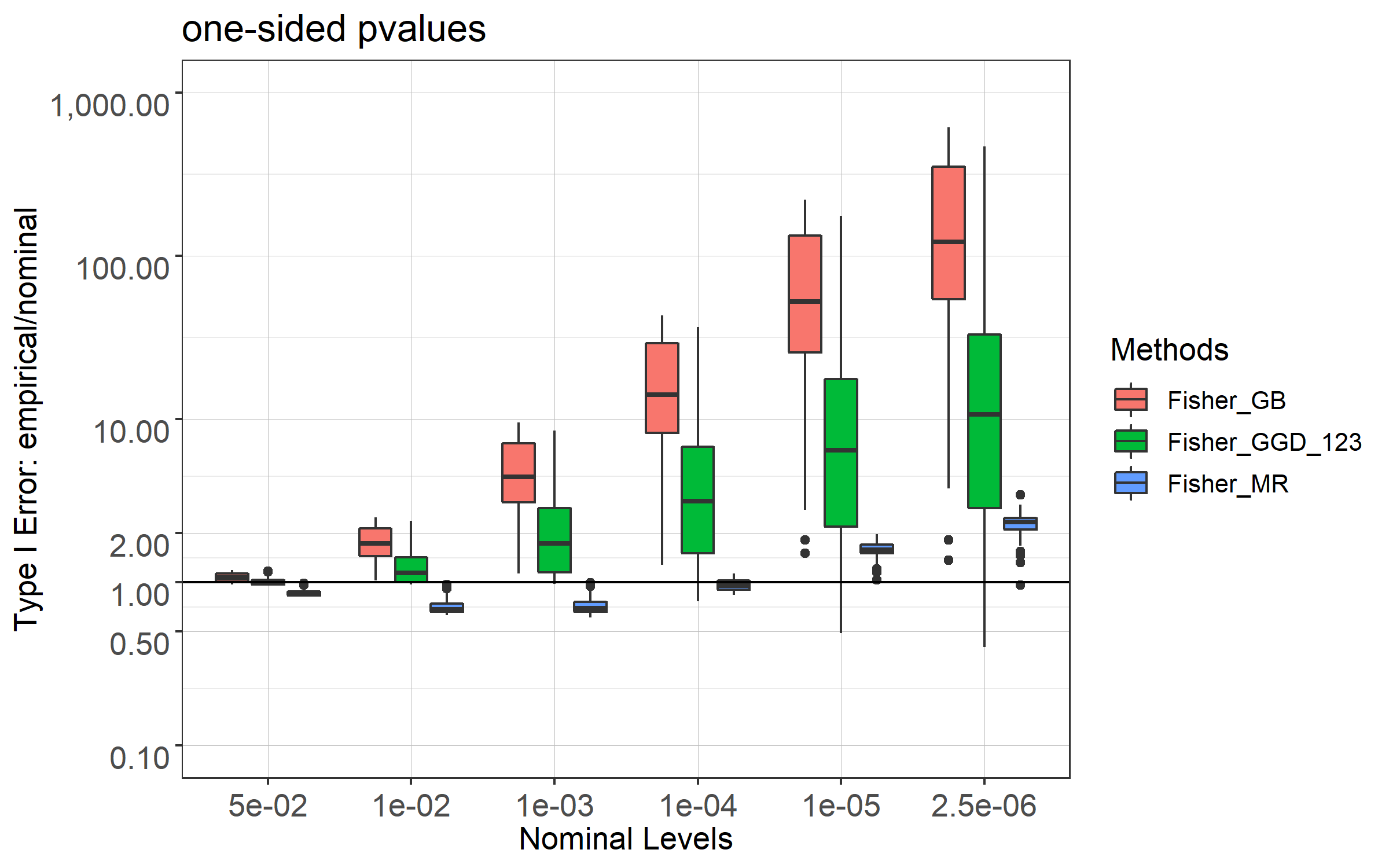}
\includegraphics[width=0.5\textwidth]{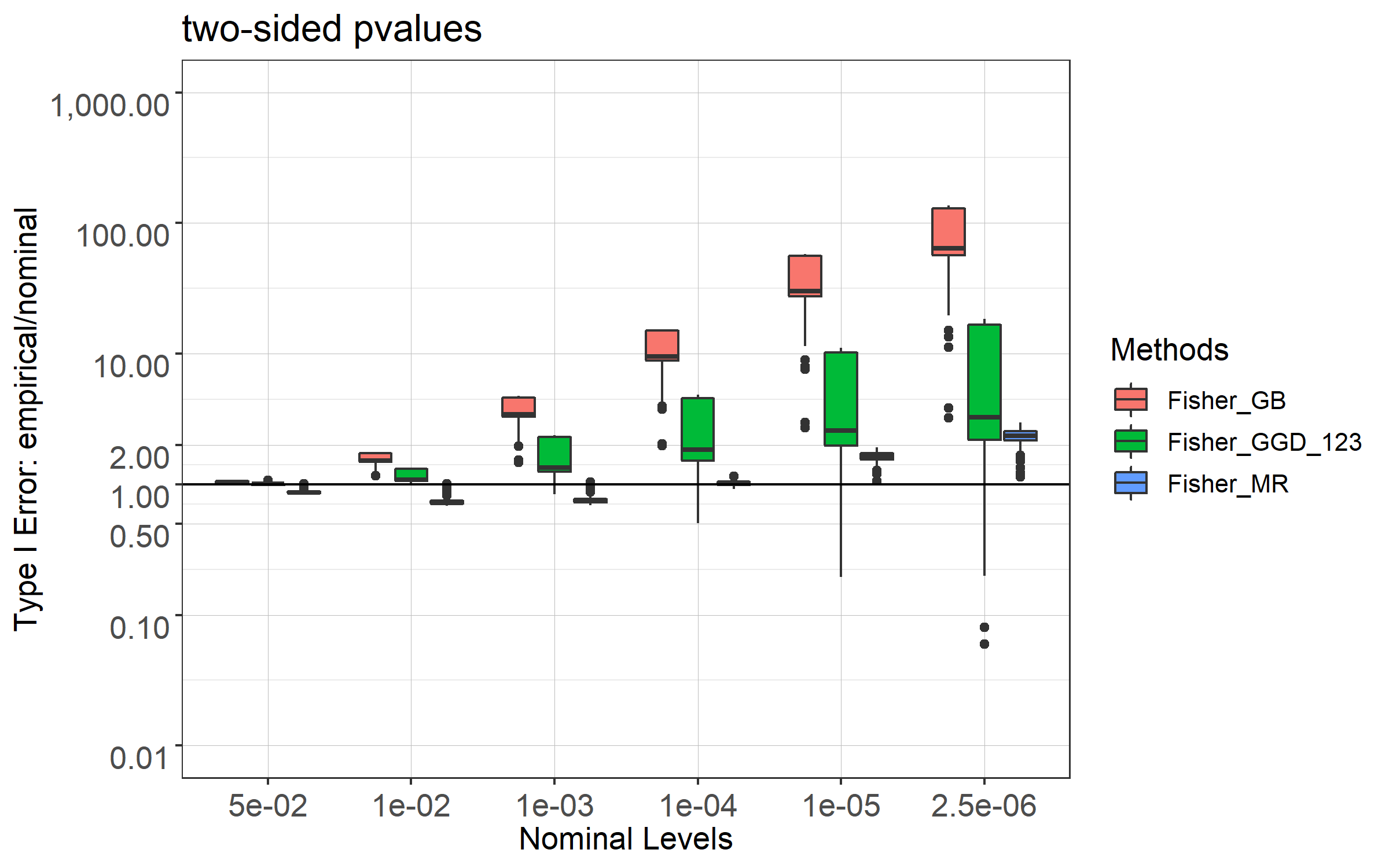}
\caption{Box-plots of the ratios between empirical type I error rates and the nominal $\alpha$ levels under the multivariate $t$-distribution.  Fisher's combination test. GB: generalized Brown's method. GGD\_123: matching the first three moments of GGD. MR: GD-based moment-ratio matching method.} \label{fig.tie_MVT_box}
\end{figure}

\section{Real data example}\label{sec:GWAS}


In this section we illustrate the application of GFisher tests and their $p$-value calculations into a gene-based rare-variant GWAS of bone mineral density (BMD). BMD is the best clinic predictor of Osteoporosis, a common disease that leads to high risk of fracture in later age of life, causing heavy public health and economic burden. Genetic factors contribute significantly to the susceptibility of Osteoporosis but a large proportion of them are yet to be discovered \citep{morris2019atlas}. Many undiscovered disease genes are likely related to rare causal variants, but rare-variant discovery is statistically challenging due to their small variation and weak signals of association at the level of population study \citep{Bodmer2008}. A simple demonstrative analysis here indicates that GFisher can provide promising results when applied in the gene-based SNP-set analysis, and that our calculation methods provide much more accurate gene $p$-values than Brown's approximation does. 

We used a publicly available data (http://www.gefos.org/?q=content/data-release-2015) of summary statistics from a large meta-study of whole genome sequencing (n=2,882 from UK10K), whole exome sequencing (n = 3,549), deep imputation of genotyped samples using a combined UK10K/1000 Genomes reference panel (n=26,534), and {\it de novo} replication genotyping (n=20,271) \citep{zheng2015whole}. We focused on the phenotype of Femoral Neck bone mineral density (FN-BMD), and considered $1,367,983$ rare variant SNPs with minor allele frequency between 0.5\% and 5\%. The data source paper indicates that the SNP $p$-values were generated by GWAMA \citep{magi2010gwama}, in which meta-analysis tests are linear combination of the Z statistics 
from independent meta studies. 
Therefore, it is reasonable to assume that the SNP test statistics asymptotically follow normal distribution and their correlations can be estimated by the genetic linkage disequilibrium (LD) among these SNPs \citep{lin2010relative}. It is also reasonable to assume the SNP $p$-values are two-sided. 

In the gene-based SNP-set association analysis, SNPs were grouped into $21,320$ genes. The number of SNPs in a gene ranges from $1$ to $5,592$ with median $22$. For each gene with $n \geq 2$ SNPs, GFisher statistics were applied to test the null hypothesis that none of the SNPs in the given gene are associated with FN-BMD, i.e., their two-tailed $p$-values in (\ref{equ.pvalues}) came from a GMM in (\ref{equ.GMM}) with zero mean vector.  We applied Fisher's combination test and oGFisher\_minp and oGFisher\_cc (both with $w_i=1$ and adapting to $d_i \in \{1, 2, 3\}$, $i=1, \cdots, n$.) 

\begin{figure}
\centering
\includegraphics[width=0.49\textwidth]{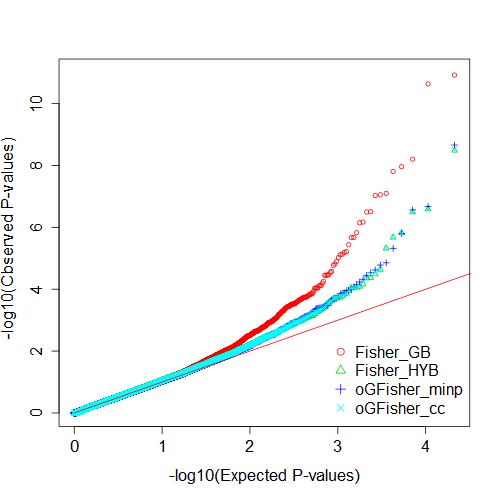}
\includegraphics[width=0.49\textwidth]{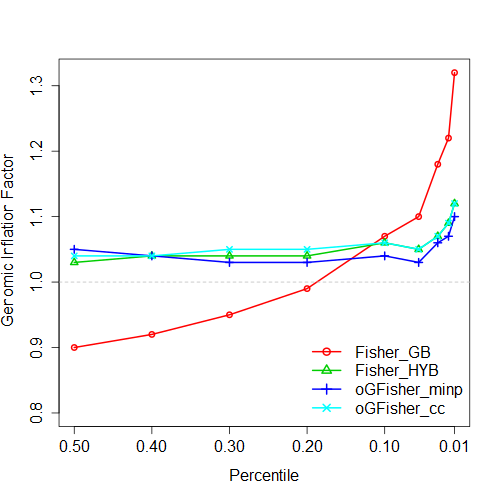}
\caption{Genome-wide type I error rate control. Left: QQ plot; right: the percentile-dependent genomic inflation factor $\lambda(p)$ over $p$ from 0.5 to 0.01. Gene $p$-values of the oGFisher tests were calculated based on the hybrid method.}
\label{fig.gwas}
\end{figure}

The quantile-quantile (QQ) plot and the genomic inflation factor are used to visualize how well the type I error rate is controlled. QQ plot compares the calculated gene $p$-values with the expected $p$-values under the null. 
Biologically, the majority of all genes are not expected to be causal to FN-BMD. Therefore, a good type I error rate control means the majority of the dots (each dot for one gene) in QQ plot should be aligned over the diagonal. QQ plot in Figure~\ref{fig.gwas} shows that Fisher's combination test $p$-values calculated by the generalized Brown's approximation method start to be inflated as early as $0.1$. The hybrid method brought down the dots much closer to the diagonal, indicating a better type I error rate control. To get a closer look at the control over different percentiles, we plot the percentile-dependent genomic inflation factor 
$$
\lambda(p) = F_1^{-1}(1-pval_p)/F_1^{-1}(1-p),
$$
where for any $p\in (0, 1)$, $pval_p$ denotes the $100p$th percentile of the calculated gene $p$-values, $F_1^{-1}$ denotes the inverse CDF of $\chi_1^2$. The most commonly used genomic inflation factor is $\lambda(0.5)$ at the median. A good type I error rate control should have $\lambda(p) \approx 1$ unless at small $p$. The right panel of Figure~\ref{fig.gwas} plots $\lambda(p)$ over $p$ from 0.5 to 0.01. It shows that the generalized Brown's method controls type I error rate poorly because $\lambda(p)$ varies from $0.9$ to more than $1.3$. The hybrid method more preferably kept $\lambda(p)$ close to 1. For clarity the figure did not include the results of the moment-ratio matching method because they are very similar to the results of the hybrid method.  
Also, oGFisher tests performed very similarly as the Fisher's combination because the statistics with $d_i=1, 2, 3$ were highly correlated and generated similar gene $p$-values. 

Results also show that GFisher can likely provide extra statistical power over the traditional single-SNP based analysis for detecting novel disease genes. Specifically, five genes passed the Bonferroni-corrected genome-wide significance level $\alpha=0.05/21,320\approx 2.35\times 10^{-6}$: \textit{LSM12} (17:42112002:42144987, $p=3.3\times10^{-9}$),  \textit{RP11-4F22.2} (17:64394186:64412972, $p=2.6\times10^{-7}$), \textit{NAGS} (17:42082031:42086436, $p=3.2\times10^{-7}$), \textit{TMC2} (20:2517252:2622430, $p=1.5\times10^{-6}$) and \textit{MIR548H4} (15:69116302:69489862, $p=2.1\times10^{-6}$). These gene level $p$-values are more significant than SNP $p$-values. For example, the smallest SNP $p$-values in \textit{LSM12}, \textit{RP11-4F22.2} and \textit{TMC2} are $1\times 10^{-7}$, $2\times 10^{-6}$, and $6\times 10^{-4}$, respectively. Such a result indicates that the individually weak signals at the SNP level could cumulatively show a strong evidence at the gene level. Therefore, GFisher is promising in detecting groups of weak genetic effects that cannot be detected individually. Furthermore, \textit{LSM12} and \textit{NAGS} are close to (within 120k base pairs) gene \textit{PYY} (17:42030106:42081837, $p=4.8\times10^{-6}$), which barely missed the genome-wide threshold. The level of Peptide YY (PYY) is well-known to be associated with BMD \citep{utz2008peptide}. Further biological validations for the functionality of these top putative genes are needed. 


\section{Discussion}\label{sec:disc}


For a general family of Fisher's combination type statistics, referred as the GFisher, this paper proposes several new methods for improving the accuracy in calculating small test $p$-values over traditional moment-matching methods. The new methods are based on two novel ideas. The first idea is a moment-ratio matching strategy, which we haven't seen applied in distribution estimation. It involves higher moments to provide better versatility to the tail property of the target distribution, while allowing fewer numbers of parameters and thus easier computation than direct moment matching.  The second idea is to component-wisely match the joint distribution of the summands of the test statistic. An analytical Q-approximation is developed for the scenario of two-sided input $p$-values. Combining the calculated higher moments based on Q-approximation and the moment-ratio matching, the hybrid method balances both accuracy and computational efficiency. These new methods facilitate the application of GFisher into large scale data analyses, where stringent type I error control is demanded. 

A few limitations of this work are to be addressed in future work. First, for one-tailed input $p$-values, current moment-ratio matching method still relies on simulated high moments (i.e., skewness and kurtosis). It would be nice to further develop an analytical method to approximate these high moments for faster computation. Second, for two-tailed input $p$-values, the Q-approximation requires statistic's degrees of freedom $d_i$'s to be integers. When any $d_i$ is non-integer, we could calculate the test $p$-value by the weighted average of the results from $\lfloor d_i\rfloor$ and $\lceil d_i\rceil$. However, a careful design of the procedure is needed. 
Furthermore, this paper focuses on controlling the type I error of GFisher. Similar as other omnibus tests in general, the proposed oGFisher procedures should in theory provide a powerful and robust test that adapts to given data. Meanwhile, a careful power study on GFisher, including the optimality of choosing $w_i$ and $d_i$, is of great theoretical and practical interests. 
We will present the relevant results of such a power study in the near future. 

\section{Conclusion}\label{sec:conc}

For the GFisher tests of correlated data, the traditional Brown's moment-matching approximate methods by either gamma distribution (GD) or the generalized gamma distribution (GGD) are not adequate to control small type I error rate $\alpha$. A moment-ratio matching method is proposed and its high accuracy is evidenced. For two-sided input $p$-values, the analytical Q-approximation and hybrid methods are further developed. They are computationally efficient and provide adequate accuracy. With these new methods, the GD model is sufficient on accuracy, and is preferred over the GGD due to computational simplicity. The new methods are developed based on Gaussian mean model, but they are reasonably robust in the scenarios of GLM and the multivariate $t$-distribution. The new developments are expected to facilitate the application of the GFisher tests into large scale data analyses for broad scientific research problems that can be addressed by global hypothesis testing.

\bigskip
\begin{center}
{\large\bf SUPPLEMENTARY MATERIAL}
\end{center}

\beginSupplFigTable 

Supplementary figures are given below for complementary support of the conclusions discussed in the manuscript. Complete settings and results are provided in Supplementary Tables 1--3 in a separate Excel file. 
 
\begin{figure}
\subfloat[Fisher; $\mathbf{\Sigma}=$ Equal(0.3)-I]{\includegraphics[width=0.49\textwidth]{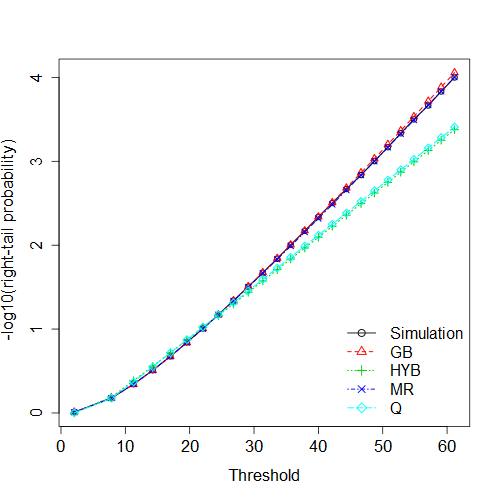}}
\subfloat[Fisher; $\mathbf{\Sigma}$= Equal(0.3)-III]{\includegraphics[width=0.49\textwidth]{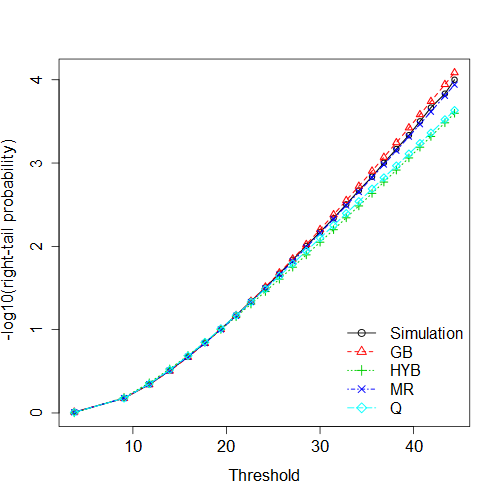}}\\
\subfloat[GFisher $d_i=i$, $w_i=2i/7$; $\mathbf{\Sigma}=$ Equal(0.3)-I]{\includegraphics[width=0.49\textwidth]{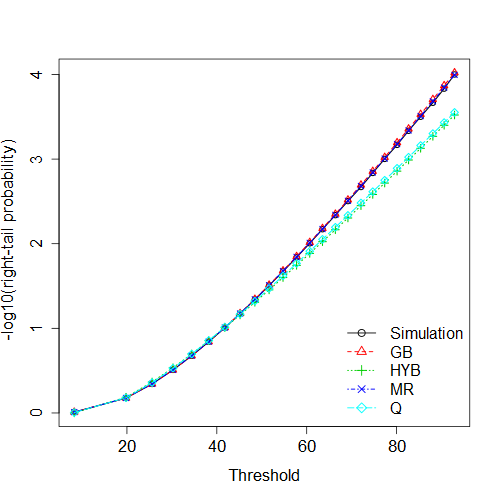}}
\subfloat[GFisher $d_i=i$, $w_i=2i/7$; $\mathbf{\Sigma}$= Equal(0.3)-III]{\includegraphics[width=0.49\textwidth]{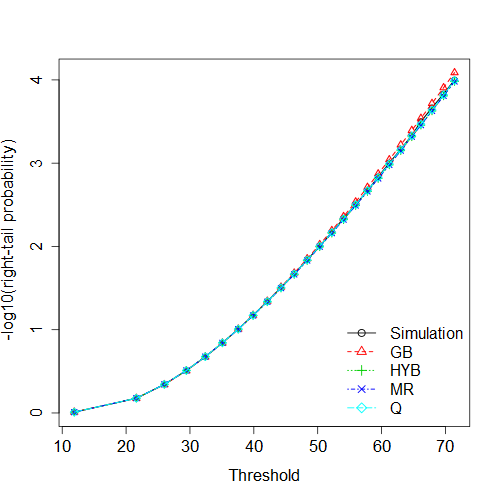}}
\caption{Right-tail probability ($-\log_{10}$) of $T$ when input $p$-values are one-sided. $n=6$. Two $\mathbf{\Sigma}$ patterns with $\rho=0.3$ are defined in Table \ref{tbl.sigma}.
}
\label{fig.survival_oneside_supp}
\end{figure}

\begin{figure}
\subfloat[Fisher; $\mathbf{\Sigma}=$ Equal(0.3)-I]{\includegraphics[width=0.49\textwidth]{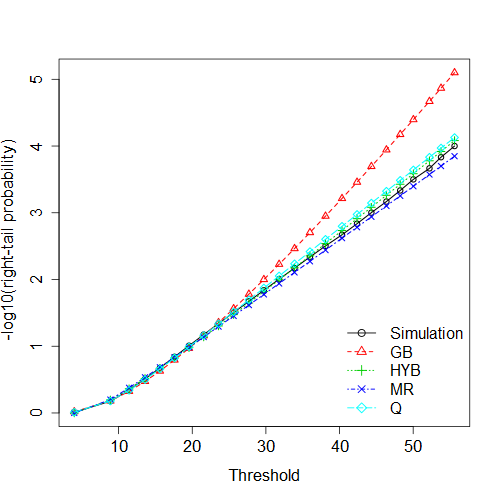}}
\subfloat[Fisher; $\mathbf{\Sigma}$= Equal(0.3)-III]{\includegraphics[width=0.49\textwidth]{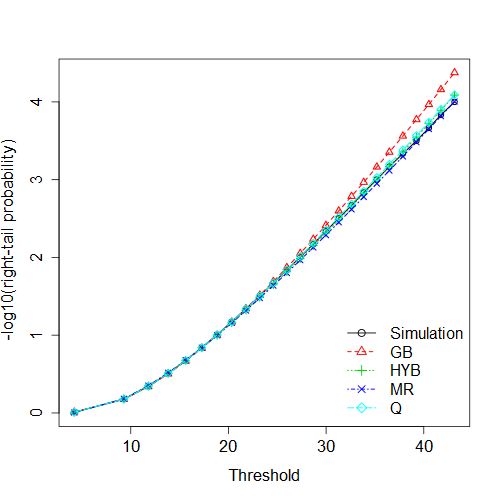}}\\
\subfloat[GFisher $d_i=i$, $w_i=2i/7$; $\mathbf{\Sigma}=$ Equal(0.3)-I]{\includegraphics[width=0.49\textwidth]{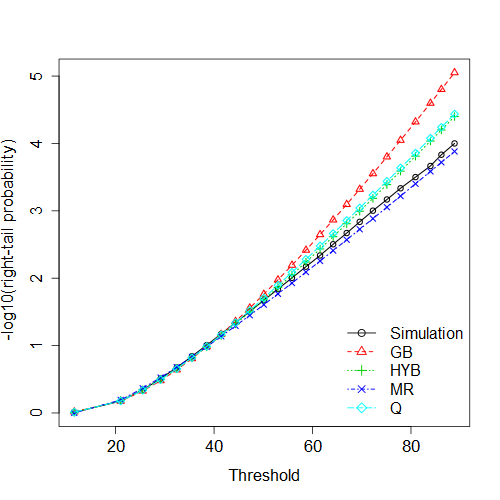}}
\subfloat[GFisher $d_i=i$, $w_i=2i/7$; $\mathbf{\Sigma}$= Equal(0.3)-III]{\includegraphics[width=0.49\textwidth]{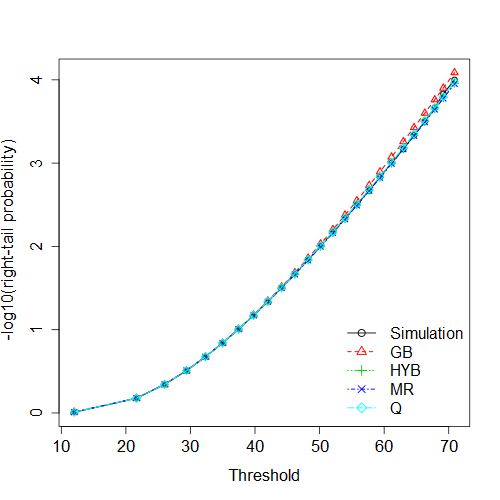}}
\caption{Right-tail probability ($-\log_{10}$) of $T$ when input $p$-values are two-sided. $n=6$. Two $\mathbf{\Sigma}$ patterns with $\rho=0.3$ are defined in Table \ref{tbl.sigma}. 
}
\label{fig.survival_twoside_supp}
\end{figure}

\begin{figure}
\centering
\includegraphics[width=0.46\textwidth]{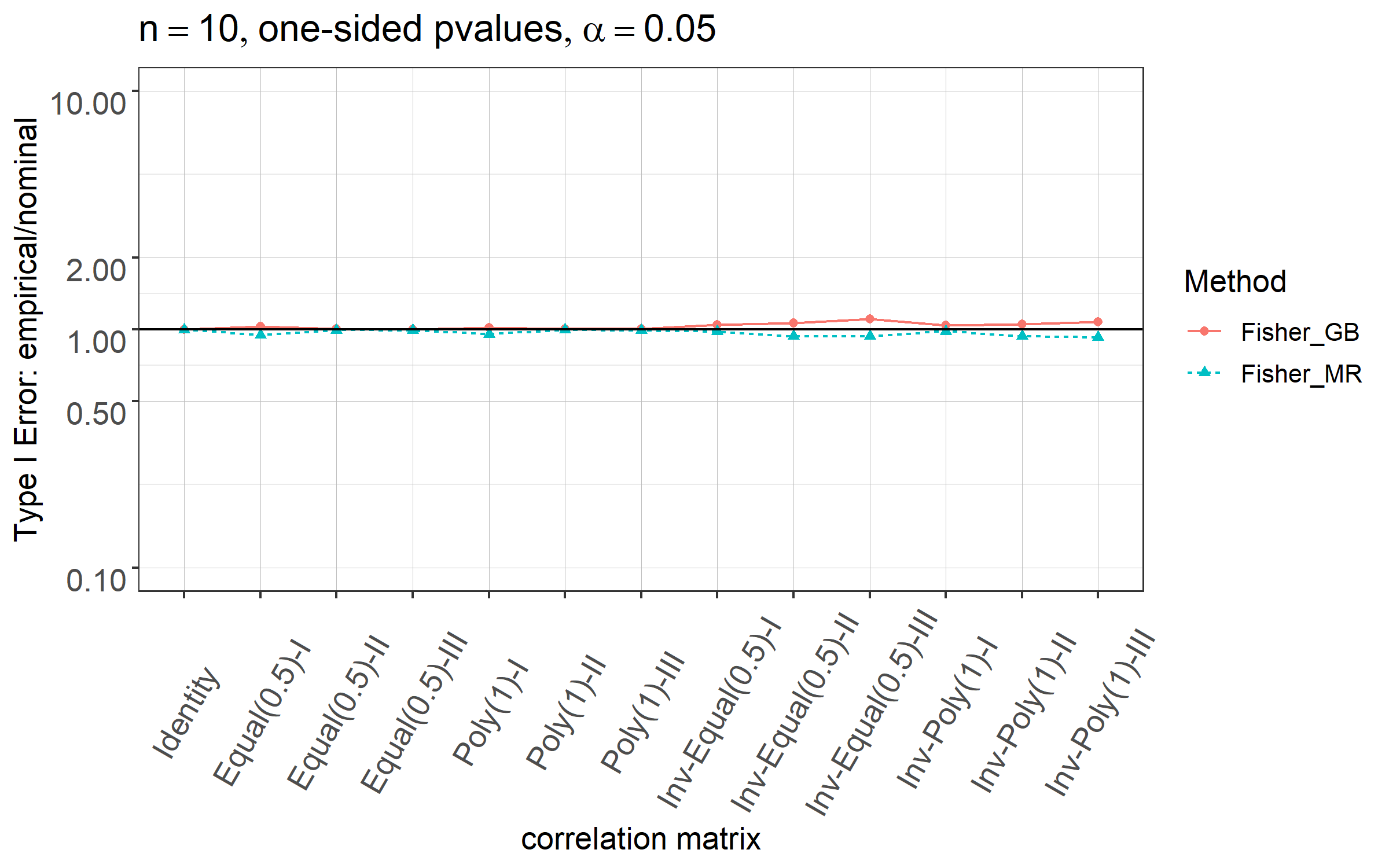}
\includegraphics[width=0.46\textwidth]{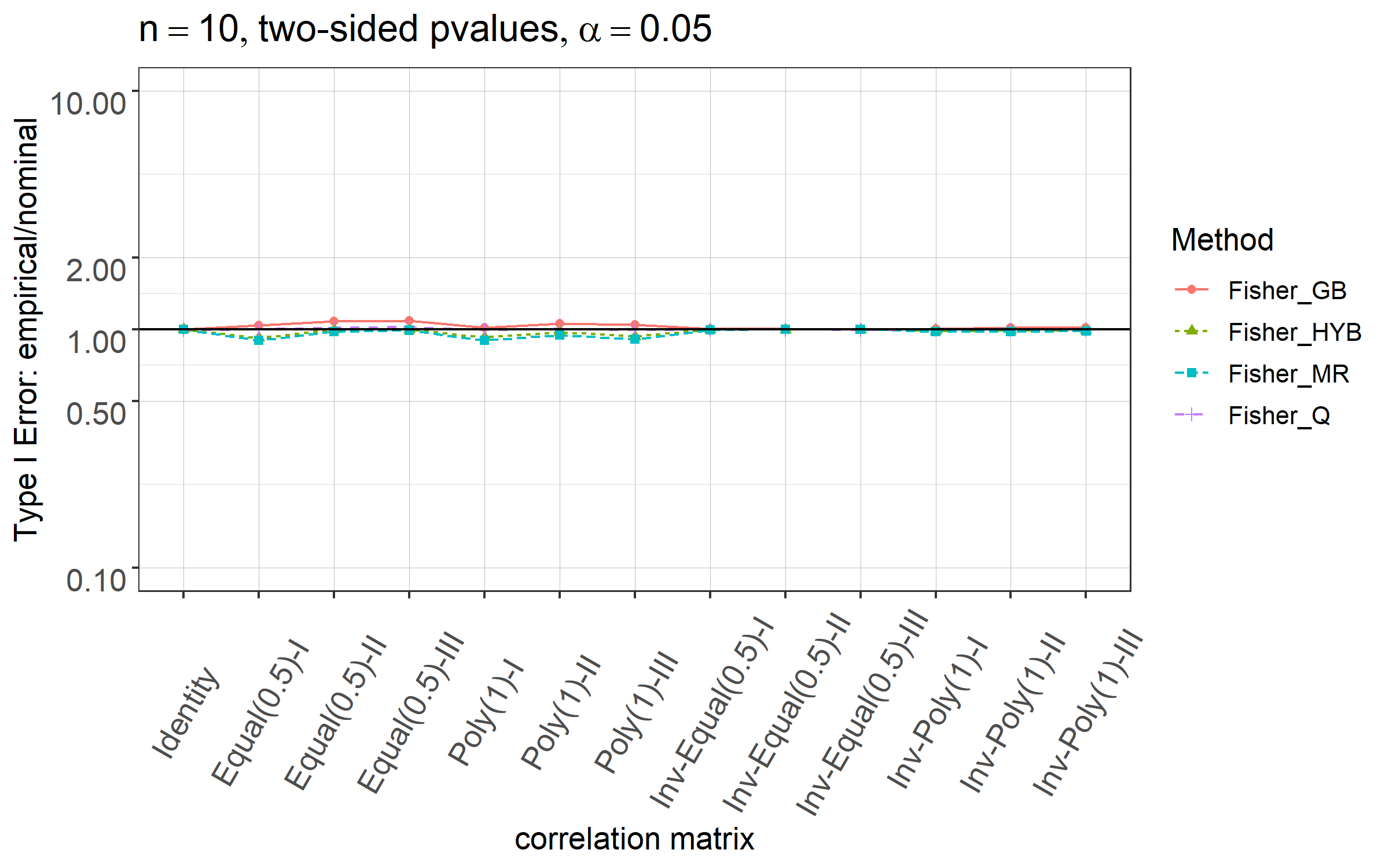}\\
\includegraphics[width=0.46\textwidth]{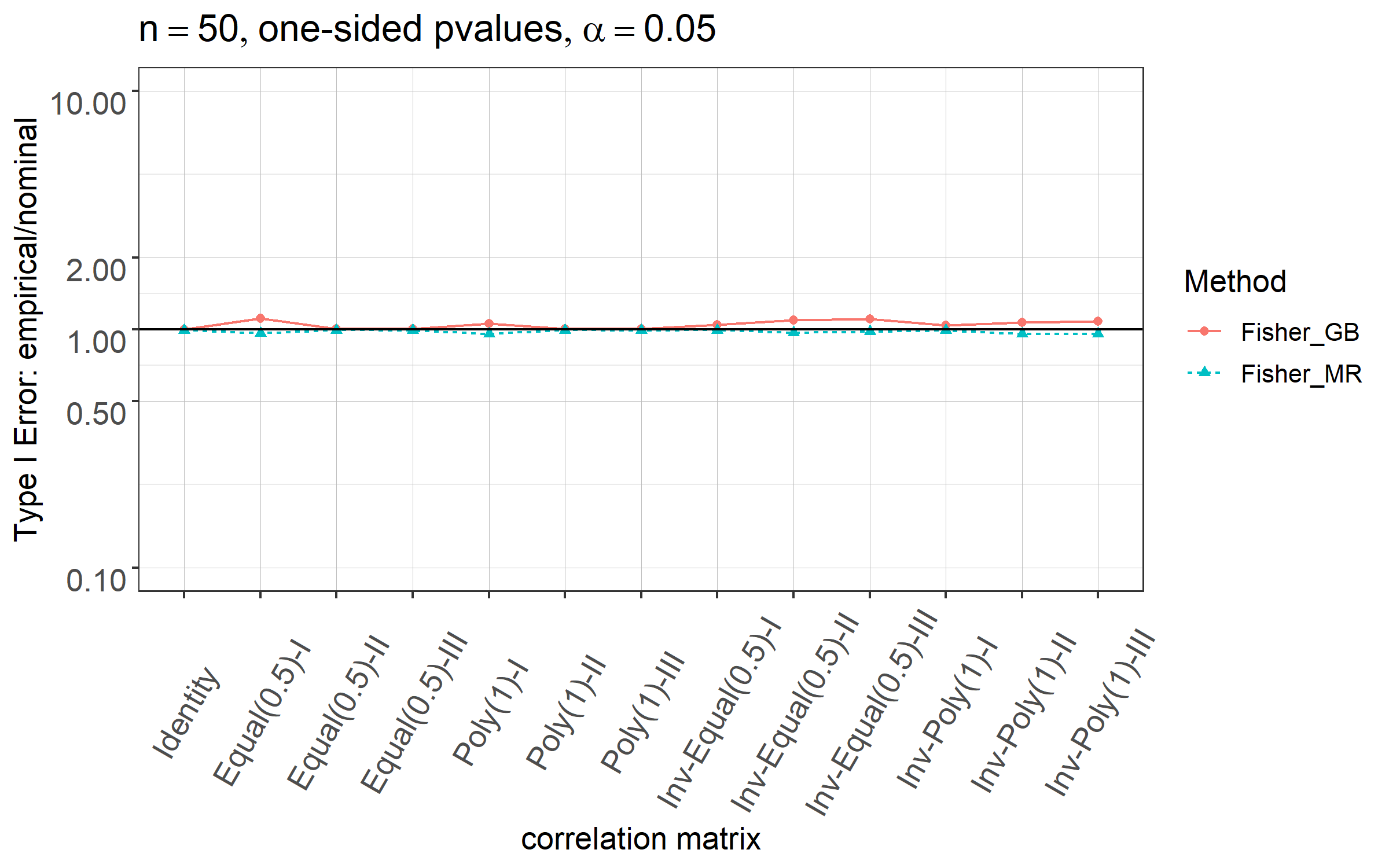}
\includegraphics[width=0.46\textwidth]{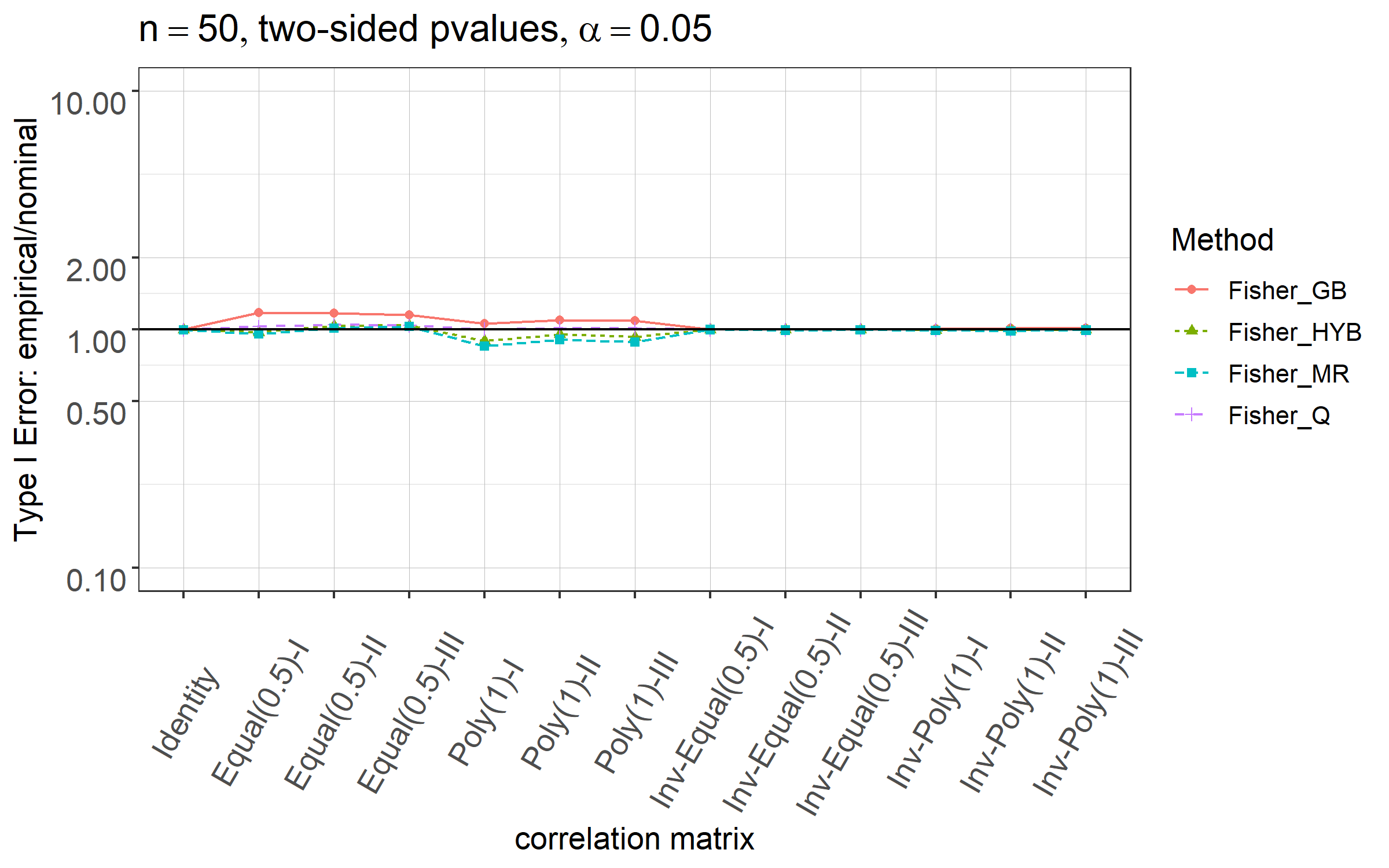}\\
\includegraphics[width=0.46\textwidth]{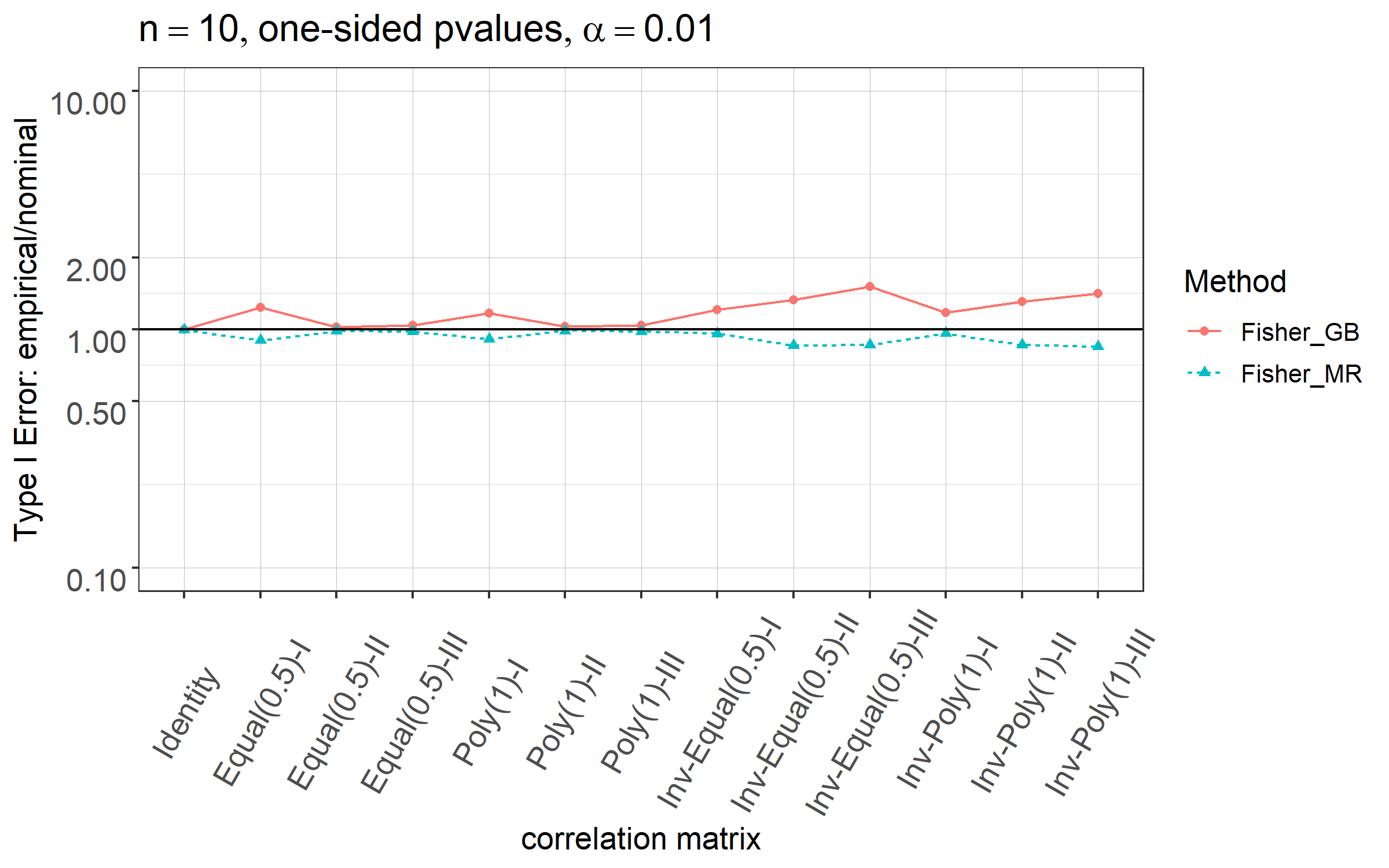}
\includegraphics[width=0.46\textwidth]{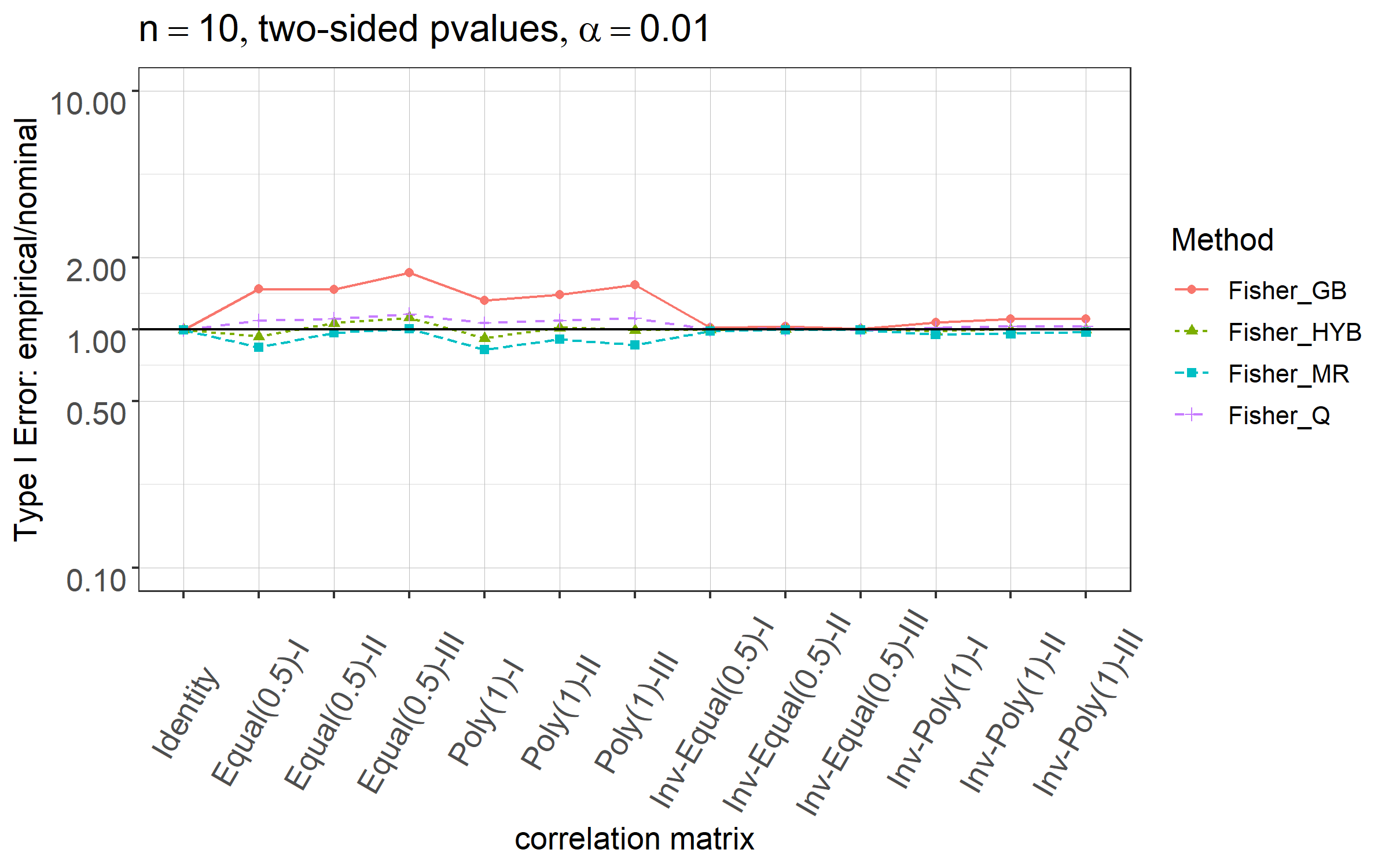}\\
\includegraphics[width=0.46\textwidth]{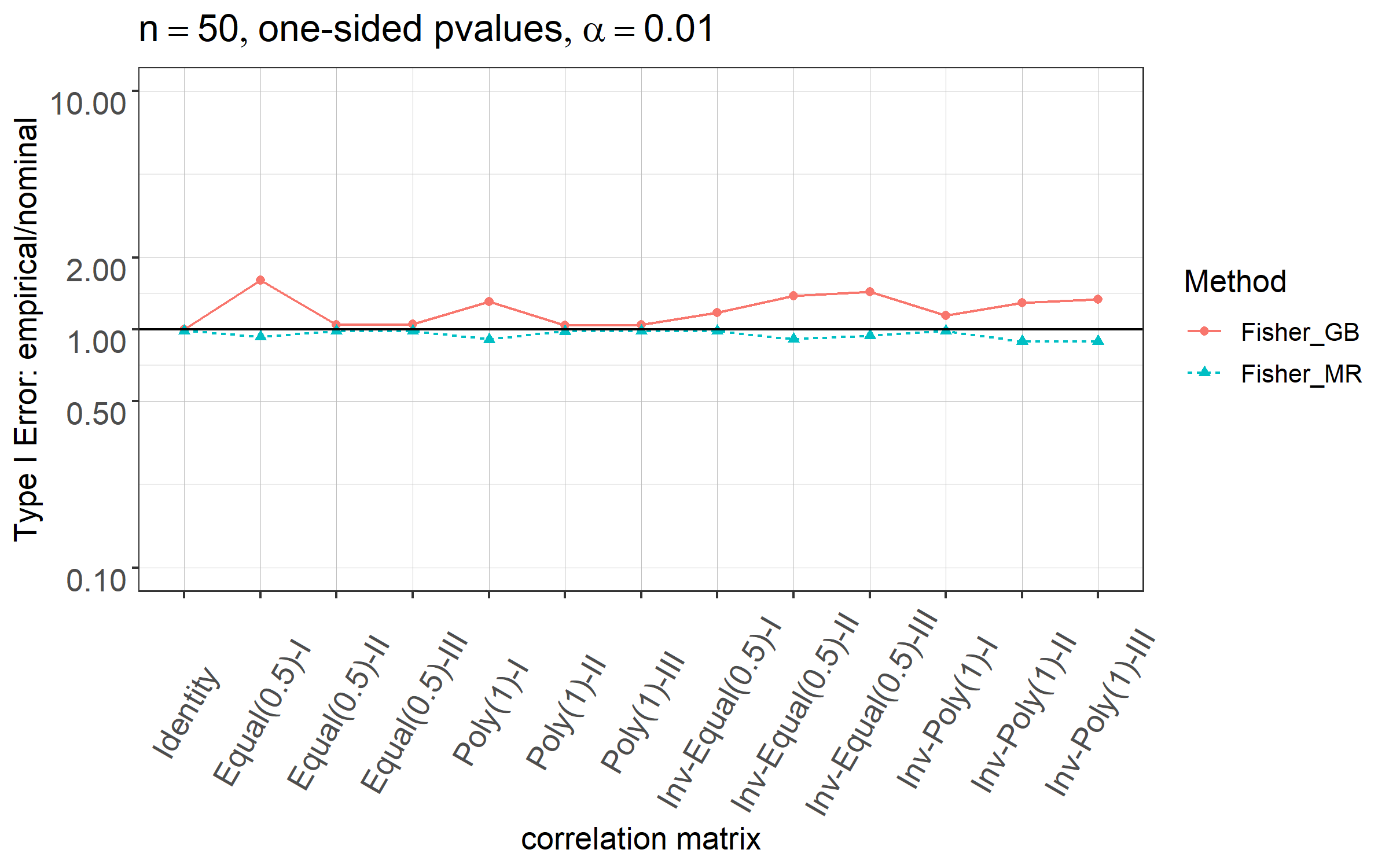}
\includegraphics[width=0.46\textwidth]{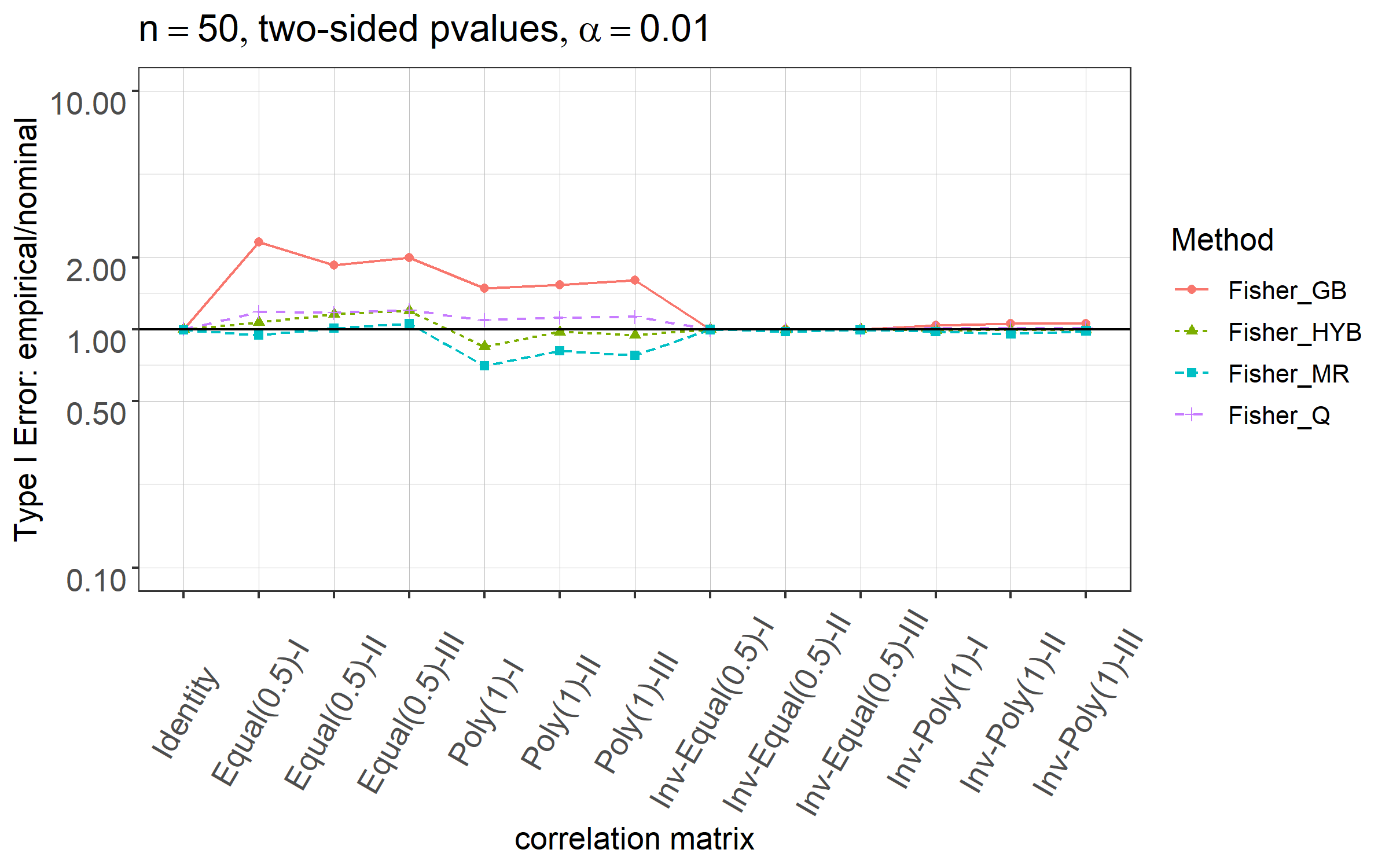}\\
\caption{Ratios between empirical type I error rates and nominal $\alpha=0.05$ or $0.01$. Fisher's combination test under independence and the 12 correlation structures in Table~\ref{tbl.sigma}. GB: generalized Brown's method. HYB: the hybrid method. MR: moment-ratio matching method. Q: Q-approximation. 
}
\label{fig.tie_GMM_cases_supp1}
\end{figure}

\begin{figure}
\centering
\includegraphics[width=0.46\textwidth]{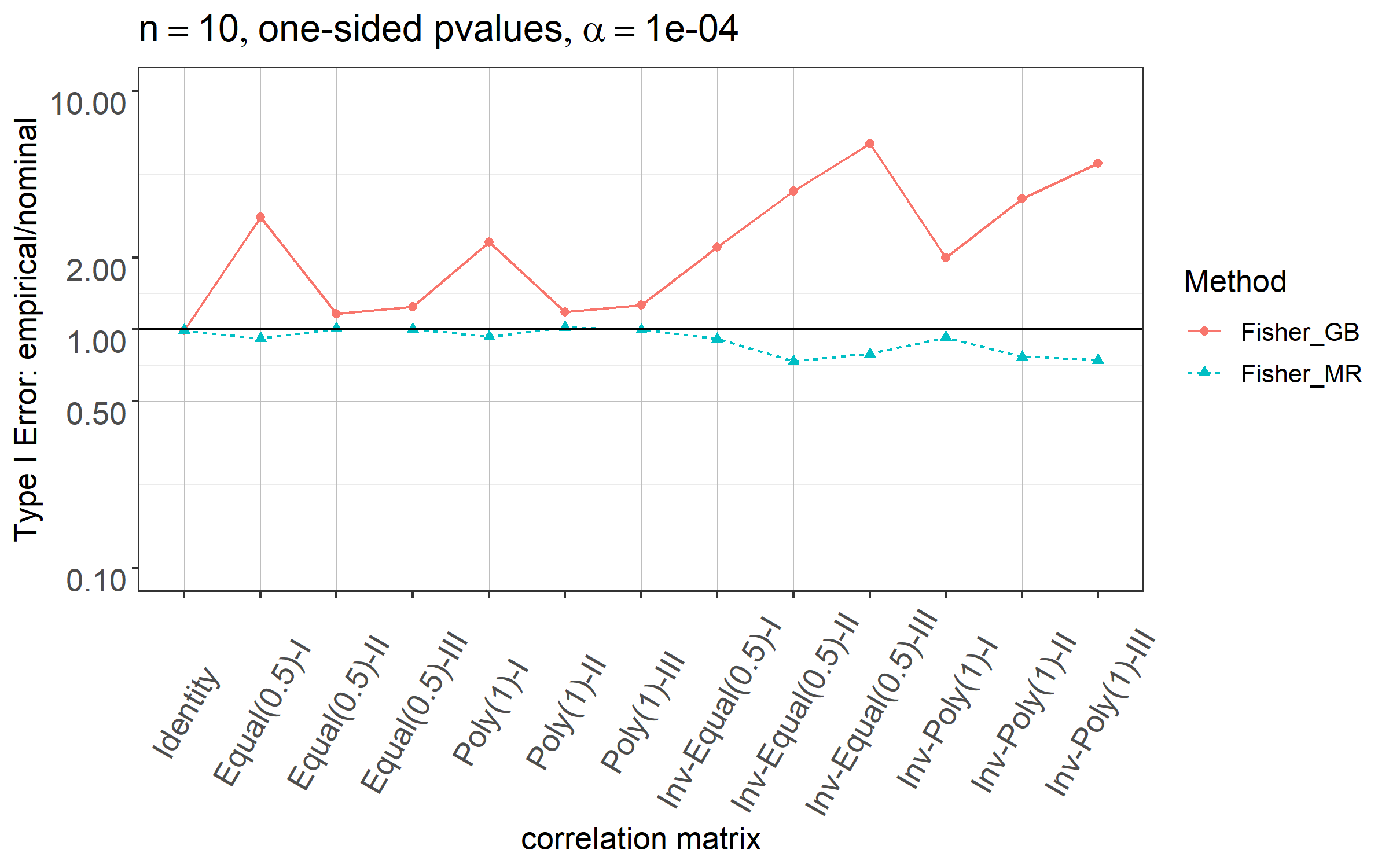}
\includegraphics[width=0.46\textwidth]{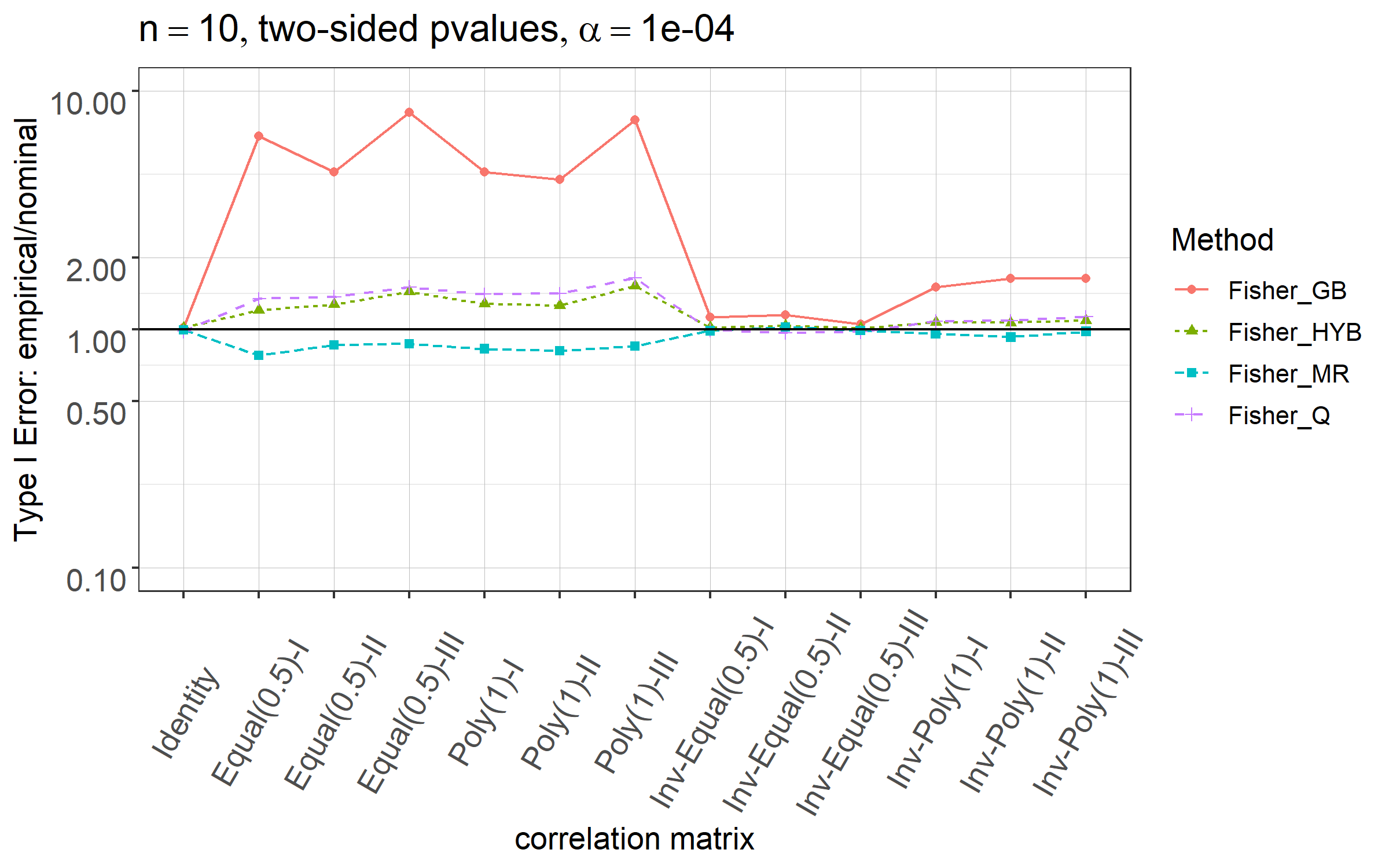}\\
\includegraphics[width=0.46\textwidth]{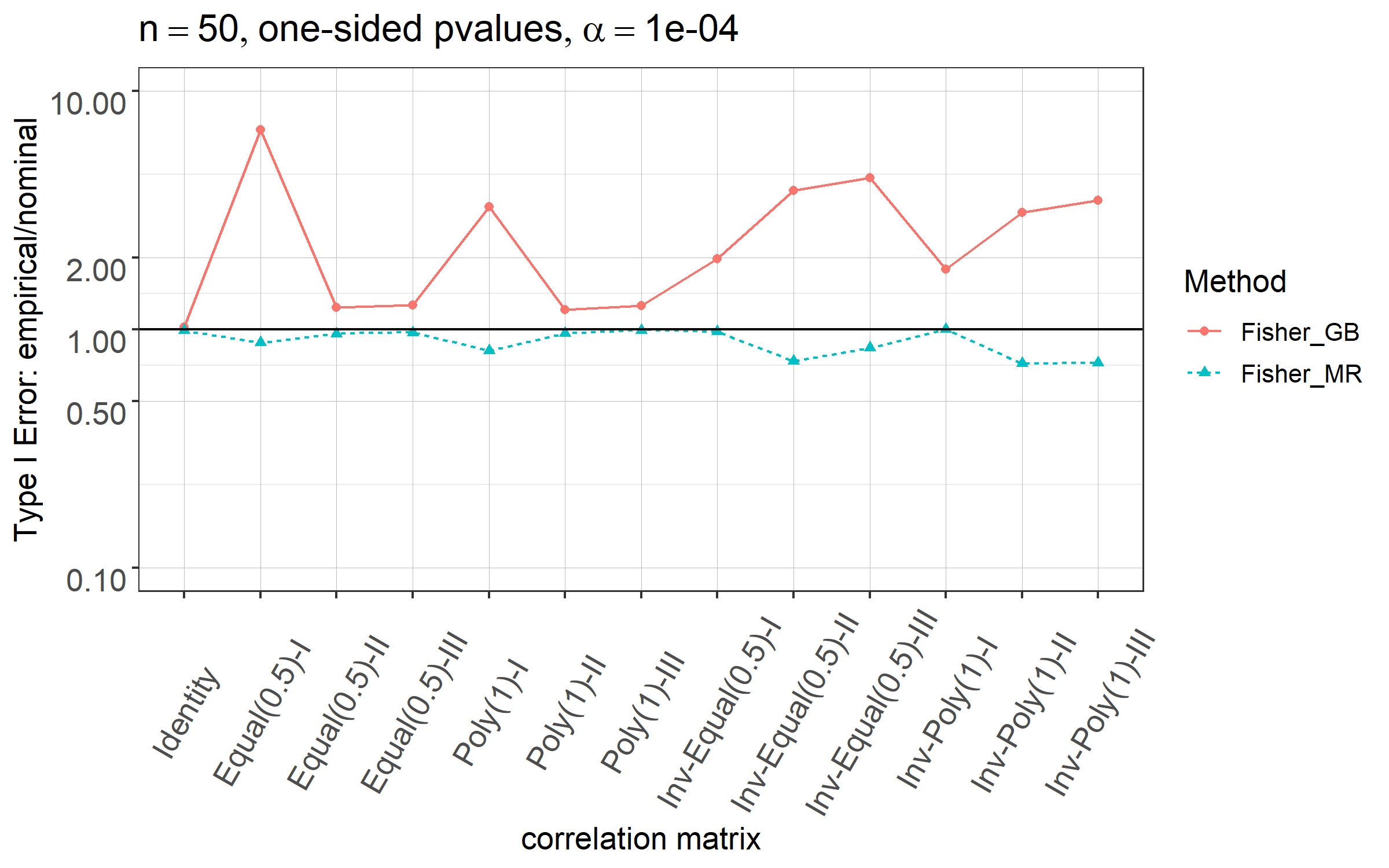}
\includegraphics[width=0.46\textwidth]{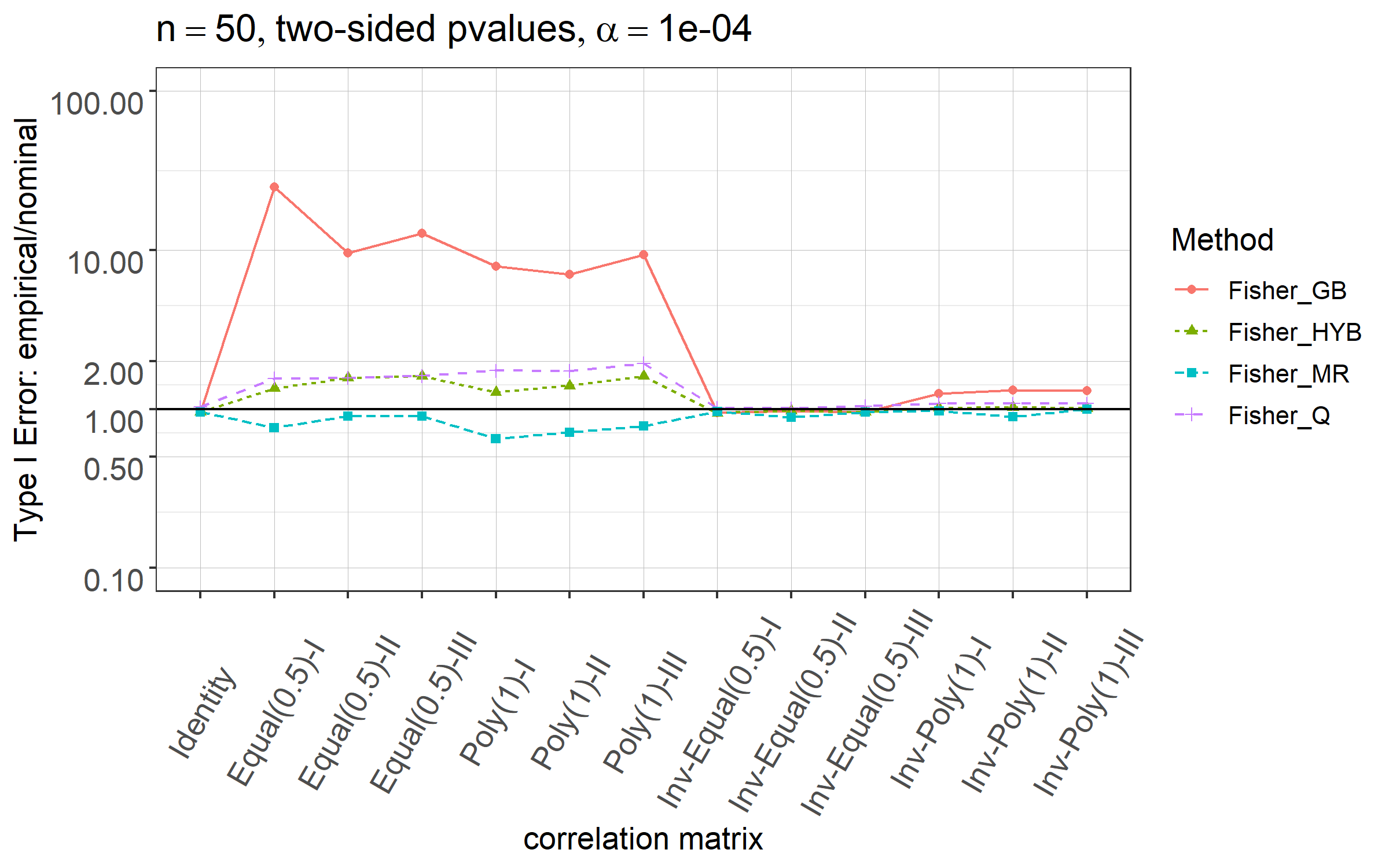}\\
\includegraphics[width=0.46\textwidth]{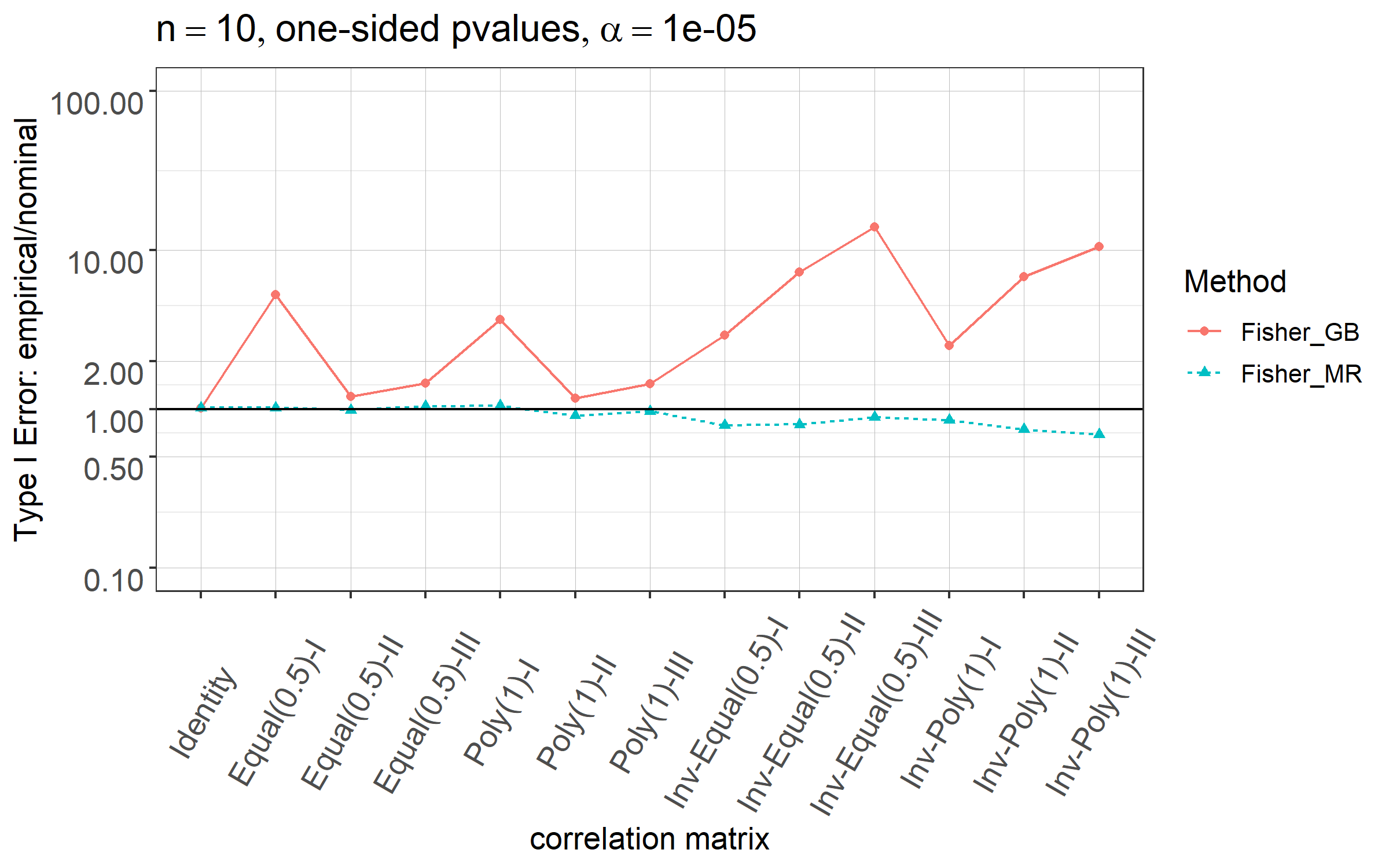}
\includegraphics[width=0.46\textwidth]{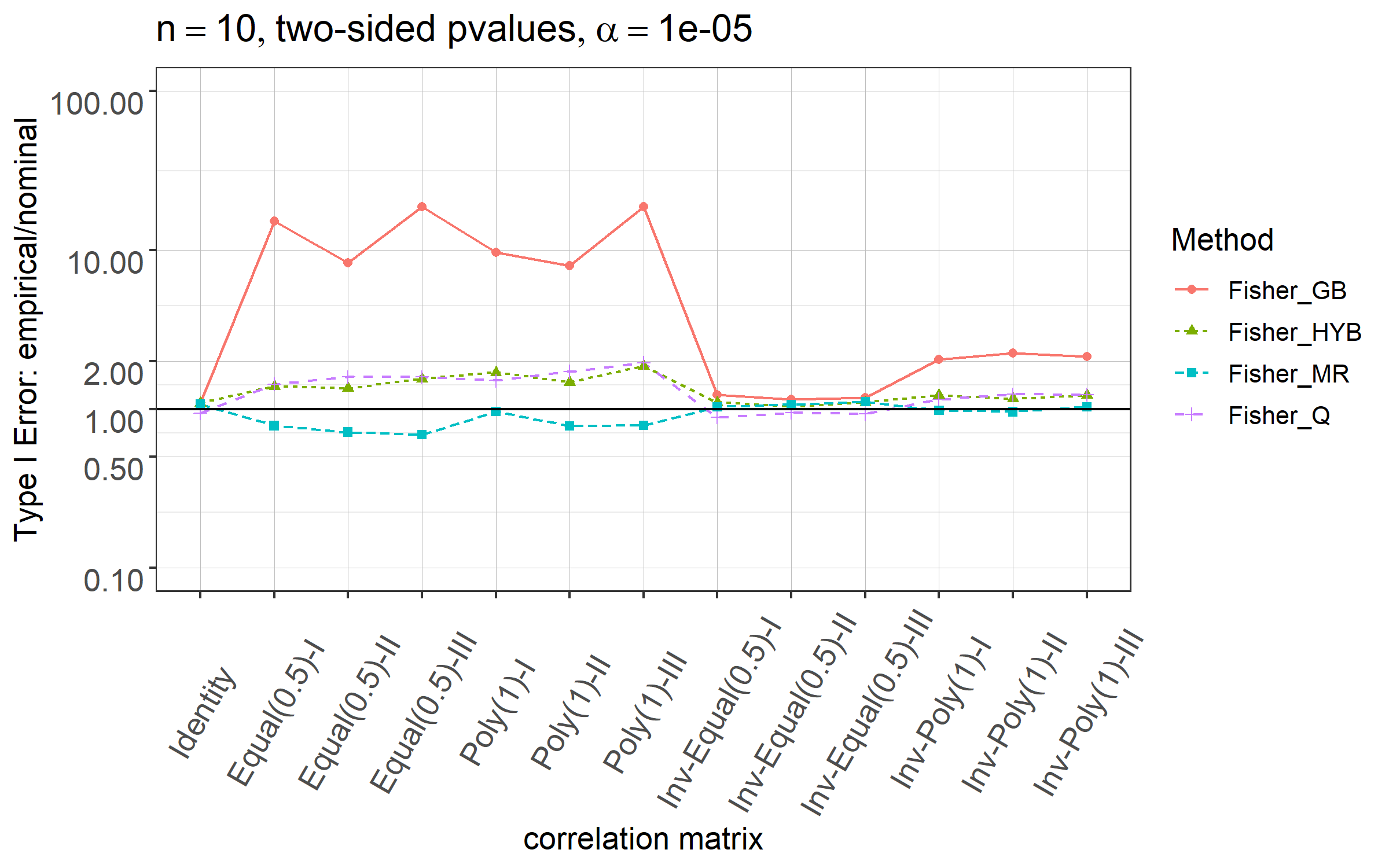}\\
\includegraphics[width=0.46\textwidth]{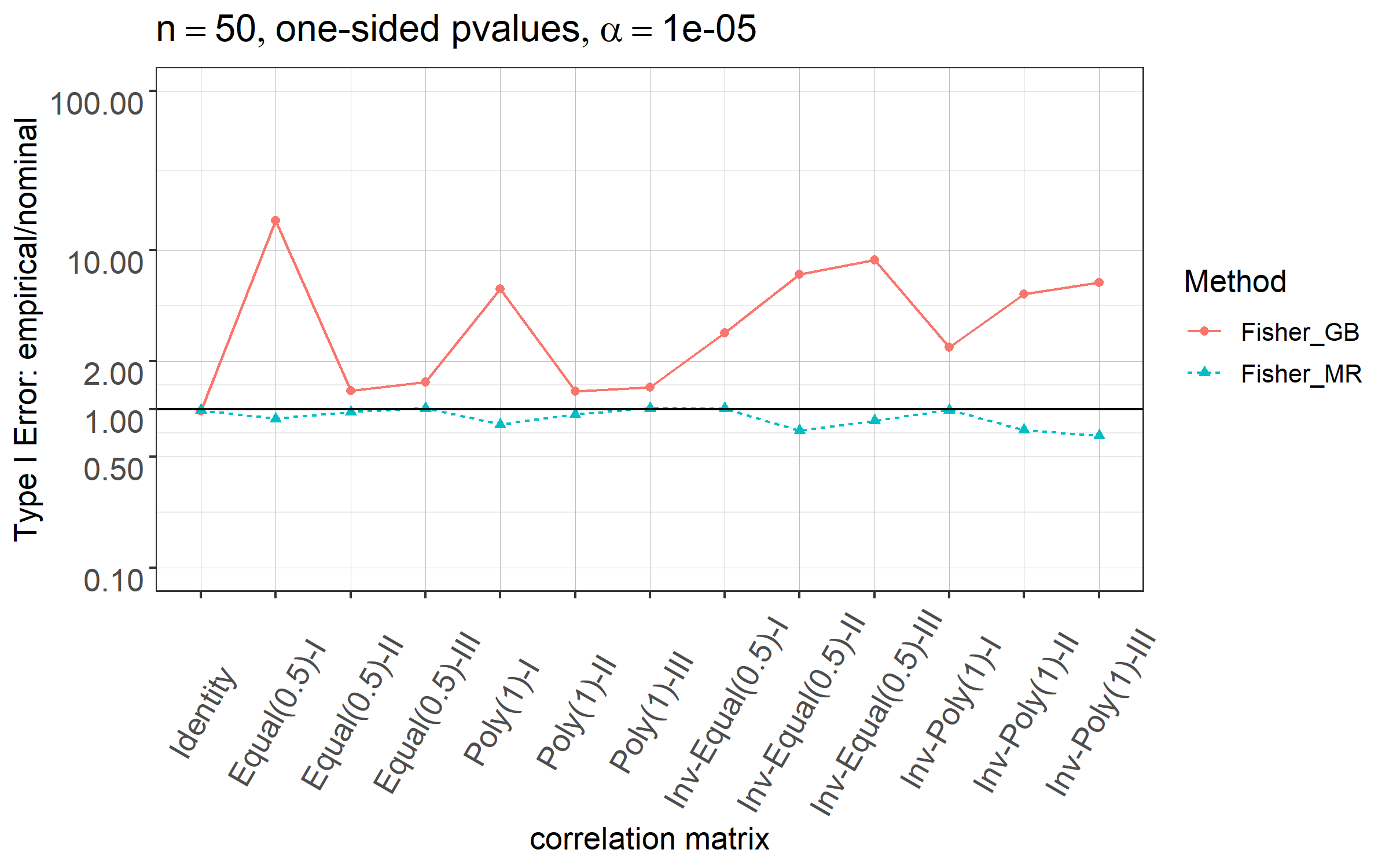}
\includegraphics[width=0.46\textwidth]{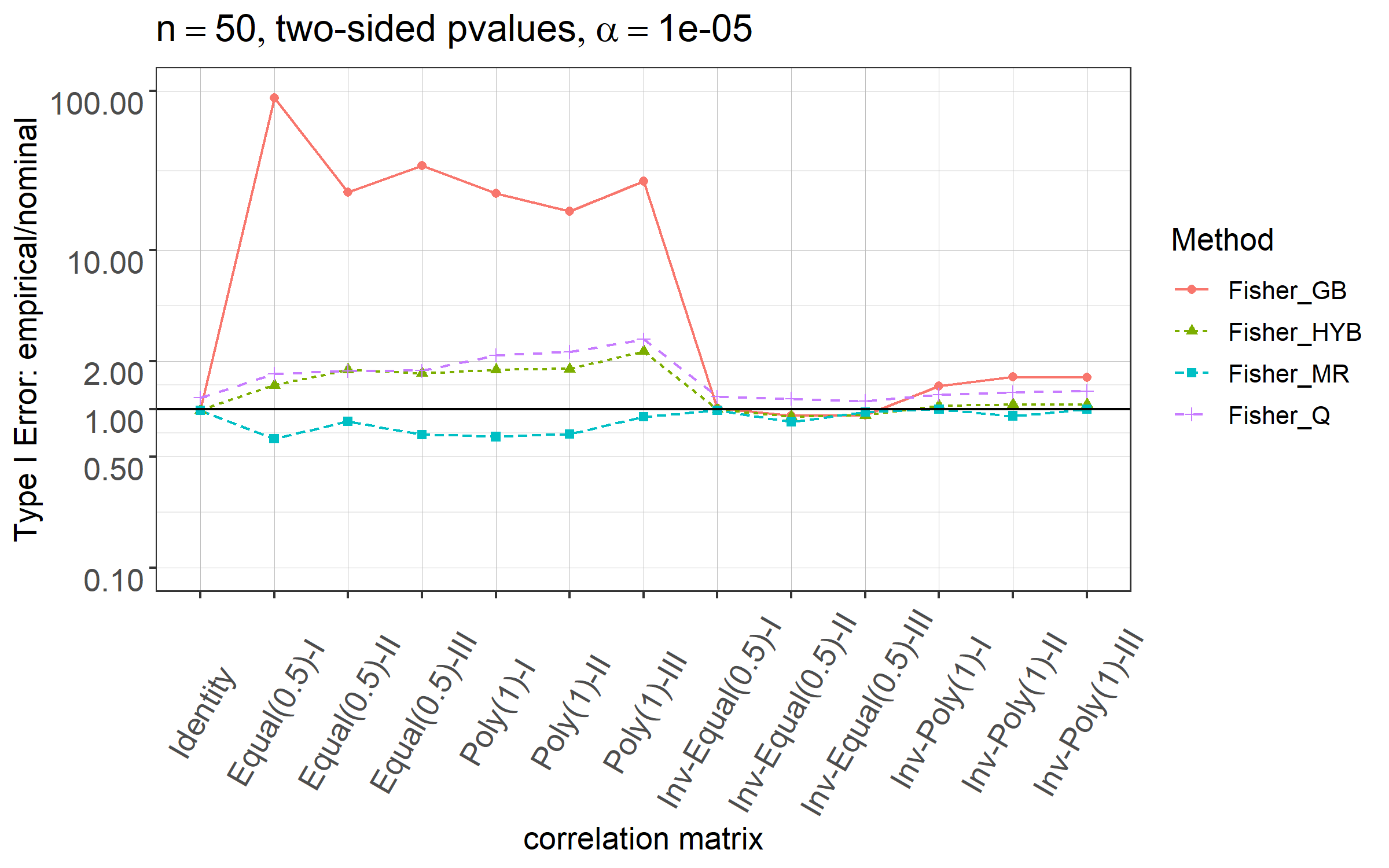}
\caption{Ratios between empirical type I error rates and nominal $\alpha=10^{-4}$ or $10^{-5}$. Fisher's combination test under independence and the 12 correlation structures in Table~\ref{tbl.sigma}. GB: generalized Brown's method. HYB: the hybrid method. MR: moment-ratio matching method. Q: Q-approximation. }
\label{fig.tie_GMM_cases_supp2}
\end{figure}

\begin{figure}
\centering
\includegraphics[width=0.45\textwidth]{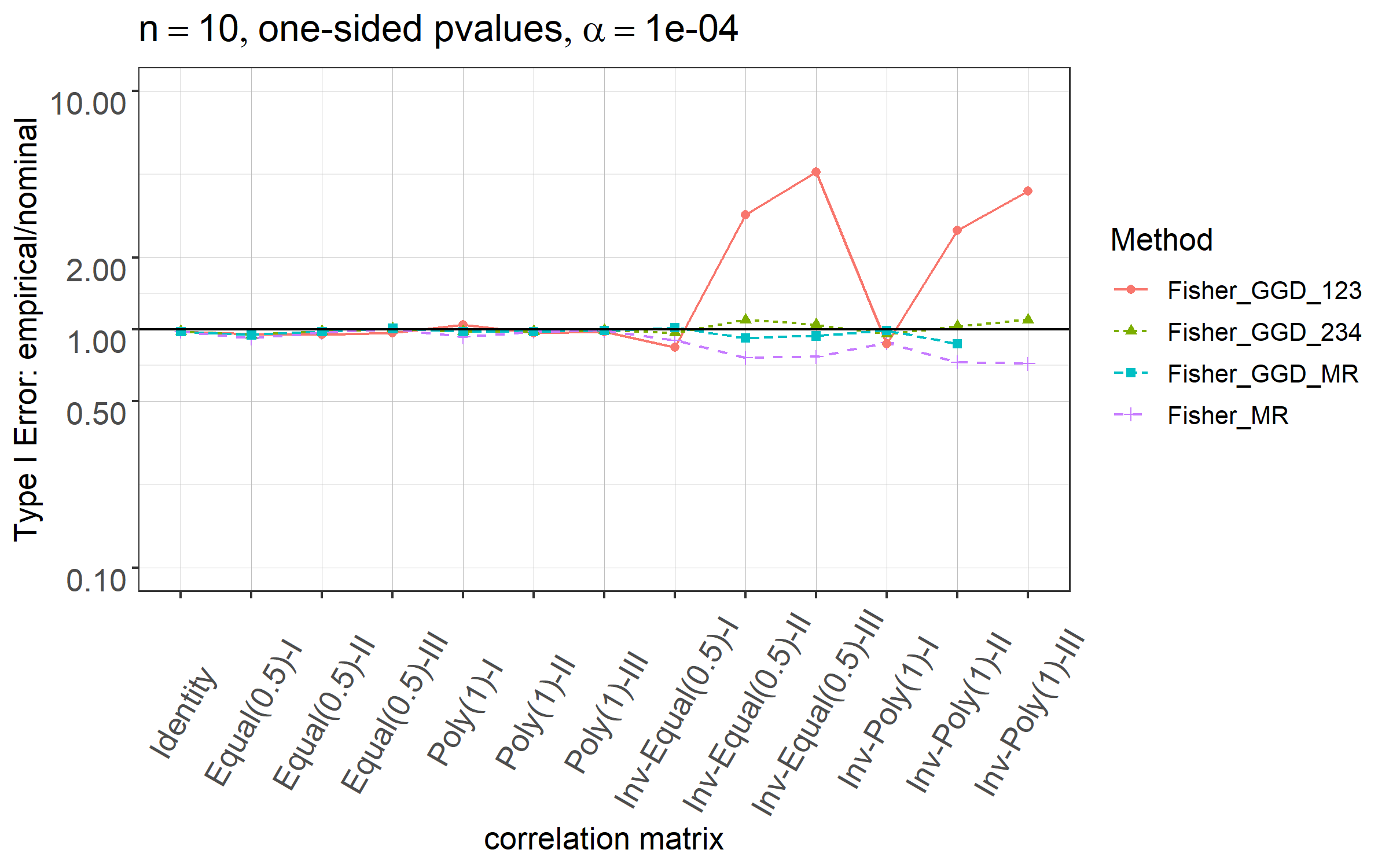}
\includegraphics[width=0.45\textwidth]{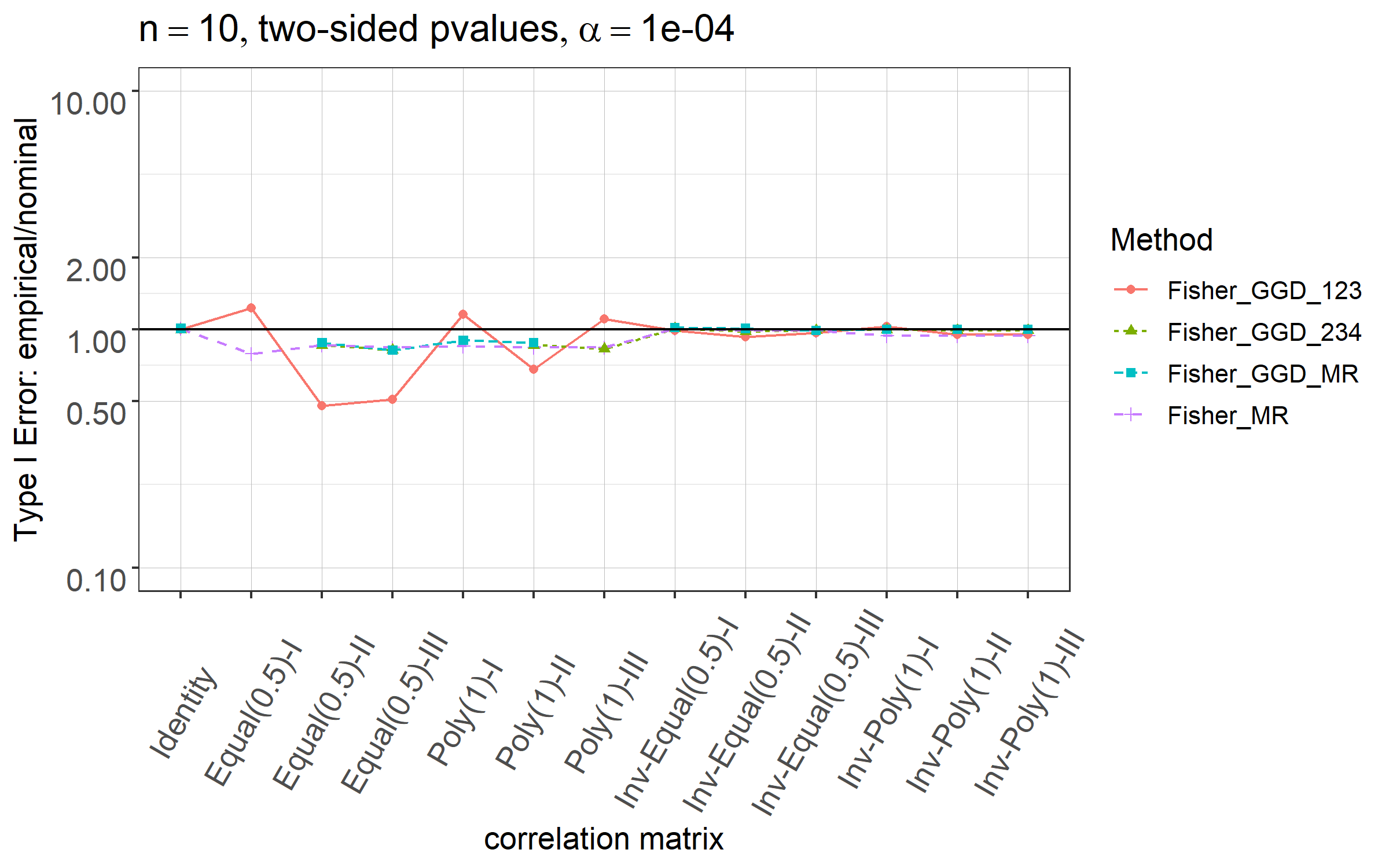}\\
\includegraphics[width=0.45\textwidth]{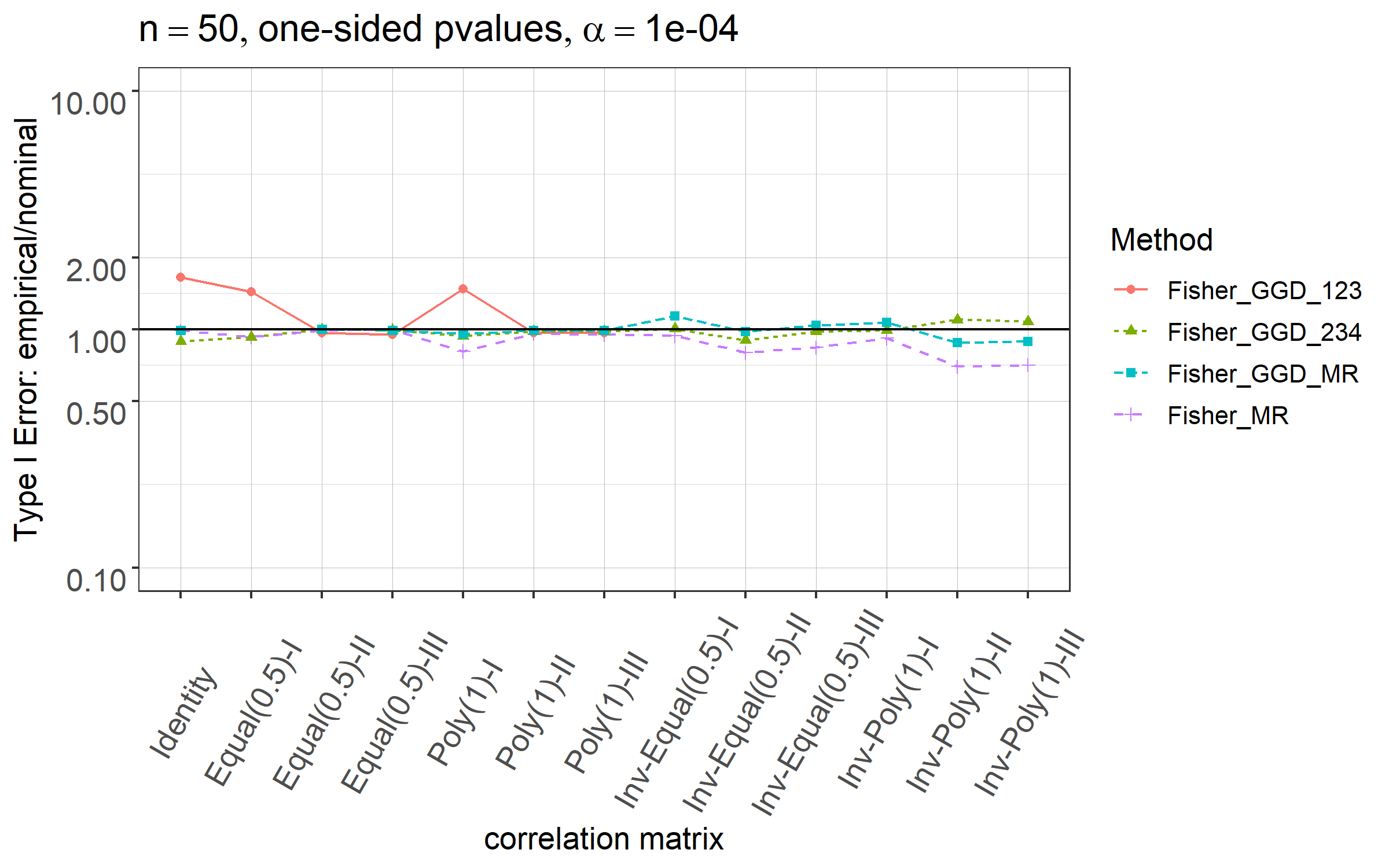}
\includegraphics[width=0.45\textwidth]{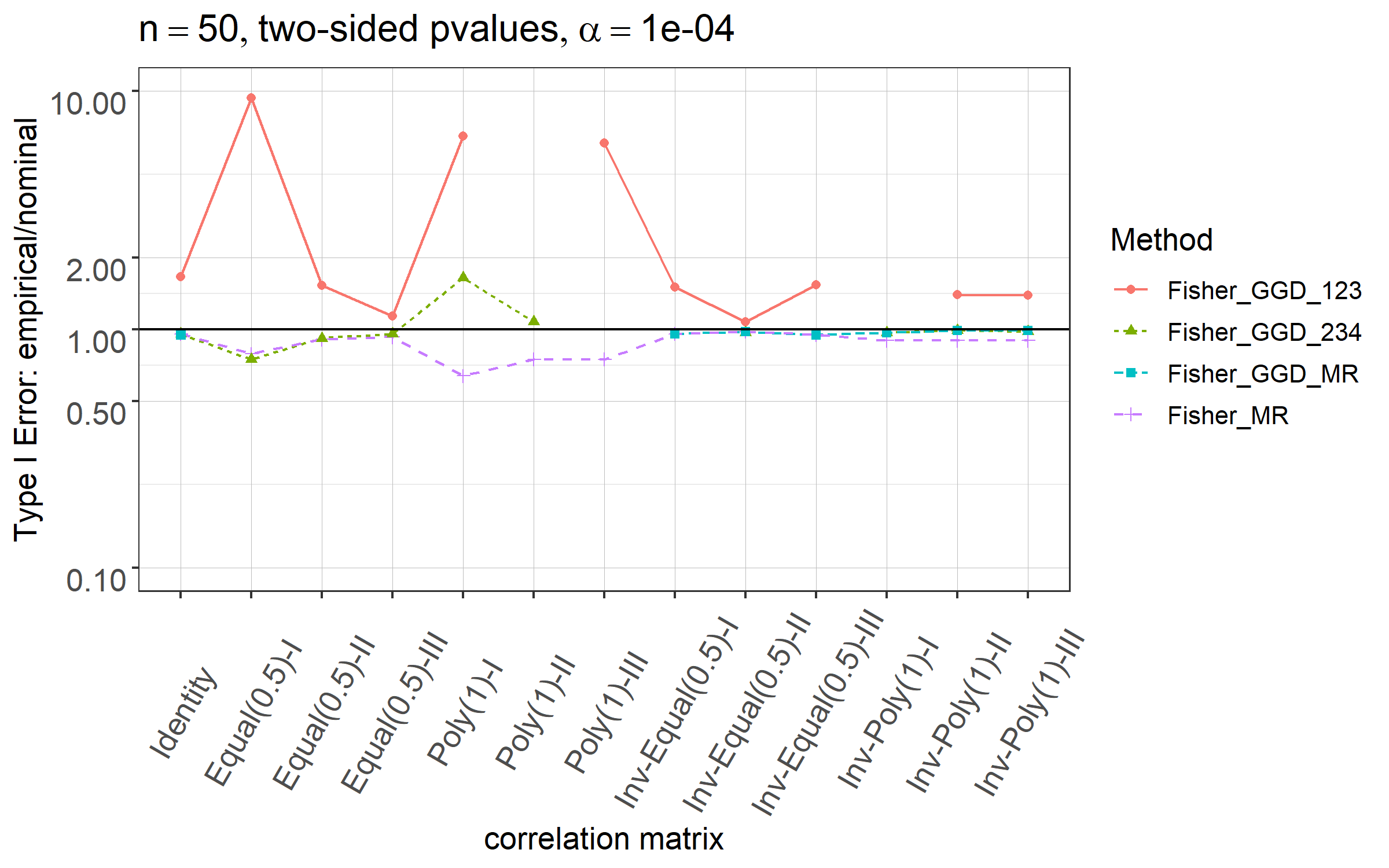}\\
\includegraphics[width=0.45\textwidth]{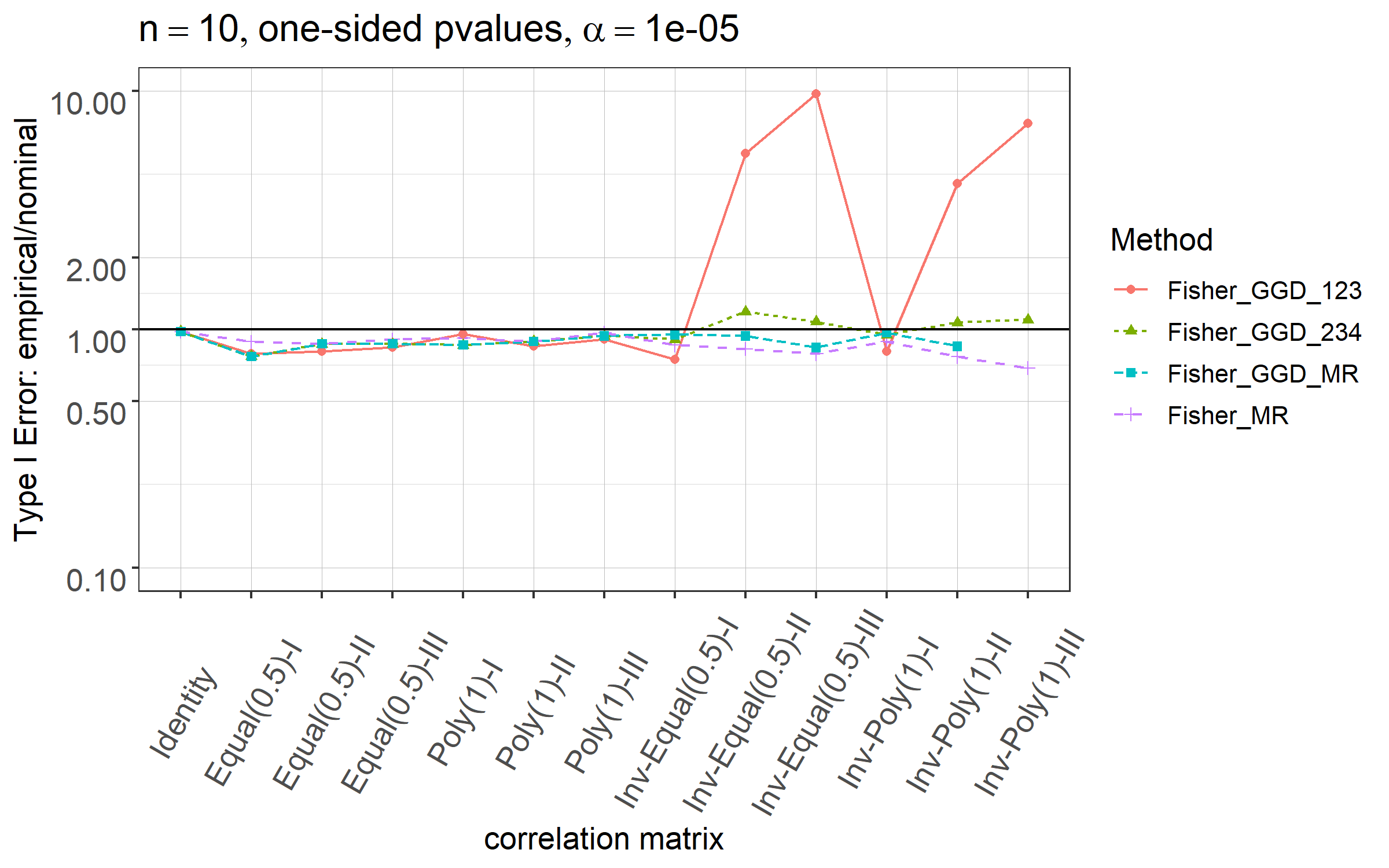}
\includegraphics[width=0.45\textwidth]{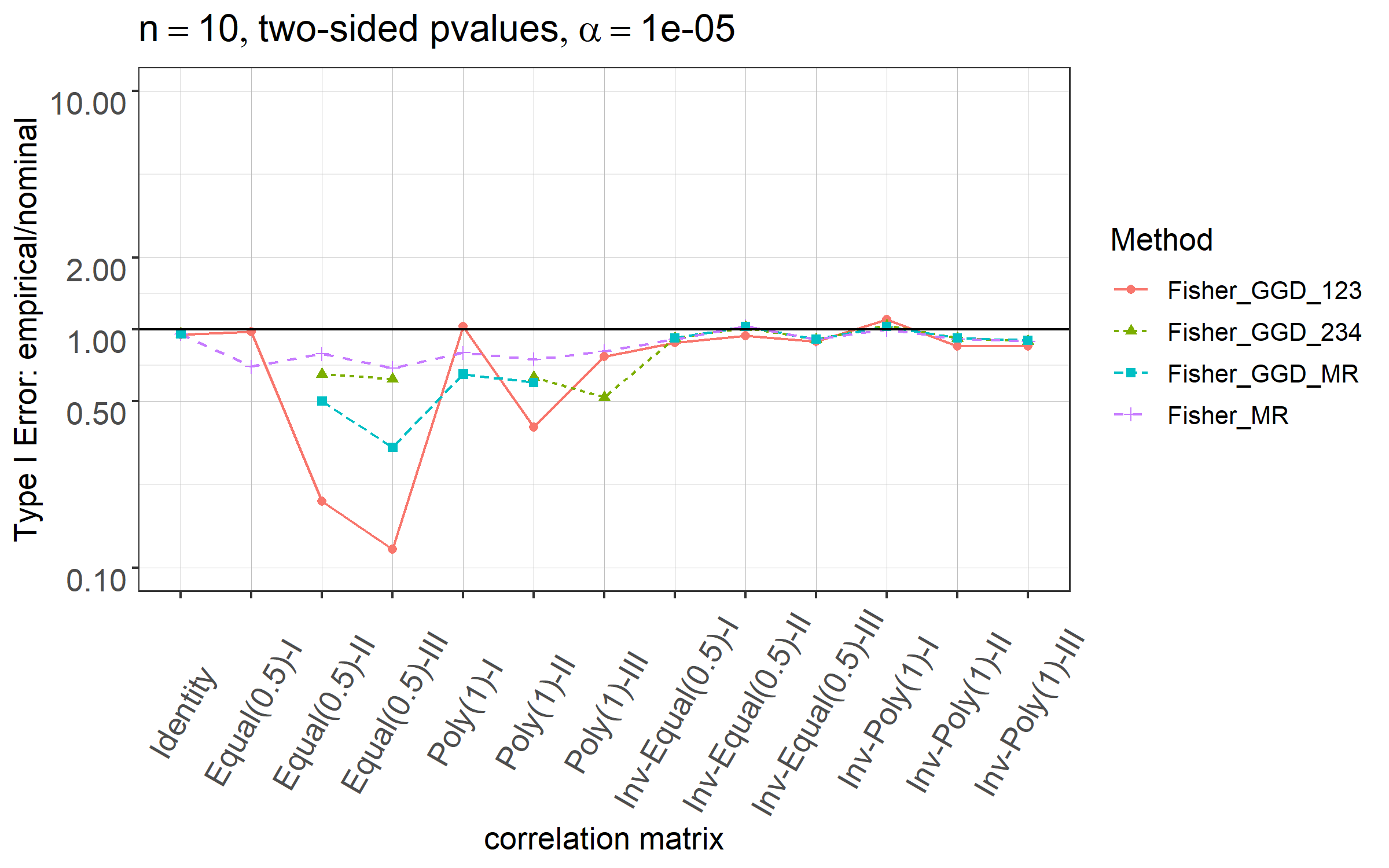}\\
\includegraphics[width=0.45\textwidth]{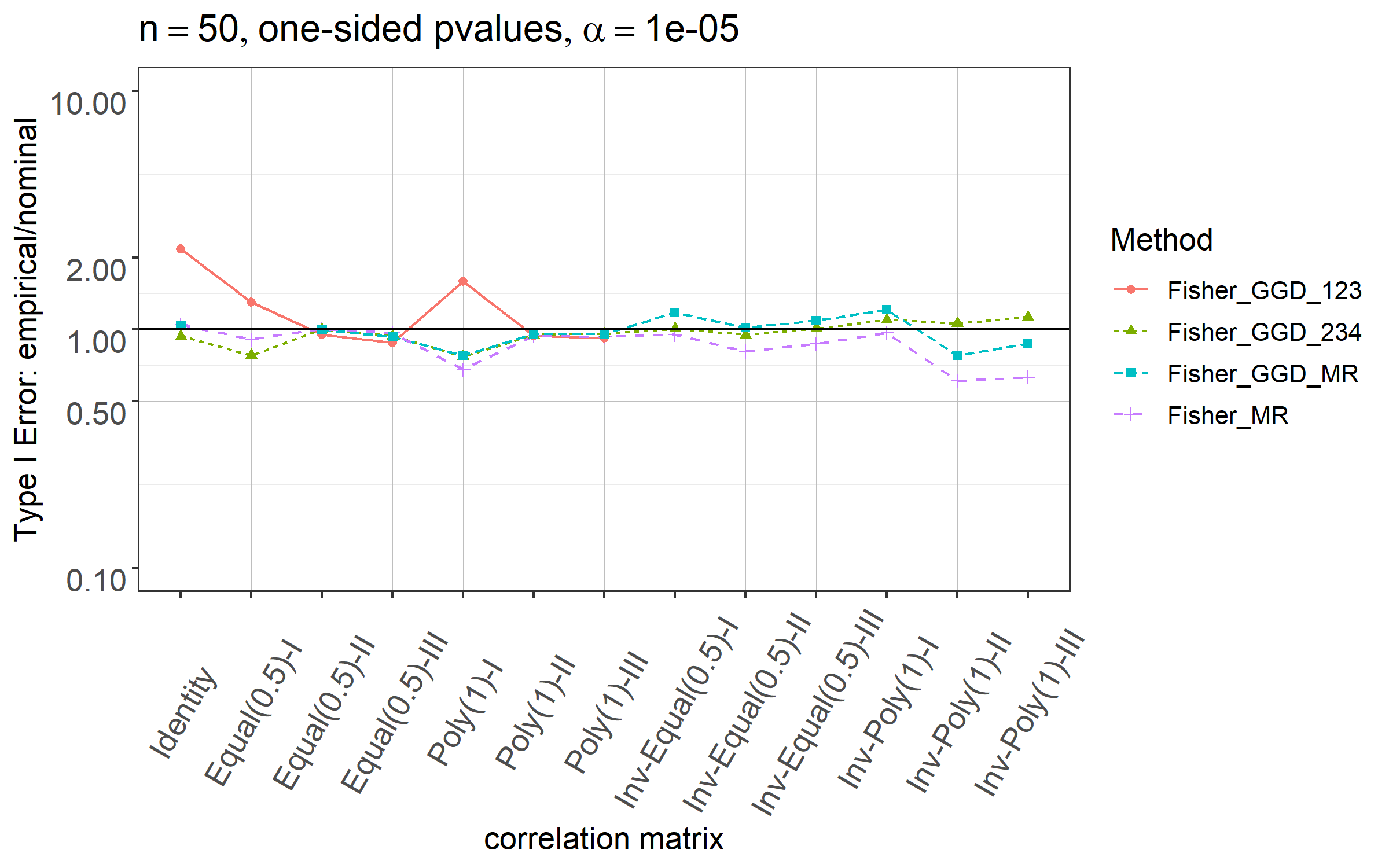}
\includegraphics[width=0.45\textwidth]{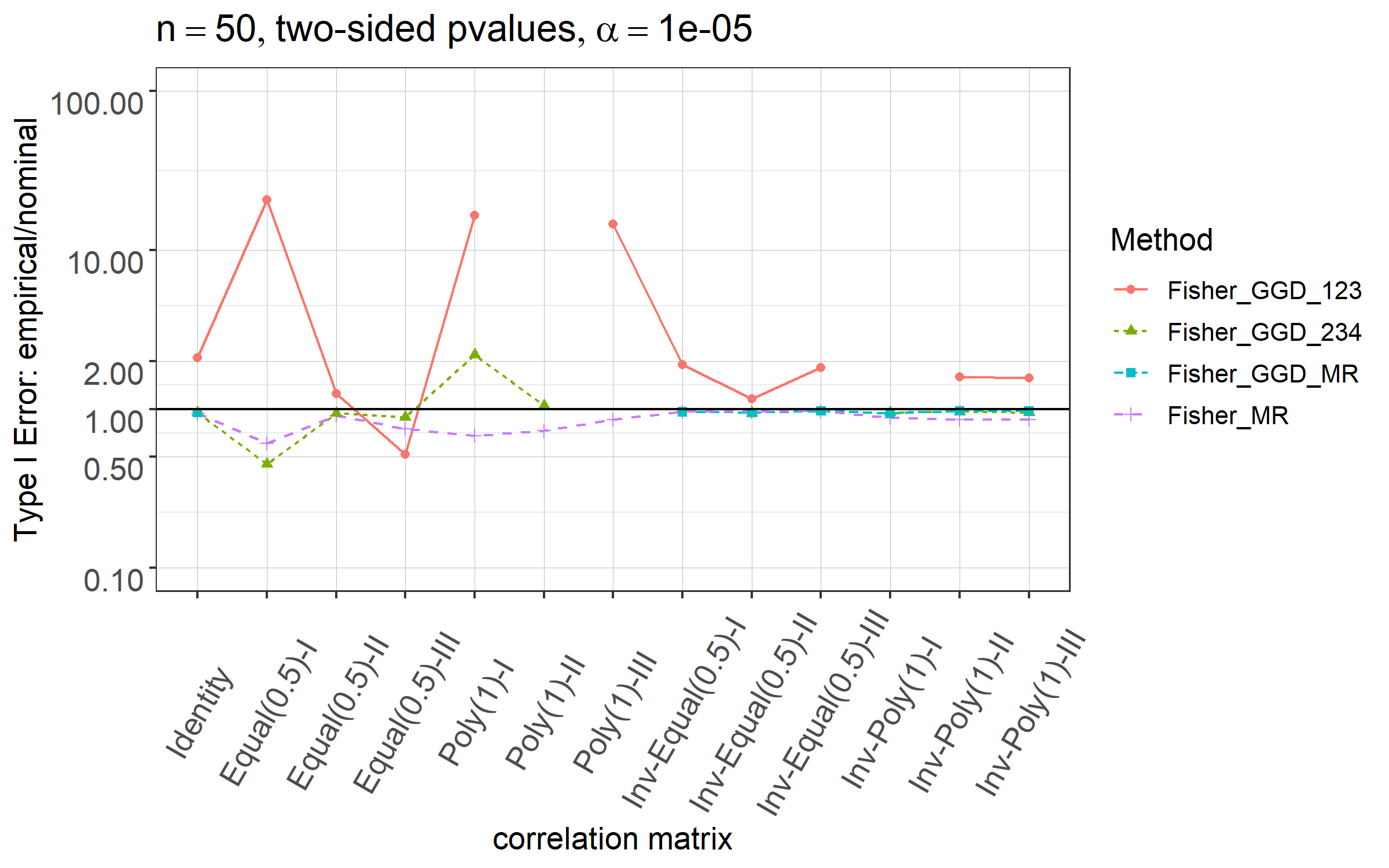}\\
\caption{Ratios between empirical type I error rates and the nominal $\alpha=10^{-4}$ or $10^{-5}$ using GGD-based methods. Fisher's combination test under independence and the 12 correlation structures in Table~\ref{tbl.sigma}. GGD\_123: matching the first three moments of GGD. GGD\_234: matching the variance, skewness and kurtosis of GGD. GGD\_MR: GGD-based moment-ratio matching method. MR:  GD-based moment-ratio matching method. Missing values indicate moment-matching equations don't have a solution. 
}
\label{fig.tie_GGD_supp}
\end{figure}

\begin{figure}
\centering
\includegraphics[width=0.45\textwidth]{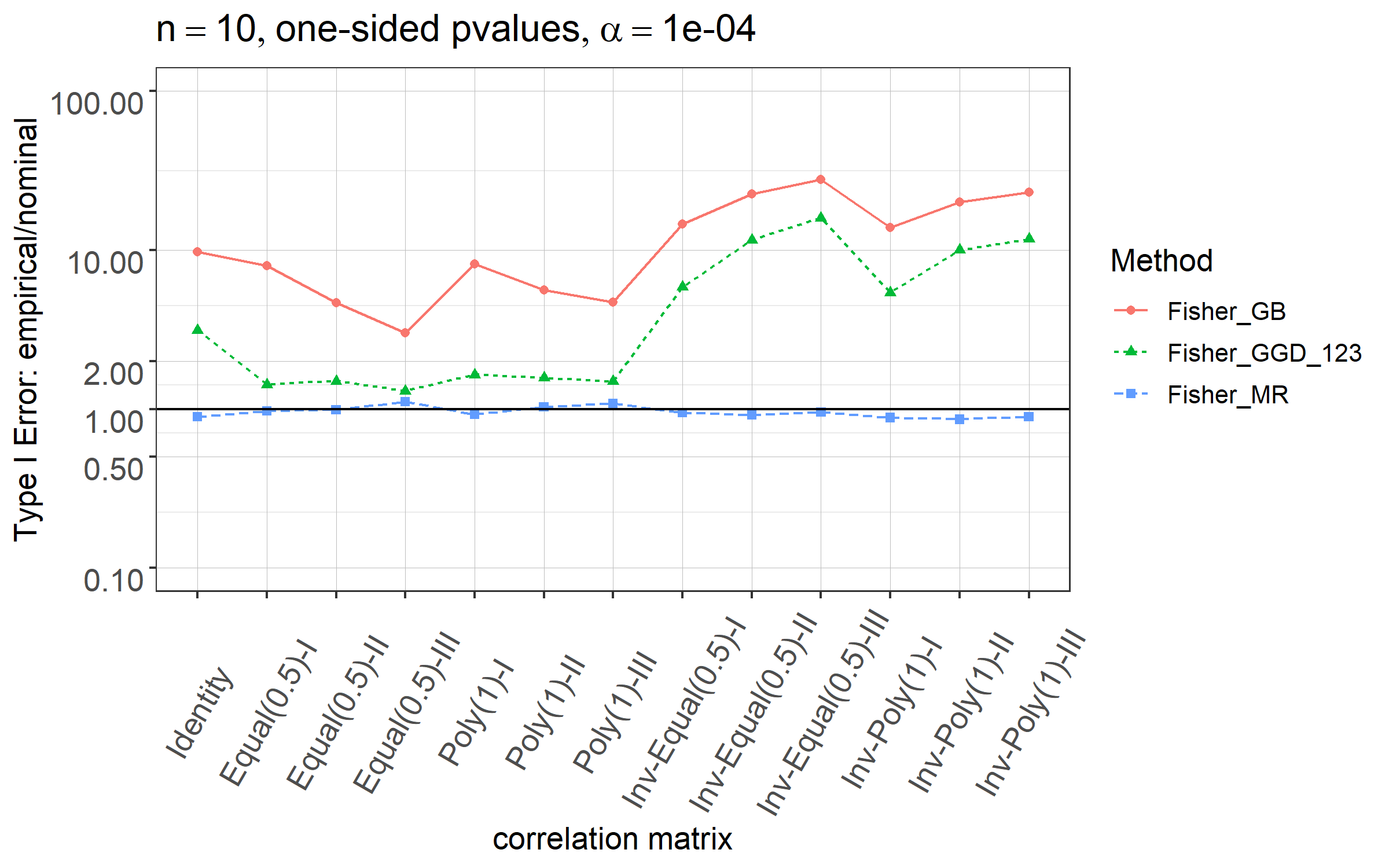}
\includegraphics[width=0.45\textwidth]{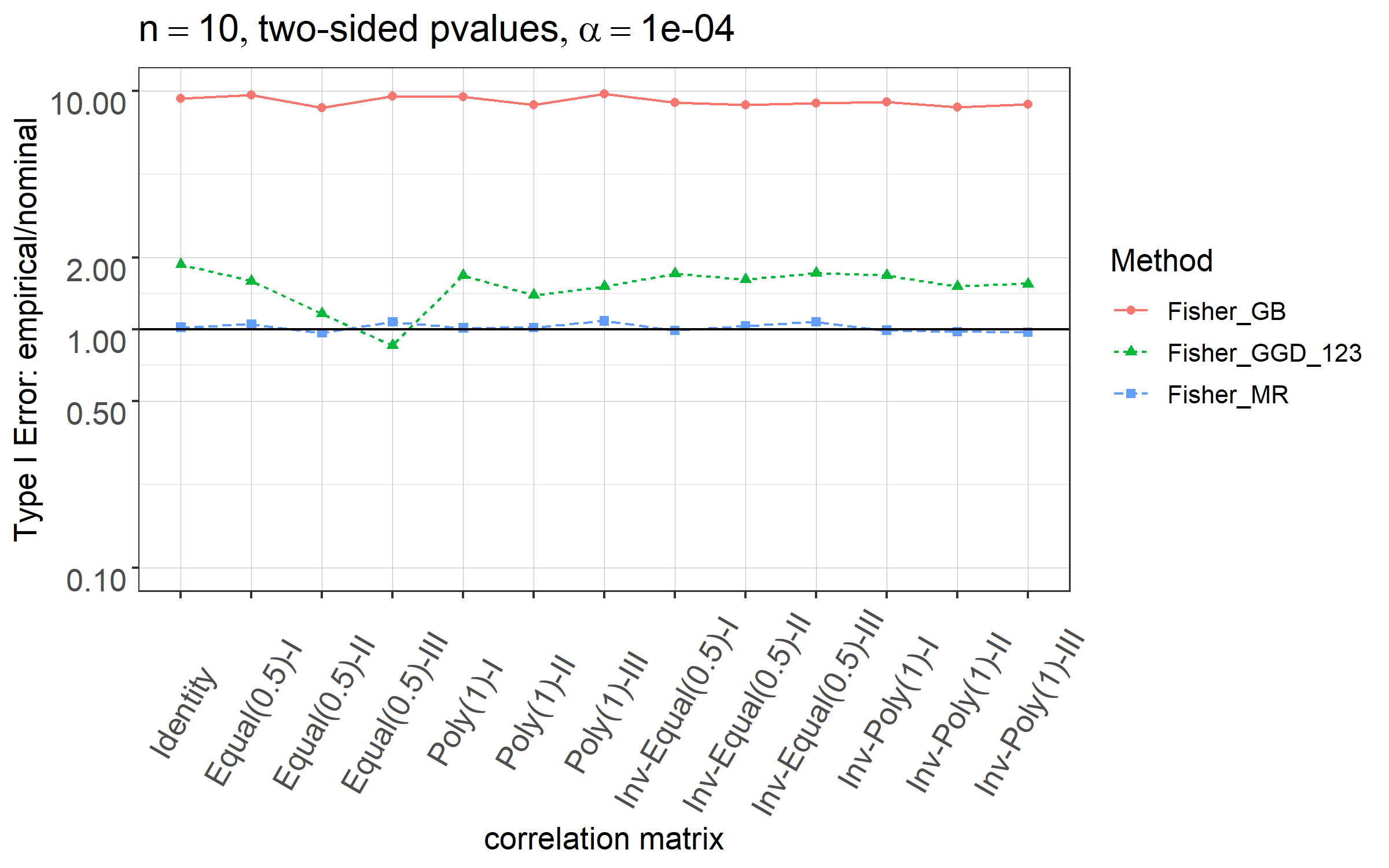}\\
\includegraphics[width=0.45\textwidth]{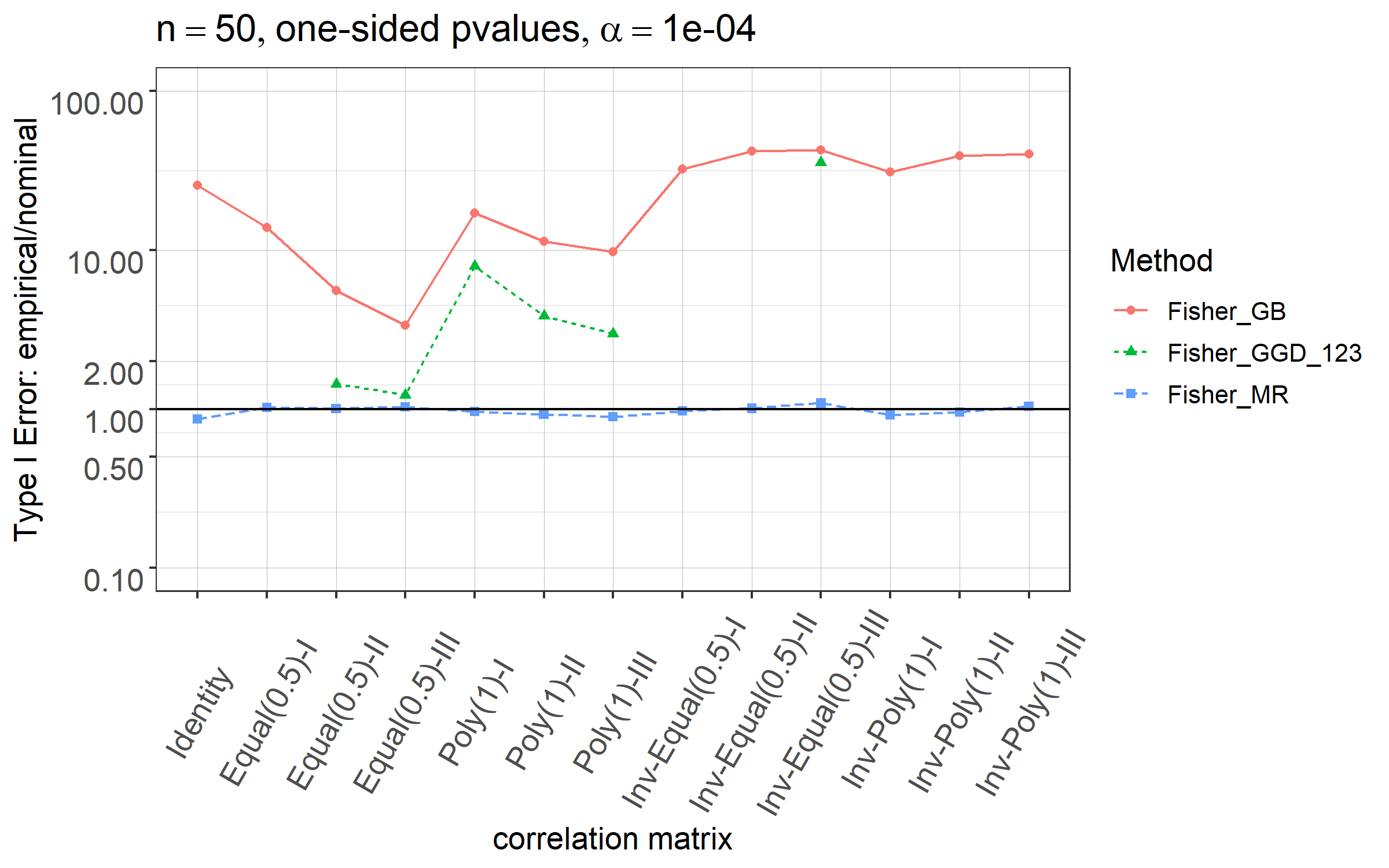}
\includegraphics[width=0.45\textwidth]{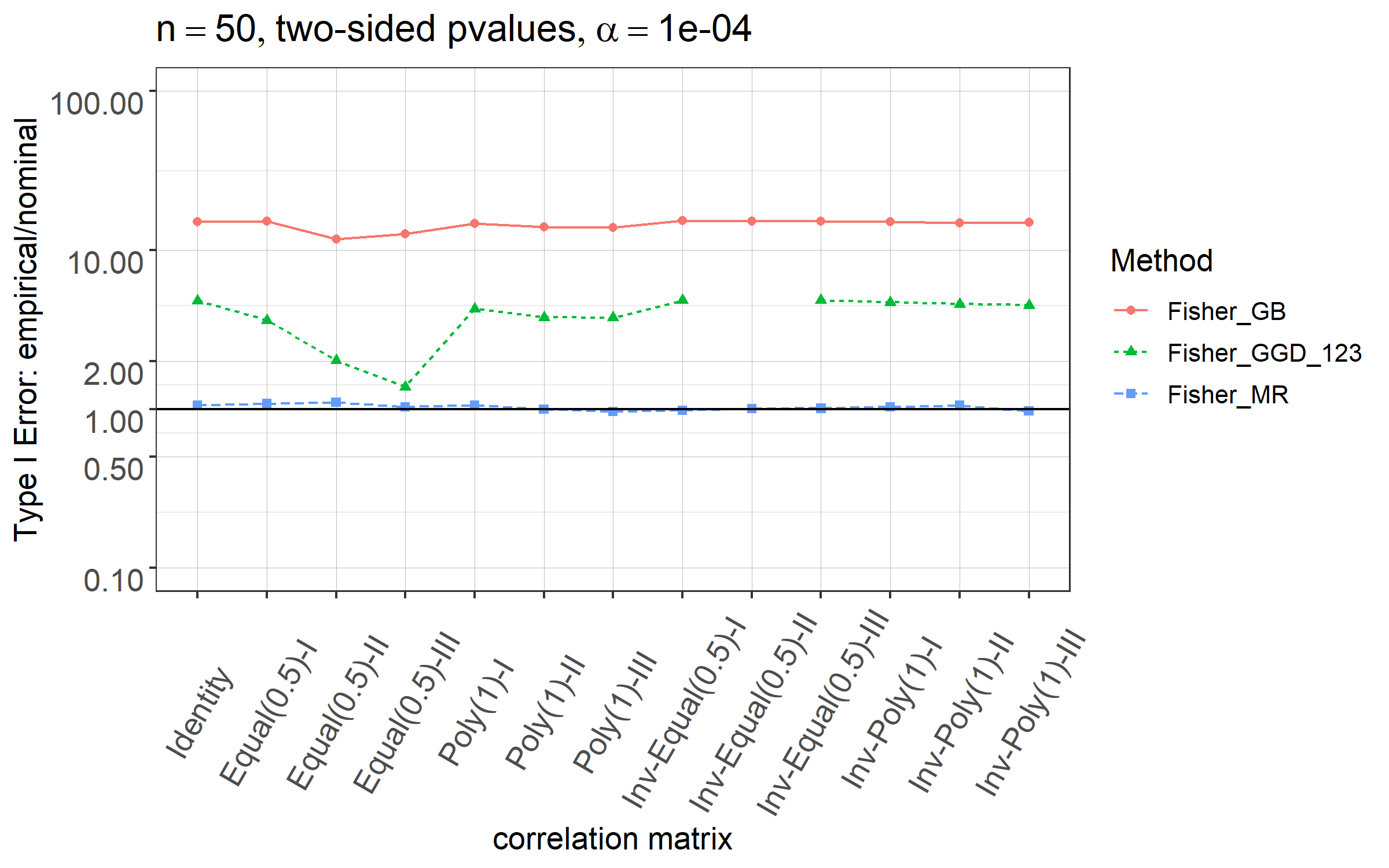}\\
\includegraphics[width=0.45\textwidth]{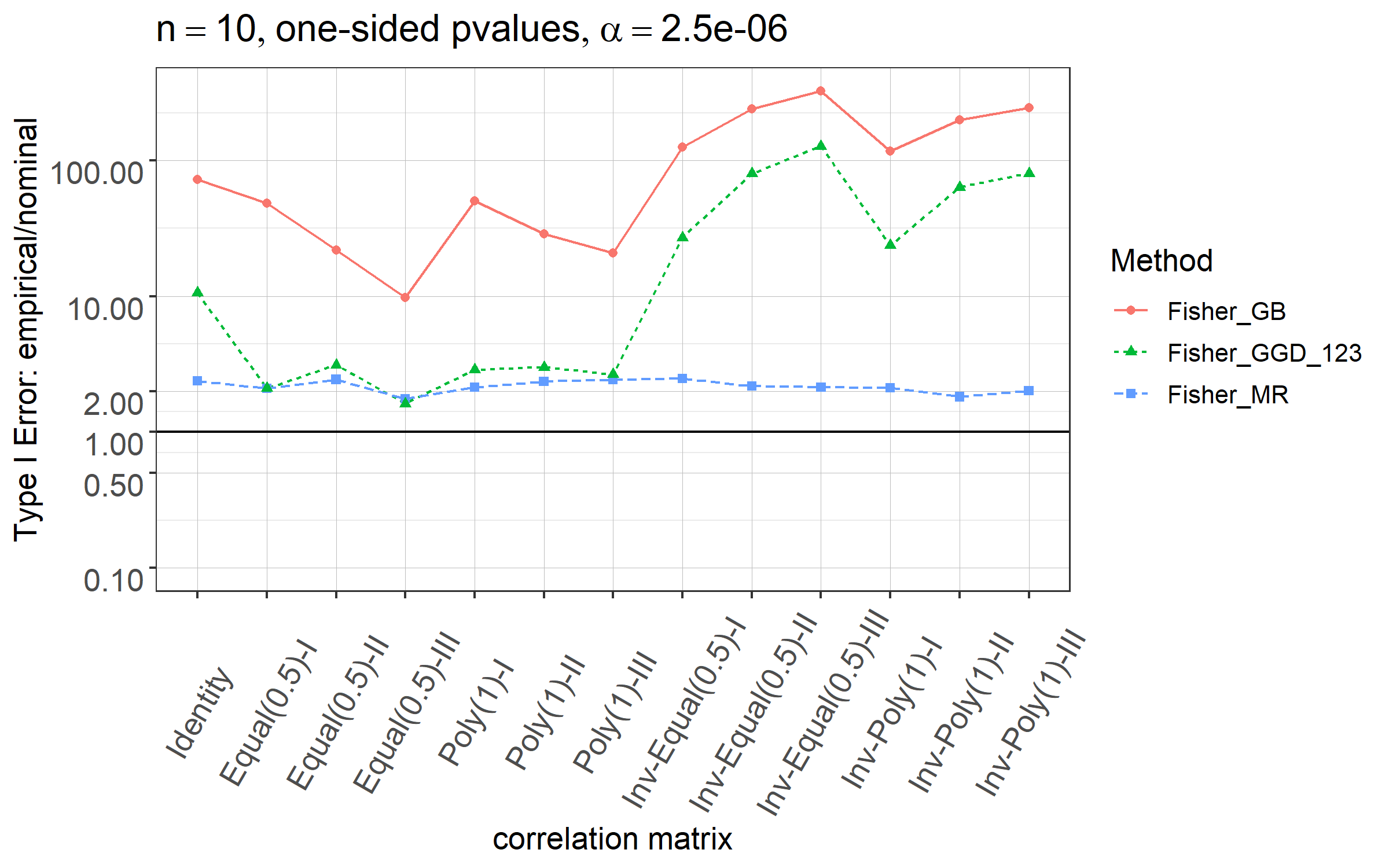}
\includegraphics[width=0.45\textwidth]{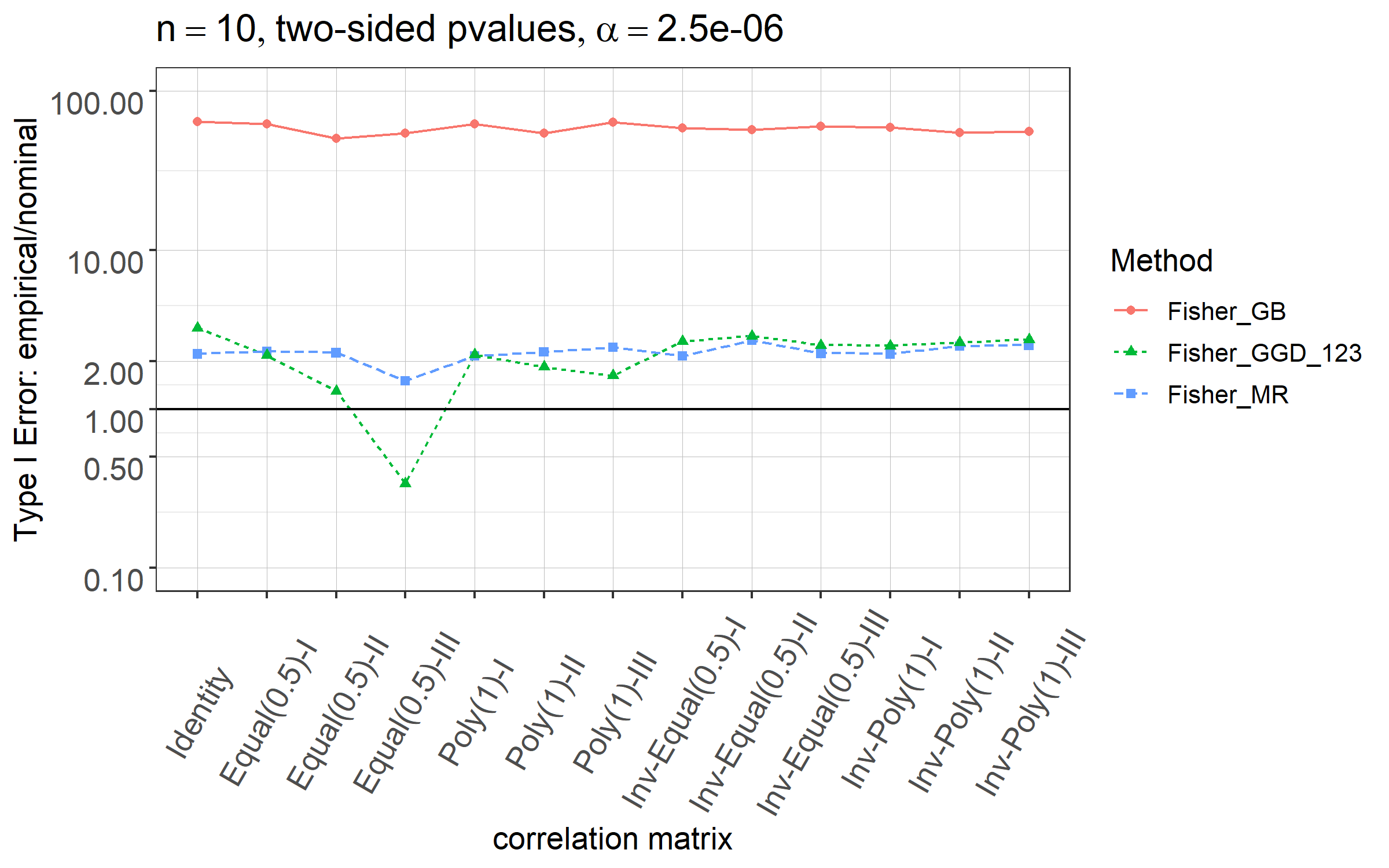}\\
\includegraphics[width=0.45\textwidth]{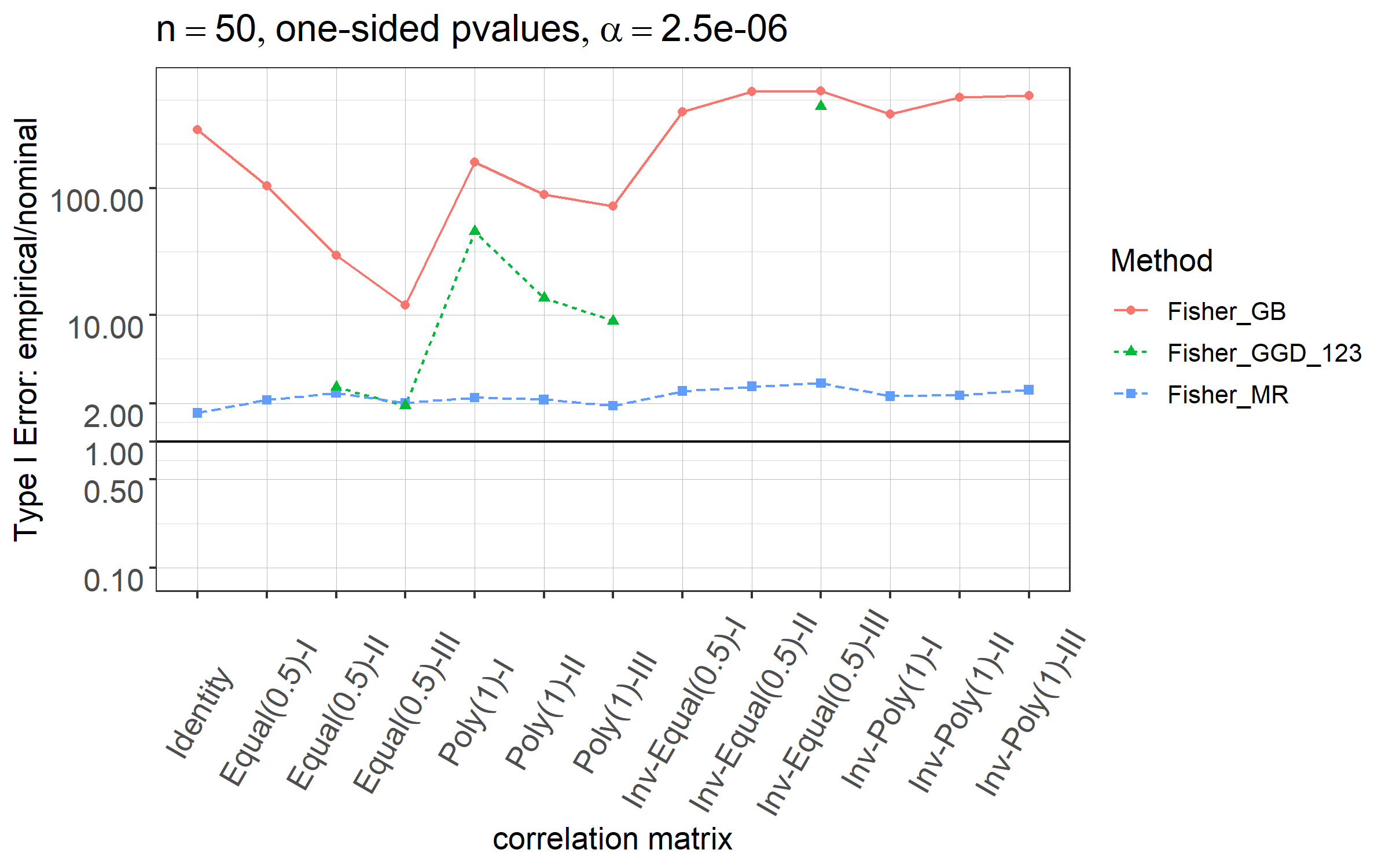}
\includegraphics[width=0.45\textwidth]{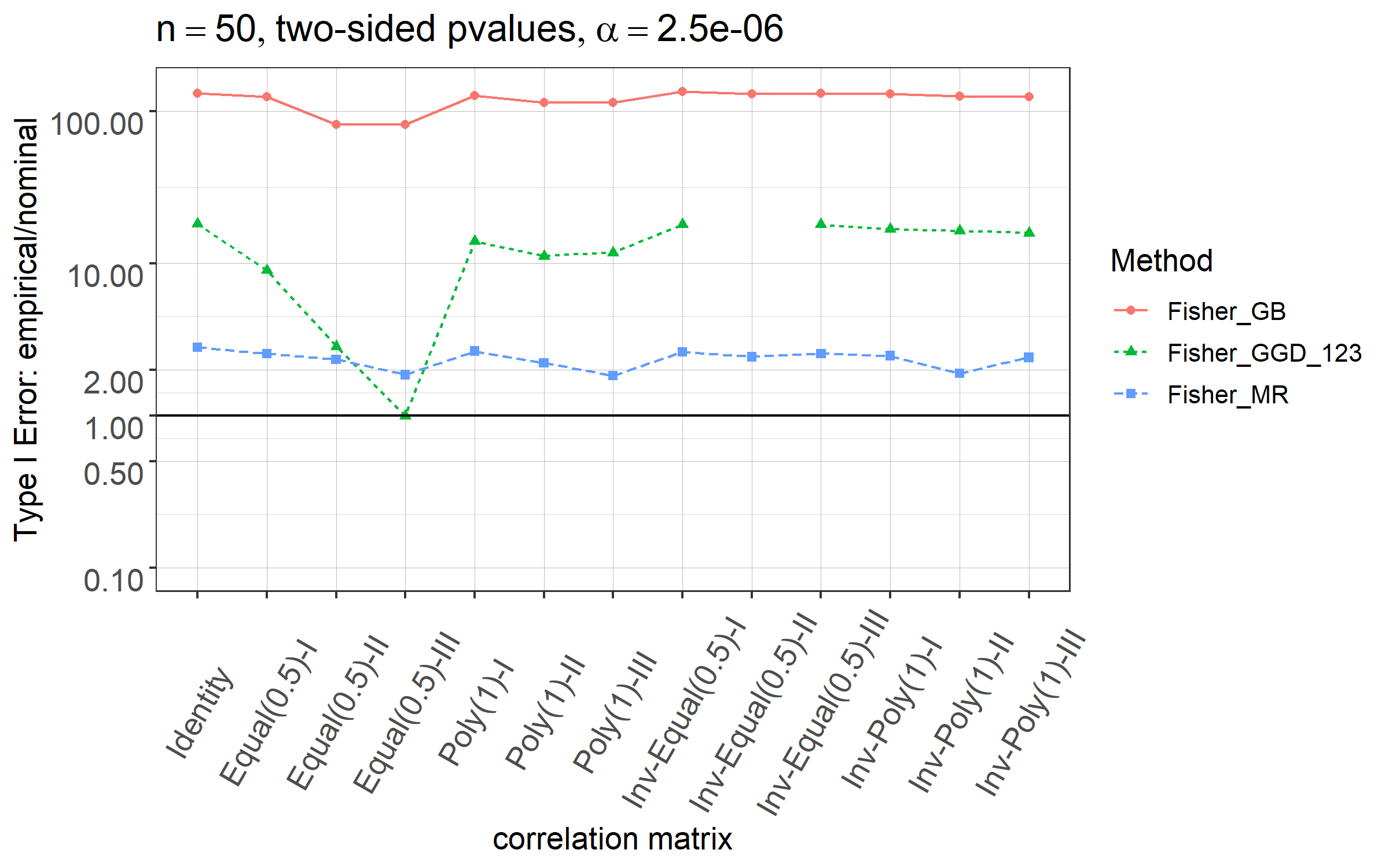}\\
\caption{Ratios between empirical type I error rates and the nominal $\alpha=10^{-4}$ or $2.5\times10^{-6}$ under the multivariate $t$-distribution. Fisher's combination test. $\mathbf{\Sigma}=\mathbf{I}$ and the 12 correlation settings in Table~\ref{tbl.sigma}. GB: generalized Brown's method. GGD\_123: matching the first three moments of GGD. MR: GD based moment-ratio matching method. Missing values indicate the moment-matching equations don't have a solution. }
\label{fig.tie_MVT_supp}
\end{figure}

\begin{description}

\item[Accuracy under GMM visualized by survival curves.] Figures \ref{fig.survival_oneside_supp} and \ref{fig.survival_twoside_supp} give survival curves of GFisher statistics obtained by simulations (the gold standard) and by relevant calculation methods. 

\item[Accuracy under GMM visualized by empirical errors.] Figures \ref{fig.tie_GMM_cases_supp1} -- \ref{fig.tie_GGD_supp} give the ratios between type I error rates and nominal $\alpha = 0.05, 0.01, 10^{-4}$ and $10^{-5}$. 

\item[Robustness under the multivariate $t$-distribution.] Figure~\ref{fig.tie_MVT_supp} shows ratios between empirical type I error rates and nominal $\alpha = 10^{-4}$ and $10^{-5}$.

\end{description}

\bibliographystyle{Chicago}
\bibliography{allMyReferences}

\begin{thebibliography}{}

\bibitem[\protect\citeauthoryear{Barnett, Mukherjee, and Lin}{Barnett
  et~al.}{2017}]{barnett2016generalized}
Barnett, I., R.~Mukherjee, and X.~Lin (2017).
\newblock The generalized higher criticism for testing snp-set effects in
  genetic association studies.
\newblock {\em Journal of the American Statistical Association\/}~{\em
  112\/}(517), 64--76.

\bibitem[\protect\citeauthoryear{Berk and Cohen}{Berk and
  Cohen}{1979}]{berk1979asymptotically}
Berk, R.~H. and A.~Cohen (1979).
\newblock Asymptotically optimal methods of combining tests.
\newblock {\em Journal of the American Statistical Association\/}~{\em
  74\/}(368), 812--814.

\bibitem[\protect\citeauthoryear{Bodmer and Bonilla}{Bodmer and
  Bonilla}{2008}]{Bodmer2008}
Bodmer, W. and C.~Bonilla (2008).
\newblock Common and rare variants in multifactorial susceptibility to common
  diseases.
\newblock {\em Nature genetics\/}~{\em 40\/}(6), 695--701.

\bibitem[\protect\citeauthoryear{Brown}{Brown}{1975}]{brown1975400}
Brown, M.~B. (1975).
\newblock 400: A method for combining non-independent, one-sided tests of
  significance.
\newblock {\em Biometrics\/}, 987--992.

\bibitem[\protect\citeauthoryear{Chen, Carter, Parla, Kramer, Goes, Pirooznia,
  Zandi, McCombie, Potash, and Karchin}{Chen et~al.}{2013}]{chen2013hybrid}
Chen, Y.-C., H.~Carter, J.~Parla, M.~Kramer, F.~S. Goes, M.~Pirooznia, P.~P.
  Zandi, W.~R. McCombie, J.~B. Potash, and R.~Karchin (2013).
\newblock A hybrid likelihood model for sequence-based disease association
  studies.
\newblock {\em PLoS genetics\/}~{\em 9\/}(1), e1003224.

\bibitem[\protect\citeauthoryear{Chen and Nadarajah}{Chen and
  Nadarajah}{2014}]{chen2014optimally}
Chen, Z. and S.~Nadarajah (2014).
\newblock On the optimally weighted z-test for combining probabilities from
  independent studies.
\newblock {\em Computational statistics \& data analysis\/}~{\em 70}, 387--394.

\bibitem[\protect\citeauthoryear{Chen, Yang, Liu, Yang, Li, and Yang}{Chen
  et~al.}{2014}]{chen2014new}
Chen, Z., W.~Yang, Q.~Liu, J.~Y. Yang, J.~Li, and M.~Q. Yang (2014).
\newblock A new statistical approach to combining p-values using gamma
  distribution and its application to genome-wide association study.
\newblock {\em BMC bioinformatics\/}~{\em 15\/}(17), S3.

\bibitem[\protect\citeauthoryear{Dai, Wu, Wu, and Zhi}{Dai
  et~al.}{2016}]{dai2016optimal}
Dai, H., G.~Wu, M.~Wu, and D.~Zhi (2016).
\newblock An optimal bahadur-efficient method in detection of sparse signals
  with applications to pathway analysis in sequencing association studies.
\newblock {\em PloS one\/}~{\em 11\/}(7), e0152667.

\bibitem[\protect\citeauthoryear{Dai, Leeder, and Cui}{Dai
  et~al.}{2014}]{dai2014modified}
Dai, H.~D., J.~S. Leeder, and Y.~Cui (2014).
\newblock A modified generalized fisher method for combining probabilities from
  dependent tests.
\newblock {\em Frontiers in genetics\/}~{\em 5}, 32.

\bibitem[\protect\citeauthoryear{Davies}{Davies}{1980}]{davies1980algorithm}
Davies, R.~B. (1980).
\newblock Algorithm as 155: The distribution of a linear combination of $\chi$
  2 random variables.
\newblock {\em Journal of the Royal Statistical Society. Series C (Applied
  Statistics)\/}~{\em 29\/}(3), 323--333.

\bibitem[\protect\citeauthoryear{Fisher}{Fisher}{1925}]{Fisher1925}
Fisher, R.~A. (1925).
\newblock {\em Statistical Methods for Research Workers\/} (1st Edition ed.).
\newblock Oliver and Boyd, Edinburgh.

\bibitem[\protect\citeauthoryear{Genz}{Genz}{1992}]{genz1992numerical}
Genz, A. (1992).
\newblock Numerical computation of multivariate normal probabilities.
\newblock {\em Journal of Computational and Graphical Statistics\/}~{\em
  1\/}(2), 141--149.

\bibitem[\protect\citeauthoryear{Good}{Good}{1955}]{good1955weighted}
Good, I. (1955).
\newblock On the weighted combination of significance tests.
\newblock {\em Journal of the Royal Statistical Society: Series B
  (Methodological)\/}~{\em 17\/}(2), 264--265.

\bibitem[\protect\citeauthoryear{Hall and Jin}{Hall and Jin}{2010}]{Hall2010}
Hall, P. and J.~Jin (2010).
\newblock Innovated higher criticism for detecting sparse signals in correlated
  noise.
\newblock {\em The Annals of Statistics\/}~{\em 38\/}(3), 1686--1732.

\bibitem[\protect\citeauthoryear{Higham}{Higham}{2002}]{higham2002computing}
Higham, N.~J. (2002).
\newblock Computing the nearest correlation matrix---a problem from finance.
\newblock {\em IMA journal of Numerical Analysis\/}~{\em 22\/}(3), 329--343.

\bibitem[\protect\citeauthoryear{Hou}{Hou}{2005}]{hou2005simple}
Hou, C.-D. (2005).
\newblock A simple approximation for the distribution of the weighted
  combination of non-independent or independent probabilities.
\newblock {\em Statistics \& probability letters\/}~{\em 73\/}(2), 179--187.

\bibitem[\protect\citeauthoryear{Imhof}{Imhof}{1961}]{imhof1961computing}
Imhof, J.-P. (1961).
\newblock Computing the distribution of quadratic forms in normal variables.
\newblock {\em Biometrika\/}~{\em 48\/}(3/4), 419--426.

\bibitem[\protect\citeauthoryear{Jarque and Bera}{Jarque and
  Bera}{1980}]{jarque1980efficient}
Jarque, C.~M. and A.~K. Bera (1980).
\newblock Efficient tests for normality, homoscedasticity and serial
  independence of regression residuals.
\newblock {\em Economics letters\/}~{\em 6\/}(3), 255--259.

\bibitem[\protect\citeauthoryear{Kost and McDermott}{Kost and
  McDermott}{2002}]{kost2002combining}
Kost, J.~T. and M.~P. McDermott (2002).
\newblock Combining dependent p-values.
\newblock {\em Statistics \& Probability Letters\/}~{\em 60\/}(2), 183--190.

\bibitem[\protect\citeauthoryear{Lancaster}{Lancaster}{1961}]{lancaster1961combination}
Lancaster, H. (1961).
\newblock The combination of probabilities: an application of orthonormal
  functions.
\newblock {\em Australian \& New Zealand Journal of Statistics\/}~{\em 3\/}(1),
  20--33.

\bibitem[\protect\citeauthoryear{Li and Tseng}{Li and
  Tseng}{2011}]{li2011adaptively}
Li, J. and G.~C. Tseng (2011).
\newblock An adaptively weighted statistic for detecting differential gene
  expression when combining multiple transcriptomic studies.
\newblock {\em The Annals of Applied Statistics\/}~{\em 5\/}(2A), 994--1019.

\bibitem[\protect\citeauthoryear{Li, Hu, Ding, and Zheng}{Li
  et~al.}{2014}]{li2014fisher}
Li, Q., J.~Hu, J.~Ding, and G.~Zheng (2014).
\newblock Fisher's method of combining dependent statistics using
  generalizations of the gamma distribution with applications to genetic
  pleiotropic associations.
\newblock {\em Biostatistics\/}~{\em 15\/}(2), 284--295.

\bibitem[\protect\citeauthoryear{Lin and Zeng}{Lin and
  Zeng}{2010}]{lin2010relative}
Lin, D. and D.~Zeng (2010).
\newblock On the relative efficiency of using summary statistics versus
  individual-level data in meta-analysis.
\newblock {\em Biometrika\/}~{\em 97\/}(2), 321--332.

\bibitem[\protect\citeauthoryear{Liptak}{Liptak}{1958}]{liptak1958combination}
Liptak, T. (1958).
\newblock On the combination of independent tests.
\newblock {\em Magyar Tud Akad Mat Kutato Int Kozl\/}~{\em 3}, 171--197.

\bibitem[\protect\citeauthoryear{Littell and Folks}{Littell and
  Folks}{1973}]{littell1973asymptotic}
Littell, R.~C. and J.~L. Folks (1973).
\newblock Asymptotic optimality of {F}isher's method of combining independent
  tests {II}.
\newblock {\em Journal of the American Statistical Association\/}~{\em
  68\/}(341), 193--194.

\bibitem[\protect\citeauthoryear{Liu and Xie}{Liu and
  Xie}{2018}]{liu2018cauchy}
Liu, Y. and J.~Xie (2018).
\newblock Cauchy combination test: a powerful test with analytic p-value
  calculation under arbitrary dependency structures.
\newblock {\em Journal of the American Statistical
  Association\/}~(just-accepted), 1--29.

\bibitem[\protect\citeauthoryear{M{\"a}gi and Morris}{M{\"a}gi and
  Morris}{2010}]{magi2010gwama}
M{\"a}gi, R. and A.~P. Morris (2010).
\newblock Gwama: software for genome-wide association meta-analysis.
\newblock {\em BMC bioinformatics\/}~{\em 11\/}(1), 288.

\bibitem[\protect\citeauthoryear{Mathai}{Mathai}{1982}]{mathai1982storage}
Mathai, A. (1982).
\newblock Storage capacity of a dam with gamma type inputs.
\newblock {\em Annals of the Institute of Statistical Mathematics\/}~{\em
  34\/}(3), 591--597.

\bibitem[\protect\citeauthoryear{Morris, Kemp, Youlten, Laurent, Logan, Chai,
  Vulpescu, Forgetta, Kleinman, Mohanty, et~al.}{Morris
  et~al.}{2019}]{morris2019atlas}
Morris, J.~A., J.~P. Kemp, S.~E. Youlten, L.~Laurent, J.~G. Logan, R.~C. Chai,
  N.~A. Vulpescu, V.~Forgetta, A.~Kleinman, S.~T. Mohanty, et~al. (2019).
\newblock An atlas of genetic influences on osteoporosis in humans and mice.
\newblock {\em Nature genetics\/}~{\em 51\/}(2), 258--266.

\bibitem[\protect\citeauthoryear{Moscuoroums}{Moscuoroums}{1985}]{moscuoroums1985distrlbution}
Moscuoroums, P. (1985).
\newblock The distrlbution of the sum of independent gamma random variables.
\newblock {\em Ann. Inst. Statist. Math\/}~{\em 37\/}(Part A), 541--544.

\bibitem[\protect\citeauthoryear{Nadarajah and Kotz}{Nadarajah and
  Kotz}{2005}]{nadarajah2005mathematical}
Nadarajah, S. and S.~Kotz (2005).
\newblock Mathematical properties of the multivariate t distribution.
\newblock {\em Acta Applicandae Mathematica\/}~{\em 89\/}(1-3), 53--84.

\bibitem[\protect\citeauthoryear{Patel and Read}{Patel and
  Read}{1996}]{patel1996handbook}
Patel, J.~K. and C.~B. Read (1996).
\newblock {\em Handbook of the normal distribution}, Volume 150.
\newblock CRC Press.

\bibitem[\protect\citeauthoryear{Poole, Gibbs, Shmulevich, Bernard, and
  Knijnenburg}{Poole et~al.}{2016}]{poole2016combining}
Poole, W., D.~L. Gibbs, I.~Shmulevich, B.~Bernard, and T.~A. Knijnenburg
  (2016).
\newblock Combining dependent p-values with an empirical adaptation of brown's
  method.
\newblock {\em Bioinformatics\/}~{\em 32\/}(17), i430--i436.

\bibitem[\protect\citeauthoryear{Rago, Willett, and Bar-Shalom}{Rago
  et~al.}{1996}]{rago1996censoring}
Rago, C., P.~Willett, and Y.~Bar-Shalom (1996).
\newblock Censoring sensors: A low-communication-rate scheme for distributed
  detection.
\newblock {\em IEEE Transactions on Aerospace and Electronic Systems\/}~{\em
  32\/}(2), 554--568.

\bibitem[\protect\citeauthoryear{Routledge}{Routledge}{1997}]{routledge1997p}
Routledge, R. (1997).
\newblock P-values from permutation and f-tests.
\newblock {\em Computational Statistics \& Data Analysis\/}~{\em 24\/}(4),
  379--386.

\bibitem[\protect\citeauthoryear{Shlyakhter, Sabeti, and Schaffner}{Shlyakhter
  et~al.}{2014}]{shlyakhter2014cosi2}
Shlyakhter, I., P.~C. Sabeti, and S.~F. Schaffner (2014).
\newblock Cosi2: an efficient simulator of exact and approximate coalescent
  with selection.
\newblock {\em Bioinformatics\/}~{\em 30\/}(23), 3427--3429.

\bibitem[\protect\citeauthoryear{Stacy et~al.}{Stacy
  et~al.}{1962}]{stacy1962generalization}
Stacy, E.~W. et~al. (1962).
\newblock A generalization of the gamma distribution.
\newblock {\em The Annals of mathematical statistics\/}~{\em 33\/}(3),
  1187--1192.

\bibitem[\protect\citeauthoryear{Stouffer, Suchman, DeVinney, Star, and
  Williams}{Stouffer et~al.}{1949}]{Stouffer1949}
Stouffer, S.~A., E.~A. Suchman, L.~C. DeVinney, S.~A. Star, and R.~M. Williams
  (1949).
\newblock {\em The American Soldier: Adjustment during Army Life}, Volume~I.
\newblock New Jersey: Princeton University Press.

\bibitem[\protect\citeauthoryear{Sun and Lin}{Sun and Lin}{2017}]{sun2017set}
Sun, R. and X.~Lin (2017).
\newblock Set-based tests for genetic association using the generalized
  berk-jones statistic.
\newblock {\em arXiv preprint arXiv:1710.02469\/}.

\bibitem[\protect\citeauthoryear{Thadewald and B{\"u}ning}{Thadewald and
  B{\"u}ning}{2007}]{thadewald2007jarque}
Thadewald, T. and H.~B{\"u}ning (2007).
\newblock Jarque--bera test and its competitors for testing normality--a power
  comparison.
\newblock {\em Journal of applied statistics\/}~{\em 34\/}(1), 87--105.

\bibitem[\protect\citeauthoryear{Tseng, Ghosh, and Feingold}{Tseng
  et~al.}{2012}]{tseng2012comprehensive}
Tseng, G.~C., D.~Ghosh, and E.~Feingold (2012).
\newblock Comprehensive literature review and statistical considerations for
  microarray meta-analysis.
\newblock {\em Nucleic acids research\/}~{\em 40\/}(9), 3785--3799.

\bibitem[\protect\citeauthoryear{Tyekucheva, Marchionni, Karchin, and
  Parmigiani}{Tyekucheva et~al.}{2011}]{tyekucheva2011integrating}
Tyekucheva, S., L.~Marchionni, R.~Karchin, and G.~Parmigiani (2011).
\newblock Integrating diverse genomic data using gene sets.
\newblock {\em Genome biology\/}~{\em 12\/}(10), R105.

\bibitem[\protect\citeauthoryear{Utz, Lawson, Misra, Mickley, Gleysteen,
  Herzog, Klibanski, and Miller}{Utz et~al.}{2008}]{utz2008peptide}
Utz, A.~L., E.~A. Lawson, M.~Misra, D.~Mickley, S.~Gleysteen, D.~B. Herzog,
  A.~Klibanski, and K.~K. Miller (2008).
\newblock Peptide yy (pyy) levels and bone mineral density (bmd) in women with
  anorexia nervosa.
\newblock {\em Bone\/}~{\em 43\/}(1), 135--139.

\bibitem[\protect\citeauthoryear{Vargo, Pasupathy, and Leemis}{Vargo
  et~al.}{2010}]{vargo2010moment}
Vargo, E., R.~Pasupathy, and L.~Leemis (2010).
\newblock Moment-ratio diagrams for univariate distributions.
\newblock {\em Journal of Quality Technology\/}~{\em 42\/}(3), 276--286.

\bibitem[\protect\citeauthoryear{Wu, Lee, Cai, Li, Boehnke, and Lin}{Wu
  et~al.}{2011}]{wu2011rarevariant}
Wu, M.~C., S.~Lee, T.~Cai, Y.~Li, M.~Boehnke, and X.~Lin (2011, Jul 15).
\newblock Rare-variant association testing for sequencing data with the
  sequence kernel association test.
\newblock {\em American Journal of Human Genetics\/}~{\em 89\/}(1), 82--93.

\bibitem[\protect\citeauthoryear{Yang}{Yang}{2010}]{yang2010distribution}
Yang, J.~J. (2010).
\newblock Distribution of fisher's combination statistic when the tests are
  dependent.
\newblock {\em Journal of Statistical Computation and Simulation\/}~{\em
  80\/}(1), 1--12.

\bibitem[\protect\citeauthoryear{Yang, Li, Williams, and Buu}{Yang
  et~al.}{2016}]{yang2016efficient}
Yang, J.~J., J.~Li, L.~K. Williams, and A.~Buu (2016).
\newblock An efficient genome-wide association test for multivariate phenotypes
  based on the fisher combination function.
\newblock {\em BMC bioinformatics\/}~{\em 17\/}(1), 19.

\bibitem[\protect\citeauthoryear{Zaykin, Zhivotovsky, Czika, Shao, and
  Wolfinger}{Zaykin et~al.}{2007}]{zaykin2007combining}
Zaykin, D.~V., L.~A. Zhivotovsky, W.~Czika, S.~Shao, and R.~D. Wolfinger
  (2007).
\newblock Combining {\it p}-values in large-scale genomics experiments.
\newblock {\em Pharmaceutical Statistics\/}~{\em 6\/}(3), 217--226.

\bibitem[\protect\citeauthoryear{Zhang, Tong, Landers, and Wu}{Zhang
  et~al.}{2019}]{hZhang2019TFisher}
Zhang, H., T.~Tong, J.~E. Landers, and Z.~Wu (2019).
\newblock Tfisher: A powerful truncation and weighting procedure for combining
  $p$-values.
\newblock {\em Annals of Applied Statistics\/}.

\bibitem[\protect\citeauthoryear{Zhang and Wu}{Zhang and
  Wu}{2018}]{zhang2018generalized}
Zhang, H. and Z.~Wu (2018).
\newblock Generalized goodness-of-fit tests for correlated data.
\newblock {\em arXiv preprint arXiv:1806.03668\/}.

\bibitem[\protect\citeauthoryear{Zheng, Forgetta, Hsu, Estrada, Rosello-Diez,
  Leo, Dahia, Park-Min, Tobias, Kooperberg, et~al.}{Zheng
  et~al.}{2015}]{zheng2015whole}
Zheng, H.-F., V.~Forgetta, Y.-H. Hsu, K.~Estrada, A.~Rosello-Diez, P.~J. Leo,
  C.~L. Dahia, K.~H. Park-Min, J.~H. Tobias, C.~Kooperberg, et~al. (2015).
\newblock Whole-genome sequencing identifies en1 as a determinant of bone
  density and fracture.
\newblock {\em Nature\/}~{\em 526\/}(7571), 112--117.

\end{thebibliography}

\end{document}